# ATLAS OF SECULAR LIGHT CURVES OF COMETS", V.2009
## Ignacio Ferrín

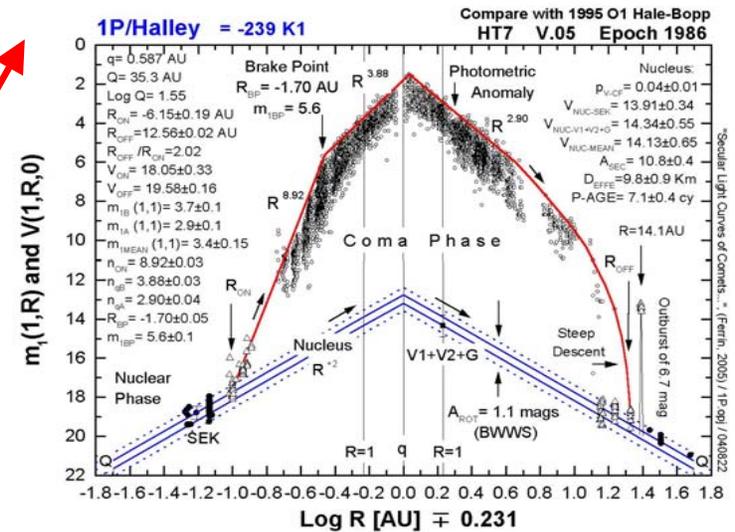

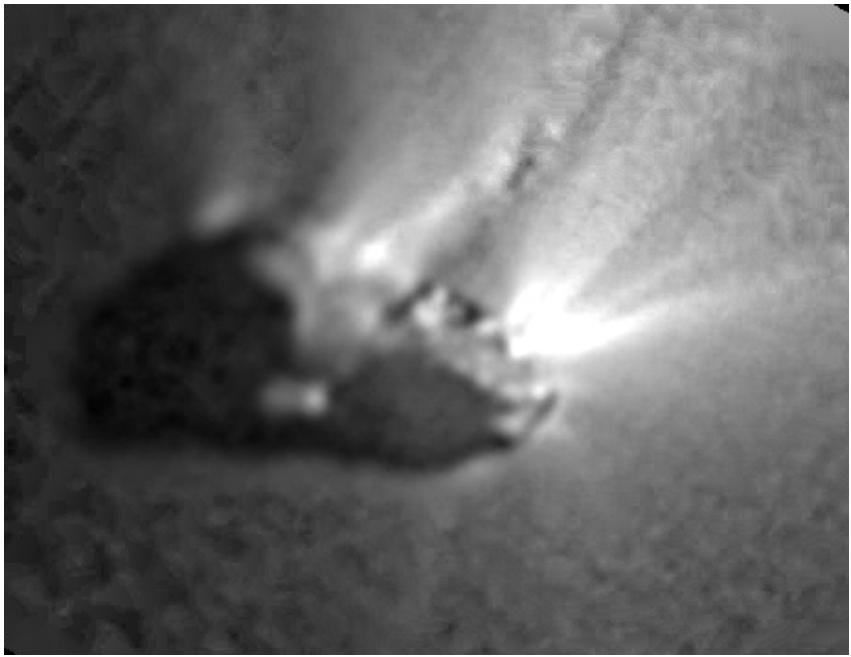

1986, March 14th. Comet 1P/Halley imaged by the Giotto Spacecraft. Photo courtesy H. Uwe Keller, Max-Planck Institut für Aeronomie, Lindau, FRG.

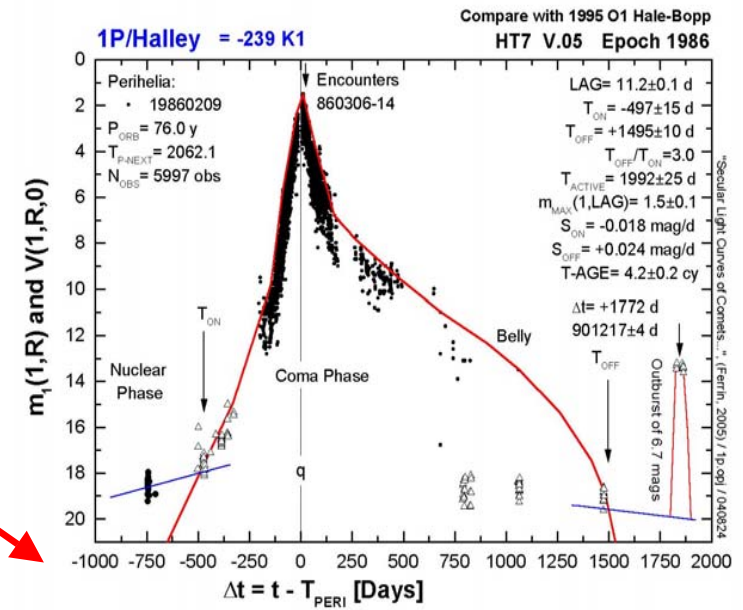

C:\COMETAS0901
/SLC-Atlas090909
/AtlasAceptado0908
/AtlasWeb
/30%AtlasWEBV15090909.doc
This work has gone through 15 revisions

# TABLE OF CONTENTS

**Comets Presented in Version 2009:**
Comet, Epoch
1P/Halley, 1986
2P/Encke, 2003, 1858
6P/D'Arrest, 1995
9P/Tempel 1, 1994
19P/Borrelly, 1994
21P/Giacobinni-Zinner, 1998
26P/Grigg-Skejellerup, 1987
28P/Neujmin 1. 1913
29P/Schwassmann-Wachmann 1, 1925-2004
32P/Comas Sola, 1996
39P/Oterma, 1958
45P/Honda-Mrkos-Pajdusakova, 1990
65P/Gunn, 1978
67P/Churyumov-Gerasimenko, 1996
73P/Schwassmann-Wachmann 3C, 1990-1995, 2001
81P/Wild 2, 1997
85P/Boethin, 1986
101P/Chernyk, 1978
107P/Wilson-Harrington, 1996
109P/Swift-Tuttle, 1992
133P/Elst-Pizarro, 1996
C/1983 H1 Iras-Araki-Alcock, 1983
C/1983 J1 Suganu-Saigusa-Fujikawa, 1983
C/1984 K1 Shoemaker, 1985
C/1995 O1 Hale-Bopp, 1997
C/1996 B2 Hyakutake, 1996
C/2002 OG108 LONEOS, 2003

**Total = 27 Comets, 54 SLCs, 70 Plots**

# "Atlas of Secular Light Curves of Comets" *

Ignacio Ferrín*,
Center for Fundamental Physics,
University of the Andes,
Apartado 700
Mérida 5101-A, Venezuela

Number of Pages:     109
Number of Figures:    70
Number of Tables:      4

*  email-address: ferrin@ula.ve

# ABSTRACT

In this work we have compiled 37,692 observations of 27 periodic and non-periodic comets to create the secular light curves (SLCs), using 2 plots per comet. The data has been reduced homogeneously. Our overriding goal is to learn the properties of the ensemble of comets. More than 30 parameters are listed, of which over ~20 are new and measured from the plots. We define two ages for a comet using activity as a proxy, the photometric age P-AGE, and the time-age, T-AGE. It is shown that these two parameters are robust, implying that the input data can have significant errors but P-AGE and T-AGE come out with small errors. This is due to their mathematical definition. It is shown that P-AGE classifies comets by shape of their light curve. The value of this Atlas is twofold: The SLCs not only show what we know, but also show what we do not know, thus pointing the way to meaningful observations. Besides their scientific value, these plots are useful for planning observations. The SLCs have not been modeled, and there is no cometary light curve standard model as there is for some variable stars (i.e. eclipsing binaries). Comets are classified by age and size. In this way it is found that 29P/Schwassmann-Wachmann 1 is a baby goliath comet, while C/1983 J1 Sugano-Saigusa-Fujikawa is a middle age dwarf. There are new classes of comets based on their photometric properties. The secular light curves presented in this Atlas exhibit complexity beyond current understanding.

# 1  INTRODUCTION

The comets that populate the solar system (SS) represents a fossil record of the formation of our planetary system, and are valuable tracers of our SS structure and dynamics.  Comets are interesting astronomical objects not only from the point of view of their physical properties, but also for their contribution to the interplanetary medium, to our own planet through collisions, and the possibility of having been carriers of water and/or primordial life. Thus the study, analysis and correlations of their properties, including their initial and end states, can yield fundamental clues needed to understand the history of the formation of our SS and their origin and evolution in the Sun's gravitational and radiation field.

The system of comets may look distant, faint, cold and uneventful, lurking into the emptiness and darkness of space.  However the picture that emerges after looking at the secular light curves (SLCs) presented in this work, is that they are active, complex, eventful, predictable at times, unpredictable at other times, inhomogeneous, and each one has a different personality tied to its composition, age and physical properties.

The first value of this Atlas is twofold: *The SLCs not only show what we know, but also show what we do not know, thus pointing the way to meaningful observations.*  Thus besides their scientific value, the SLCs are useful for planning observations.

The second value of this Atlas is that it combines nuclear and whole coma magnitudes, thus creating the photometric history of the object.  Whole coma magnitudes are determined mostly by amateurs, while faint nuclear magnitudes are measured mostly by professional astronomers.  Thus the SLCs successfully combine the work of professionals and amateurs.

The third value of this Atlas is that each apparition is identified by an Epoch label.  The importance of this label is that future apparitions will be plotted in a new Epoch plot, to be compared with the former one.  After many apparitions, a movie of the secular light curve could be built with the individual plots, showing evolutionary changes and the photometric history of the comet in two phase spaces.  At the present time only comet 2P has two frames of the movie, in 1858 and 2003.  Thus, *each plot of this Atlas is a frame of a movie.*

It is an illuminating experience to print out the plots and to lay large copies side by side to create a mega-plot, and this exercise is enhanced if they are organized by increasing photometric age, P-AGE (defined in section 4.1-12).  We can then see the richness of shapes, the information and the complexity that the secular light curves exhibit.  It is important to realize that the variety of these shapes does not have as of now a theoretical explanation.  In this regard we are much farther behind than the corresponding field of variable stars where many of them already have complex theoretical models capable of explaining and extracting numerous physical parameters of the stars, solely from their light curves. Just consider the amount of information derived from the light curves of RR Lyrae, Cepheids and eclipsing binary stars. It is hoped that the present work will help to achieve this goal for comets.

Modeling these SLCs with their different shapes, is a challenge. The observed SLCs are the sum of light given off by gas and dust, thus really needing two models. Comparing with the field of asteroidal research, we can say that currently *there is no standard thermal model* of cometary secular light curves. The SLCs are so precisely known in some cases, that there is very little margin of error for fitting theoretical parameters. The ones that come to mind are visual geometric albedo, infrared albedo, bond albedo, and thermal conductivity, thickness of the crust layer, pole orientation (obliquity of the nucleus and longitude of the pole), ice composition, and stratigraphic distribution of ices.

We will present the observational evidence with a minimum of interpretation. Statistical analysis, correlations and ensemble properties will be presented later on. Otherwise this work would be prohibitively lengthy.

An explanation of the objectives and methodology of this project has previously been given in Papers I, II, III, IV and V (Ferrín 2005a, 2005b, 2006, 2007, 2008). The secular light curves previously published are included in the present compilation for reasons of completeness and because some of them have been updated and/or corrected.

In these series of papers we have tried to include as many as possible observations available in the literature. The Version.Year plots presented in this Atlas provide fundamental information on the object, such as the distances of turn on and turn off ($R_{ON}$, $R_{OFF}$) (defined in section 4.1-1 and 4.1-2), the lag at perihelion (LAG) (defined in section 4.2-19), the photometric age (P-AGE) (defined in section 4.1-12), the amplitude of the SLC, $A_{SEC}$ (defined in section 4.1-10), and the general shape of the SLC. In the present compilation, 1) there are enough objects to start a statistical analysis of the comet population. Ensemble properties can be assessed. 2) We get a bird's eye view of the shape of the SLCs.

Kamel's (1992) compilation on photometric data of comets included plots of their light curves. However he did not derive any photometric parameters from his plots and the nuclear line is not plotted either, making his compilation of limited value. Additionally he chose to present his plots in one single phase space.

In this work we have compiled 37.692 observations of 27 periodic and non-periodic comets, to create 54 secular light curves, using two plots per comet.

## 2  THE PLOTS

The magnitude at $\Delta$, R, $\alpha$, is denoted by $m_1(\Delta,R)$ for visual observations and $V(\Delta,R,\alpha)$ for instrumental magnitudes, where $\Delta$ is the comet-Earth distance, R is the Sun-comet distance, and $\alpha$ is the phase angle (Earth-comet-Sun). The information on the brightness of the comet is presented in two plots, the Log R plot and the time plot. The reason to select the two plots is because they give independent and different parameters. The Log

R plot may be reflected at R= 1 AU or may be reflected at q, the perihelion distance:

**1)** The 'reflected double-Log R plot' presents the reduced magnitude vs Log R reflected at R= 1 AU. Reduced means reduced to $\Delta$ = 1 AU, m(1,R)= m($\Delta$,R) - 5 Log $\Delta$. In this plot time runs horizontally from left to right, although non linearly. Negative Logs indicate observations pre-perihelion. *The value of the reflected double-Log R diagram is that power laws on R (~$R^{+n}$) plot as straight lines.* The slope 5 line at the bottom of the plot in the form of a pyramid is due to the atmosphereless nucleus. Additionally the inclusion of the R= 1 AU line allows the determination of the absolute magnitude by extrapolation (actually by interpolation of the pre and post-perihelion intervals).

It will be shown that the determination of the absolute magnitude of the comet is not always a simple matter as has been assumed in the past. Thus previous determinations of absolute magnitudes should be revised, especially if they are based on the H10 system (see section 3 below on photometry).

**2)** The 'time plot' presents the reduced magnitude vs time to perihelion. This is the most basic and simple plot. Negative times are pre-perihelion. The advantage of this plot is that time runs horizontally linearly from left to right thus showing the time history of the object.

**3)** The third type of plot is a 'reflected at q, double-Log R plot'. Comets with extreme LAG values only plot well in the reflected at q diagram. As a consequence it is not a simple matter to determine the absolute magnitude m(1,1). Comet 101P had to be presented with this plot because of its extreme LAG value that distorts the normal Log R plot reflected at R= 1 AU. Also, comets with q < 1 AU have to be presented with this type of plot to make space for observations inside the Earth's orbit.

The three plots give an unprecedented amount of information on the population of comets, and allow the elaboration of statistics. Much of this information was not previously available.

The fit to the envelope (defined in Section 3.2) and the determination of parameters of the secular light curves in general, are the author's best assessment of the situation. However the reader might choose to apply his own fittings and assessments or re-determine some values of parameters or plot his own observations. This author's interpretation of the SLCs can be found in Section 7.

## 3  PHOTOMETRIC SYSTEM

In this section we will describe the methods and mathematical functions and procedures that we have adopted to analyze these data.

### 3.1  Old H10 photometric system

In the old literature the brightness of a comet was defined using the law

$$H = H0 + 5 \text{ Log } \Delta + 2.5 \text{ n Log R} \qquad (1)$$

where n is a parameter that describes the dependence of the brightness with heliocentric distance, and H0 is the absolute magnitude at $\Delta$ = R = 1. The old photometric system H10 assumed n = 4 (Vsekhsvyatskii 1964).

From the shapes of the secular light curves presented in this work, it is clear that most comets do not follow power laws, and the 6 that do differ from the n = 4 value used in the H10 system (see Table 1). The other comets exhibit complex secular light curves that have to be defined by more complex laws or even polynomials. In summary we have to conclude that the H10 system is unable to explain the observed secular light curves. Thus all previous values based on this system should be revised.

## 3.2 Envelope

A visual observation of a comet is affected by several effects, all of which decrease the perceived brightness of the object: moon light, twilight, cirrus clouds, dirty optics (dust on the mirror), dirty atmosphere (pollution), low altitude (haze), excess magnification, the Delta-effect, etc. All of these wash out the outer regions of the comet, decreasing its intensity and thus its magnitude. The Delta-effect (Kamel 1997) is produced when a comet gets very near to the Earth. Then our planet may be immersed in the outer regions of the coma. The coma extends out so much in space, that it is impossible to derive a reliable total magnitude. Again some light is left out, decreasing the observed magnitude. An example in modern times of this happening may have been comet C/1983 H1 IRAS-Araki-Alcock. *Thus it is a fundamental premise of this work that the envelope of the observations defines the secular light curve, since there is no known physical effect that could increase the perceived brightness of a comet measured by two different observers, at the same instant of time.* It is significant to point out that even CCD observation may suffer from this complication. If the exposure is not deep enough, the outer parts of the coma will not be recorded, again resulting in an incomplete magnitude.

If as in the past the mean values were used, they would create a significant systematic error of 2-4 magnitudes. Thus previously published absolute magnitude determinations have to be revised *upwards*.

To define the envelopes we followed the procedure suggested by Sosa and Fernandez (2009). We define 10-20 days time bins (or $\Delta$Log = 0.1-0.2 in the log R plot), and select data points in the 95% upper percentile. These selected data points are fitted by least squares with a polynomial of degree 1 to 4. The fitted polynomials define the envelope.

Kamel's (1992) compilation on photometric data of comets included numerical corrections for several effects. In a review of Kamel's work, Green (1991) cites the opinion of Morris on some of his reduction procedures, and concludes that some observations had corrections of 2-6 magnitudes. There is little scientific value in a correction of 6 magnitudes. Since our target error is ~0.1 magnitudes, we decided to use only uncorrected magnitudes compiled by Kamel, and let the brightest observations define the envelope. Corrections of the data are not needed if several apparitions are combined. In a given apparition the object may be too low in the sky thus making all the observations systematically low, and thus requiring an uncertain correction. These same circumstances are unlikely to be repeated in the next apparition, and the new observations will likely be above the old ones defining a nice envelope of the points. Most of the light curves presented in this work are the combination of several apparitions and thus do not require corrections. However as comets get older they get less active so it is also possible that they will not be brighter even at a higher air mass.

### 3.3 Conversion from $m_1$ to V

$m_1$ observations are mostly taken by amateurs using the naked eye and results are published in the ICQ (International Comet Quarterly, Green 2007), while V magnitudes are measured by professional astronomers using CCDs and published in other journals. Ferrín (2005b) has studied this problem and found that the mean conversion from $m_1$ to V for three comets (1P/Halley, 2P/Encke and C/1973 XII Kohoutek) is V = $m_1$ - 0.026±0.008 mag. This value is so small that we will take the two systems as identical.

### 3.4 Conversion from R to V

Corrections due to band passes are still needed, however. Many nuclear magnitudes are measured in R photometric system, and they must be reduced to the V system. In Paper II we compiled a table of V, R and I magnitudes for several comets and we found that the mean <V-R> correction is <V-R> = 0.48 ± 0.13 mag (mean of 77 observations). In this work the mean value of variable x will be denoted by <x>.

### 3.5 Infinite Aperture Magnitudes, IAM

Many observed comets exhibit a prominent coma, so no nuclear magnitude can be derived. Instead, *infinite aperture magnitudes* have to be measured (Ferrín 2005b).

It has been known for some time (Green 1997) that there is a lack of flux in many CCD observations. The reason for this lack of flux was not always clear. Fink et al. (1999) have considered this problem and have found an explanation: *'It is rather well known that visual observations in general, are several magnitudes brighter than CCD measurements and reconciling CCD visual magnitudes is 'more of an art than a science'. Figure 4* [of their work] *shows why this is the case. As we increase the aperture size on our CCD images we get brighter magnitudes. It is evident by the progression of magnitudes for increasing aperture size that if we integrate farther out we would readily reach the magnitudes reported by visual observers. The eye is essentially a logarithmic detector and has a larger dynamic range than a CCD so that it effectively includes more of the coma than an individual non-saturated CCD exposure'.*

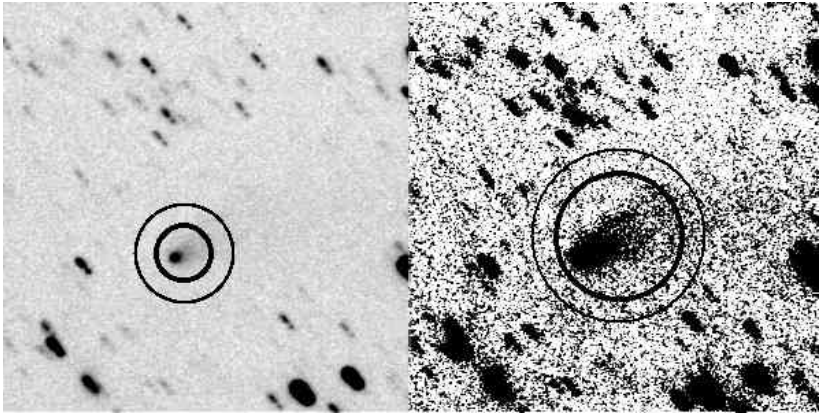

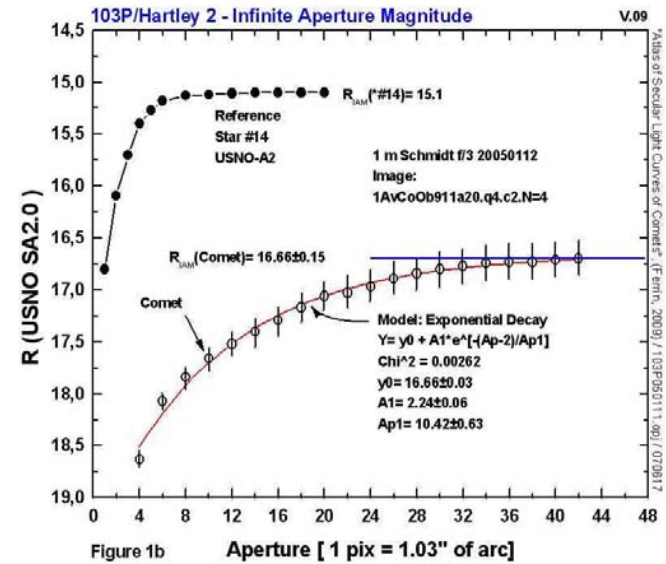

Figure 1a. Comet 103P/Hartley 2 imaged with the 1 m Schmidt telescope of the National Observatory of Venezuela, at f/3, on 2005, January 12$^{th}$. Since large photometric apertures have to be used to extract the whole flux, the image has previously been cleaned of nearby stars using a cloning tool. Left: A normal stretching of the image shows that an aperture of radius 20 pixels is apparently enough to extract a total magnitude. Stretching refers to the maximum and minimum pixel intensity to display the image in the computer monitor. The standard stretching is selected by the computer, who does not know about preserving the flux. Thus the monitor displays a deceivingly faint image. Right: A forced stretching reveals a much larger tail than expected. At least a 44 pixel radius is needed to extract a total magnitude. This has been called the insufficient CCD aperture error (Ferrín, 2005b), and manifest itself by producing measurements that lie much below the envelope of the observations. The comet had a magnitude of R= 16.66±0.15, so it is easy to conclude that the effect is much larger for brighter comets.

Figure 1b. Derivation of an infinite aperture magnitude for Figure 1a.. It can be seen how the calibration star and the comet increase in flux with increasing photometric apertures. The asymptotic value is the infinite aperture magnitude. Notice the huge aperture needed to extract the whole flux of the comet, much larger than typically used apertures. Notice also how the calibration star #14 converges rapidly to its infinite aperture magnitude but still needs a 8 pixel radius to retrieve an IAM. If a comet of magnitude 16 needs an aperture of 2 arc minutes in diameter to extract the whole flux, then a brighter comet will need a much larger aperture. Usually an exponential decay fits the data. USNO-A2 is the US Naval Observatory Astrometric Catalog 2.

To avoid this *insufficient CCD aperture error* (Ferrín 2005b) we have to measure all the light from the comet. The problem can be illustrated looking at Figure 1a where we show an image of comet 103P/Hartley 2 with a normal stretching of the image, and with a forced stretching. Since large photometric apertures have to be used to extract the whole flux, the image has previously been cleaned of nearby stars, using a background cloning tool.

Stretching refers to the maximum and minimum pixel intensity to display the image in the computer monitor. The normal stretching is selected by the machine, who does not know about preserving the flux. The computer monitor displays a *deceivingly faint image*. It turns out that a forced stretching reveals a much larger image than expected.

Figure 1a reveals that without proper image enhancement it is easy to use photometric CCD apertures that do not extract the whole flux from the comet producing measurements that lie much below the envelope of the observations.

Of course the optimal extraction aperture increases with the brightness of the comet. We see that a comet of magnitude 16.6 requires an aperture of 2 arc minutes in diameter. Ferrín (2005b) shows that a comet of magnitude 13.3 requires an aperture of 5 arc minutes in diameter. Brighter comets will require even larger apertures but this is distance dependent. Unfortunately, many published photometry use apertures in the range of arc seconds. Many more examples of this error can be found at the web site cited at the end of this paper.

In Figure 1b we show the method we used to extract an infinite aperture magnitude, IAM, from now on denoted as $R_{IAM}$, $V_{IAM}$. The flux is measured with increasing CCD photometric apertures, and *the asymptotic value at infinity is the infinite aperture magnitude*. Usually an exponential decay fits the data.

Notice how the calibration star #14 converges very rapidly to an infinite aperture magnitude. Not so the comet, that requires an aperture with more than 40 pixels in radius to extract a total flux. Since the scale of this telescope is ~1"/pixel, 40 pixels represent 80" of diameter = 1.33 minutes of diameter, clearly much larger than the usual apertures published in the literature.

Infinite aperture magnitudes should be adopted whenever the comet exhibits a coma. IAM are used in several comets of this Atlas and should be adopted by all observers from now on to secure extracting the whole flux of the comet.

### 3.6 Reduction procedures

In the reduction we must distinguish between the coma observations, and the nucleus observations. The observed visual magnitude of a comet with coma, $m_1(\Delta,R)$ is given by

$$m_1(\Delta,R) = m(1,1) + 5 \log \Delta + 2.5 \, n \log R \qquad (2)$$

where m(1,1) is the absolute magnitude of the comet at $\Delta$ = 1 AU and R = 1 AU. The $\Delta$ correction is a purely geometric parameter due solely to the changing Earth-comet distance that must be subtracted.

$$m_1(1,R) = m_1(\Delta,R) - 5 \Log \Delta \qquad (3)$$

For nuclear observations an additional term is needed because we are then seeing the bare nucleus under different phase angles:

$$V_N(1,R,\alpha) = V_N(1,1,0) + 5 \Log \Delta + 5 \Log R + \beta.\alpha \qquad (4)$$

where $V_N(1,1,0)$ is the absolute nuclear magnitude, $\beta$ is the phase coefficient, and $\alpha$ the phase angle (Earth-comet-Sun). Then

$$V_N(1,R,0) = V_N(1,1,0) - 5 \Log \Delta - \beta.\alpha \qquad (5)$$

$m_1(1,R)$ and $V_N(1,R,0)$ are plotted vs heliocentric distance, R, in the Log R plot, and vs time in the time plot.

### 3.7 Rotational error

Most comets have highly elongated nuclei and thus have rotational light curves of large amplitude, $A_{ROT}$, reaching to 1.1 magnitudes in the case of comet 1P/Halley (Belton et al. 1986) which may represent an extreme. Thus snapshot observations are affected by the 'rotational error'. We have assigned to snapshot observations a mean error of ±0.63 magnitudes to take into account this effect. ±0.63 mag represents the 68 percentile of 32 rotational light curve amplitudes presented in the web site given at the end of this paper.

### 3.8 Photographic measurements

We do not use photographic measurements, and some remaining from Kamel's (1992) compilation have been deleted. The only exceptions are when there are not enough visual observations (28P/Neujmin, Epoch 1913 and 39P/Oterma, Epoch 1958). The reason for not using photographic magnitudes is that they are contaminated by the CN line at 3883 Å which is the most intense line of the whole cometary spectrum.

### 3.9 Phase curve, phase coefficient, and absolute nucleus dimension

The best way to derive the absolute nuclear magnitude of a comet is by plotting the phase diagram, $V(1,1,\alpha)$ vs $\alpha$. The slope of this curve gives the phase coefficient, $\beta$, which is needed to reduce nuclear observations to a common phase angle of 0°, and the interception of this line with the y-axis gives $V_N(1,1,0)$.

Some researchers still use a $\beta = 0.035$ mag/°, which is an old estimate not supported by current values of $<\beta> = 0.046$ mag/° (Paper I).

### 3.10 Data sets

We will use the data set of Kamel (1992) without corrections for observations prior to 1990, and the ICQ (Green 2007) for observations after that date. The bulk of data comes from these two authors, but many other visual and CCD observations can be found in web sites scattered all around the world. A few observations are by this author and are denoted by TW (This Work). Faint and nuclear observations are compiled from the literature. Many scientific references give nuclear diameters for comets. These values can be converted to absolute magnitude using equations (11) and (12) and assuming a geometric albedo $p_V = 0.04$. Particularly useful are the compilations of magnitudes and/or diameters of Chen & Jewitt (1994), Scotti (1995), Lowry et al. (1999), Tancredi et al. (2000, 2006), Neslusan (2003), Lowry & Weissman (2003), Lowry et al. (2003), Meech et al. (2004), Lamy et al. (2005), Lowry & Fitzsimmons (2005), Hicks et al. (2007).

Objects that have been denoted as asteroids (107P, 133P) may have information in the Asteroid Dynamic Site ( http://hamilton.dm.unipi.it/astdys/).

### 3.11  Data prior to 1950

Tom Gehrels (1999) had something relevant to say about former asteroidal photometry (equally valid for cometary photometry): *'Before the 50s, asteroid magnitudes were not usable for statistics. They would be off by as much as 3 magnitudes: fainter than the 11$^{th}$ magnitude, they tended to be all the same because the Bonner Durchmusterung had been used for calibration and it did not go fainter'*. Consequently, it is not advisable to use faint observations before the 50s. However there are several cases in which we do not have any other choice.

### 3.12  Absolute calibration

P-AGE and T-AGE (discussed in Section 4) are given in comet years to emphasize that they are not calibrated in Earth's years. Thus there is a need for an absolute calibration. At the present time it is possible to calibrate only comet 2P/Encke in current years (Paper V), using the fact that its SLC has been obtained for the Epochs 1858 and 2003. A plot of Earth's years vs P-AGE and T-AGE using a linear extrapolation shows that the comet had P-AGE= 0 and T-AGE= 0 in 1645±40 AD. This is not the birth date of the comet but the 'effective' date at which the comet began sublimating. However the interpretation and implications of this zero-age-date is much more complex that can be analyzed here and is beyond the scope of this paper. The important point to emphasize is that an absolute calibration *has been achieved*.

## 4  PARAMETERS MEASURED FROM THE PLOTS

The Log R and the time plots provide a wealth of new information. There are ~40 parameters listed, of which ~20 are new and measured from the plots (Figures 2a and 2b).

Figure 2a. Key to the Log R plot. The parameters are explained in Section 4.1.

15## 4.1 Log R Plots

The symbol legend for the log R plot can be found on the time plot. The title of each plot identifies the comet in the new and old nomenclature system to allow going back to historical references. The first designation of the comet gives the first apparition (the discovery year). The label at the top right indicates if the comet belongs to the Jupiter family (JF), to the Saturn family (SF), to the Oort Cloud (OC), to the Asteroidal Belt (ABC), to the Encke type (ET), if it is a Centaur (CEN) or if it belongs to the Halley type (HT). This classification follows closely the one given by Levison (1996), except that Chiron type and Centaur are synonymous, ABC is included in his ecliptic comets and finally he did not consider the existence of Saturn family comets (28P, 85P). The two numbers following the family are the photometric age, P-AGE, measured in units of comet years (cy) and the diameter measured in km. The reason for using P-AGE as a label here is that the definition of P-AGE is robust, as will be demonstrated later on. When reading these numbers it is important to keep in mind that ~95% of comets have 0 < P-AGE < 100 cy and 0 < $D_{EFFE}$ < 10 km. Any object above 100/10 or near 0/0 is exceptional.

Next is the Version of the plot. Although most plots have gone through many versions in the course of arriving at a final solution (usually more than 10 versions), all plots are identified as Version 1 (if they are preliminary) or V.09 (the year data completion). The upper left hand side of each plot gives the perihelion distance, q, the aphelion distance for that epoch, Q (where Q= q (1+e) / (1-e) where e is the eccentricity), and Log Q to identify the extent of the plot. Q and q can be located at the bottom of the plots and come in units of AU. At the extreme lower right of the plots, the date line tilted 90º is the last update of observations.

The Epoch label identifies the apparition that has contributed most significantly to the definition of the envelope. The following parameters are deduced from the Log R plots. See Figure 2a for a graphical explanation of these parameters.

**1) $R_{ON}$ [AU].** The turn-ondistance of the coma. The negative sign in this parameter in $R_{ON}$ and in Log R in the plots indicates values before perihelion but it does not mean that the log is negative. Physically $R_{ON}$ corresponds to the onset of *steady* activity. It is the interception of the nuclear line and the coma envelope. Browsing the secular light curves it can be seen that the turn-on and turn-off points are very sudden affairs. When there are enough data points these parameters can be measured easily and accurately because of the sharp change of slope. Usually $R_{ON}$ takes place *before* perihelion, but for comet 133P/Elst-Pizarro takes place *after* perihelion.

**2) $R_{OFF}$.** The turn-off distance of the coma, usually larger than $R_{ON}$. This is the interception of the coma envelope and the nuclear line post-perihelion. $R_{OFF}$ takes place *after* perihelion.

**3) $R_{SUM} = -R_{ON} + R_{OFF}$.** This is equivalent to the $T_{ACTIVE} = -T_{ON} + T_{OFF}$ parameter defined in the time domain. It is a measure of activity.

**4) $V_{ON}$.** The magnitude at which the nucleus turns on.

**5) $V_{OFF}$.** The magnitude at which the nucleus turns off.

**6) $-R_{OFF} / R_{ON}$.** An asymmetry parameter for the Log R plot. The mean value for the current dataset is $< -R_{OFF} / R_{ON} > = 1.22 \pm 0.34$.

**7) $m_{1B}(1,1)$.** The absolute magnitude of the coma, measured *before* perihelion by extrapolation to Log R = 0.

**8) $m_{1A}(1,1)$,** the absolute magnitude of the coma *after* perihelion. This parameter is needed for comets with q < 1 AU. When the secular light curve is highly asymmetric, the absolute magnitude after perihelion is different from the value before perihelion. In this work the mean value of variable x will be denoted by <x>. Thus $<m_1(1,1)>$ = the mean value of the absolute magnitude before and after perihelion, is used frequently.

**9) $V_N(1,1,0) = V_{NUC}$.** Absolute nuclear magnitude measured by different authors. The authors are listed in the individual comments of each comet.

**10a) $A_{SEC}(1,1) = V_N(1,1,0) - m_1(1,1)$** = amplitude of the secular light curve above the nuclear magnitude measured at 1 AU from the Earth, and 1 AU from the sun. $A_{SEC}(1,1)$ is a measure of the activity of the nucleus and thus of age. The value is calculated at (1,1) to allow comparison to different comets. Do not confuse with $A_{ROT}$ the amplitude of the *rotational* light curve. For comets with q < 1 AU, there are two values, $A_{SEC}(1,-1)$ and $A_{SEC}(1,+1)$. The listed value is the mean of the pre and post-values.

**10b) $A_{SEC}(1,q) = V_N(1,q,0) - m_1(1,q)$** = amplitude of the secular light curve above the nuclear magnitude measured at 1 AU from the Earth, and at the perihelion distance from the sun. Since most comets do not reach to q = 1 AU, $A_{SEC}(1,1)$ is different from $A_{SEC}(1,q)$. For comparison with the definition of the absolute magnitude m(1,1), we could say that *$A_{SEC}(1,1)$ is the absolute amplitude of the secular light curve*. In this sense, absolute means that it is calculated at (1,1) to allow comparison with other comets that also do not reach to (1,1). Then we could say that *$A_{SEC}(1,q)$ is the current amplitude of the secular light curve.*

**11) $D_{EFFE}$,** the effective diameter of the comet. It is common to see the effective diameter defined in the literature as $D = (a.b.c)^{1/3}$. However the mean nuclear magnitude of a nucleus of semi-axis $a \geq b \geq c$ is defined in terms of the *mean V value* of the *rotational* light curve, thus it is important to see to what diameter this magnitude corresponds:

$$V_{MAX} = C + 2.5 \text{ Log } a.c \qquad (6)$$

$$V_{MIN} = C + 2.5 \text{ Log } b.c \qquad (7)$$

$$V(1,1,0) = V_{NUC} = (V_{MAX} + V_{MIN})/2 = C + 2.5 \text{ Log } (a.b.c^2)^{1/2}$$

$D_{EFFE} = 2 \ (a.b.c^2)^{1/4}$ \hfill (8)

where C is the zero point constant. Notice the weight of semi-axis c which is usually poorly determined and thus assumed equal to b. Then

$D_{EFFE} \sim 2 \ (a.b^3)^{1/4}$ \hfill (9)

These values of the effective diameter correspond to the middle of the rotational light curve. The diameter of the nucleus is customarily derived from the formula (Jewitt 1991)

$p_V \, r_N^2 = 2.24 \times 10^{22} \, R^2 \, \Delta^2 \, 10^{0.4(m_\odot - m + \beta \alpha)}$ \hfill (10)

where $p_V$ is the geometric albedo, $r_N$ is the nuclear radius, $m_\odot$ is the solar magnitude ($V_\odot = -26.74$, $R_\odot = -27.10$), m is the observed magnitude, R is the Sun-comet distance, $\Delta$ is the Earth-comet distance, $\alpha$ is the phase angle, and $\beta$ the phase coefficient. In this equation r is given in meters, and R and $\Delta$ are in AU.

Equation (10) can be simplified considerably if a) we use the absolute nuclear magnitude $V_N(1,1,0)$, b) if we replace the radius by the effective diameter, $D_{EFFE}$, and c) if we calculate this value in km not in m. The result is a much more compact and amicable formula:

$\text{Log} \, [ \, p_V \, D^2_{EFFE} / 4 \, ] = 5.654 - 0.4 \, V_N(1,1,0)$ \hfill (11)

if the absolute nuclear magnitude in the visual, $V_N(1,1,0)$, is used. If instead, the absolute nuclear magnitude in the red is used, $R_N(1,1,0)$, then:

$\text{Log} \, [ \, p_R \cdot D^2_{EFFE} / 4 ] = 5.510 - 0.4 \, R_N(1,1,0)$ \hfill (12)

Notice that now in these two formulas $D_{EFFE}$ (in km), $V_N(1,1,0)$, and $R_N(1,1,0)$ *all correspond to the mean value of the rotational light curve*.

**12a) P-AGE (1,1)** = Photometric Age measured at 1 AU from the Earth, 1 AU from the sun. It is an objective of this work to be able to define a parameter that measures the age of a comet solely from the secular light curves. Although it is not possible in most cases to assign an absolute physical age (exception 2P/Encke), it is nevertheless possible to define a parameter related to activity that ranks the comets by age. We call it P-AGE to distinguish it from a real age. It should be emphasized that P-AGE is not a dynamical age (although it may be related to it), but rather it is related to the loss of volatiles as a proxy for age. The ability to order comets according to their relative ages could be a useful tool to understand a number of events in the history of these objects.

Consider the three parameters $A_{SEC}$, $R_{ON}$ and $-R_{ON}+R_{OFF}$. As a comet ages, the amplitude of the secular light curve, $A_{SEC}$, must decrease. In fact $A_{SEC}$ must be zero for an inert nucleus. Thus $A_{SEC}$ must be related to activity and age. *In this work we take both as synonymous. In fact, activity is a proxy for age.* $R_{ON}$ is also related to age. As the comet ages, the crust on the nucleus increases in depth, sublimating ices must recede inside the nucleus, sustained sublimation is quenched, and the comet needs to get nearer to the Sun to be activated (Yabushita & Wada, 1988; Meech 2000). Thus $R_{ON}$ decreases with age. On the other hand, $R_{SUM}$ = $-R_{ON}+R_{OFF}$ measures the total space of activity of the comet. Comets that have exhausted their CO and $CO_2$, must get nearer to the Sun to be active. Comets whose activity is dominated by water ice become active much nearer to the Sun than CO or $CO_2$ dominated comets (Delsemme 1982; Meech 2000). Thus a parameter that measures age and activity at the same time, and that includes the three above quantities could be $A_{SEC}$ * ( $-R_{ON}$ + $R_{OFF}$ ). This value defines the area of a rectangle in the phase space $A_{SEC}$ vs $R_{SUM}$.

So defined, P-AGE would give small values for old comets and large values for new comets, inverted from what we would like. It would be interesting to scale these values to human ages. We will call these *'comet years (cy)'* to reflect the fact that they have not yet been scaled to Earth's years. To calibrate the scale, we will arbitrarily set to 28P/Neujmin 1 an age of 100 cy. With this calibration we define P-AGE thus:

P-AGE (1,1) = 1440 / [$A_{SEC}$(1,1)*($-R_{ON}$ + $R_{OFF}$)]
comet years (cy)  (13)

The value of the constant has been chosen to force comet 28P to an age of 100 cy. Scaling to human ages may seem naïve and unorthodox. However it places the comets in perspective and provides a scale for comparison. This enhances the usefulness of P-AGE, and when the evolution of $A_{SEC}$, $R_{ON}$ and $R_{OFF}$ with time is studied and calibrated with a suitable physical model, it will be possible to convert these values to a real physical age, thus achieving the objective we have set in this paper. This parameter classifies the secular light curves by shape, an interesting property and a proof of its validity (see Section 6.1.6 and Figure 6).

Notice that the photometric age, P-AGE, is measured at ∆=1 AU, R=1 AU, thus the correct notation is P-AGE(1,1). This is done to be able to compare comets with different q. For comparison with the definition of absolute magnitude m(1,1), P-AGE (1,1) could be interpreted as the *absolute photometric age*, and its main value is that it allows the comparison of comets with different qs.

We have shown in Paper I that the definition of P-AGE is robust. Even if there are large errors in the input parameters, P-AGE comes out with a small error due to its mathematical definition (Equation 13). The same argument can be applied to the definition of T-AGE, given in the next section.

The definition of P-AGE fails for two comets, 107P and 133P because the turn on point is past perihelion. However the definition of T-AGE (section 4.2-24) is still valid.

**12b) P-AGE (1,q)** = Photometric Age measured at 1 AU from the Earth, perihelion distance q, from the sun. It is necessary to introduce P-AGE (1,q) because for comets with q > 1 AU the object never gets to R = 1 AU, and thus P-AGE (1,1) is not the correct interpretation of the photometric age. To evaluate it we replace $A_{SEC}$ (1,1) by $A_{SEC}$ (1,q) in equation (13). P-AGE(1,q) could then be interpreted as the *current photometric age*. Notice however that if the comet exhibits a significant LAG (defined in 4.2-19), then the correct value to use is $A_{SEC}$ (1,LAG) to calculate P-AGE(1,q).

**13) $n_{ON}$,** the slope parameter of the secular light curve near the turn on point.

**14) $n_{qB}$,** the slope of the curve just before perihelion.

**15) $n_{qA}$,** the slope of the curve just after perihelion.

**16) $R_{BP}$** = Heliocentric distance of the break point in AU. Young objects exhibit two linear laws before perihelion, and these laws change slope at the break point (see the plots at the end of this paper).

**17) $m_{BP}$** = magnitude of the break point.

**18) $A_{ROT}(PTV)$** = peak to valley amplitude of the *rotational* light curve. This information is needed because all nuclear observations should lie inside the rotational amplitude. The largest value of the rotational amplitude available in the literature is chosen.

## 4.2   Time plots

The following parameters are listed in the time plots. See Figure 2b for a graphical explanation of these parameters. The symbol legend for the plots can be found on the time plot.

**19) LAG,** the shift in maximum brightness measured from perihelion in days, positive after perihelion.

**20) $T_{ON}$ [days]** = the time at which the nucleus turns on. The negative sign in this parameter indicates pre-perihelion quantities. It corresponds to $R_{ON}$ but in the time domain.

**21) $T_{OFF}$** = the time after perihelion at which the nucleus turns off.

**22) $-T_{OFF}/T_{ON}$,** an asymmetry parameter but in the time domain.

**23) $T_{ACTIVE}$ = ( $-T_{ON}$ + $T_{OFF}$ ),** in days. It is a measure of the total time that the comet is active.

**24a) T-AGE (1,1)** = time-age. It is possible to define an age from the time plot in the same way we did for P-AGE:

$$\text{T-AGE} = 90240 / [A_{SEC}*(-T_{ON} + T_{OFF})] \text{ comet years (cy)} \tag{14}$$

The value of the constant has be chosen to force comet 28P to an age of 100 cy.

**24b) T-AGE (1,q)** = time-age measured at perihelion. This definition is equivalent to the P-AGE(1,q) definition but in the time domain.

**25) Perihelia.** The perihelia that contributed to the SLC, in the format YYYYMMDD.

**26) $P_{ORB}$ ,** the orbital period around the Sun, in years.

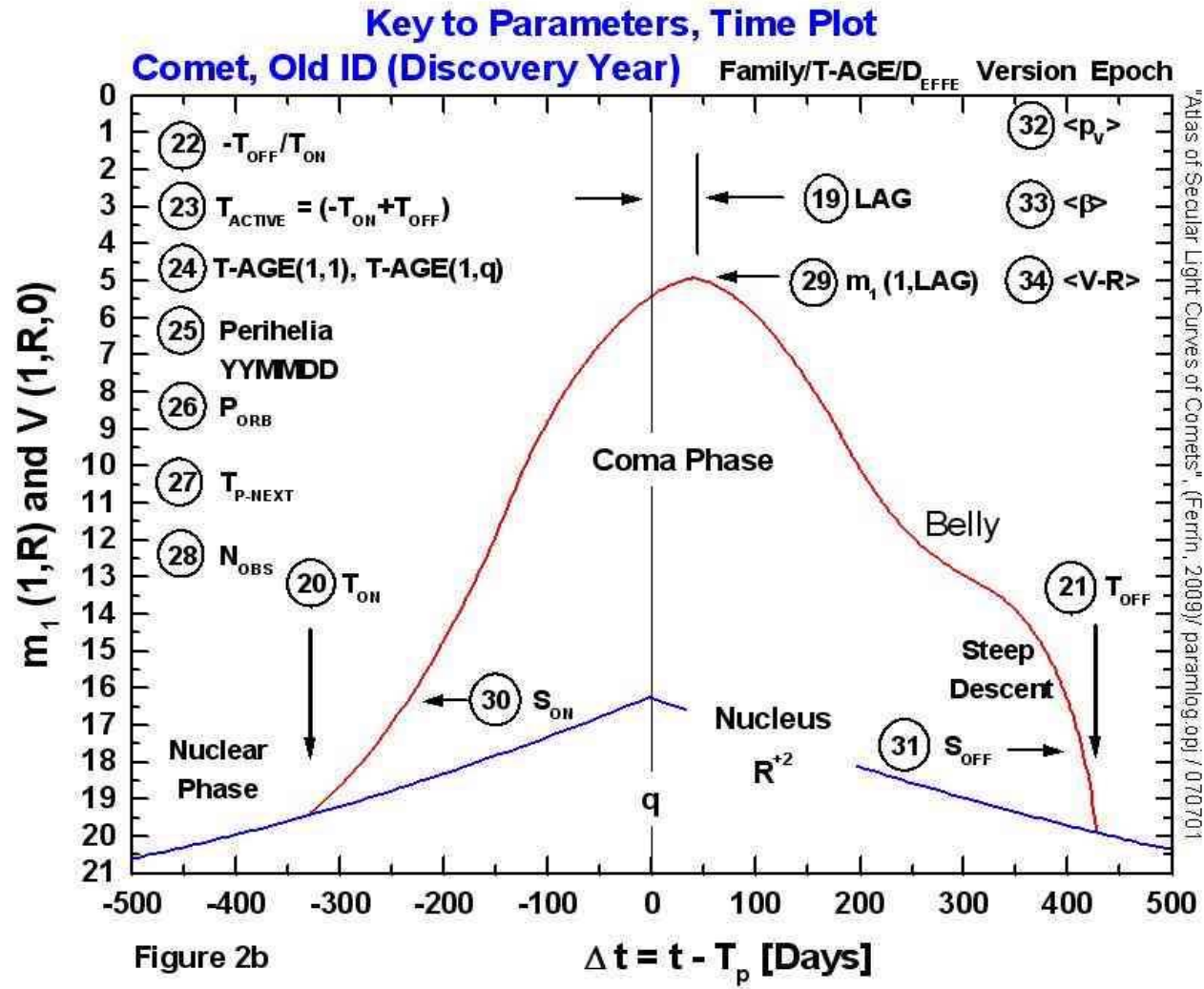

Figure 2b. Key to the time plot. The parameters are explained in Section 4.2.

27) $T_{PERI-NEXT}$ , the *approximate* date of the next apparition for planning purposes.

28) $N_{OBS}$ , the total number of observations used in the secular light curve.

29) $m_{MAX}(1,LAG)$ = maximum reduced magnitude measured at the time LAG. This value should change with orbital and time evolution.

30) $S_{ON}$ = The slope of the envelope at $T_{ON}$, for planning purposes, measured in mag/day.

31) $S_{OFF}$ = The slope of the envelope at $T_{OFF}$, for planning purposes, measured in mag/day.

32) $<p_V>$ = the mean value of the geometric albedo in the visual taken from the literature.

33) $<\beta>$ = the mean value of the phase coefficient. A table of phase coefficients was published in Paper II, and

:

an updated list is kept at the web site cited at the end of this paper.

34) **<V-R>** = the mean value of the color index V-R. A table of color indexes was published in Paper II, and an updated list is kept at the web site cited at the end of this paper.

## 5  AGE-SIZE CLASSIFICATION

It is useful to classify comets by age and size. An histogram of diameters (see Figure 3) shows that most JF comets are less than 10 km in diameter. Thus it is possible to define size classes thus:

   0 < D < 1.5 km, dwarf comet, d.
  1.5 < D < 3, small comet, s.
   3 < D < 6 km, medium size comet, ms.
   6 < D < 10 km, large comet, L.
  10  < D < 50 km, extra large comet, XL.
   50 < D , goliath comet, G

The same can be done for the age of a comet. Most comets have photometric ages less than 100 cy.

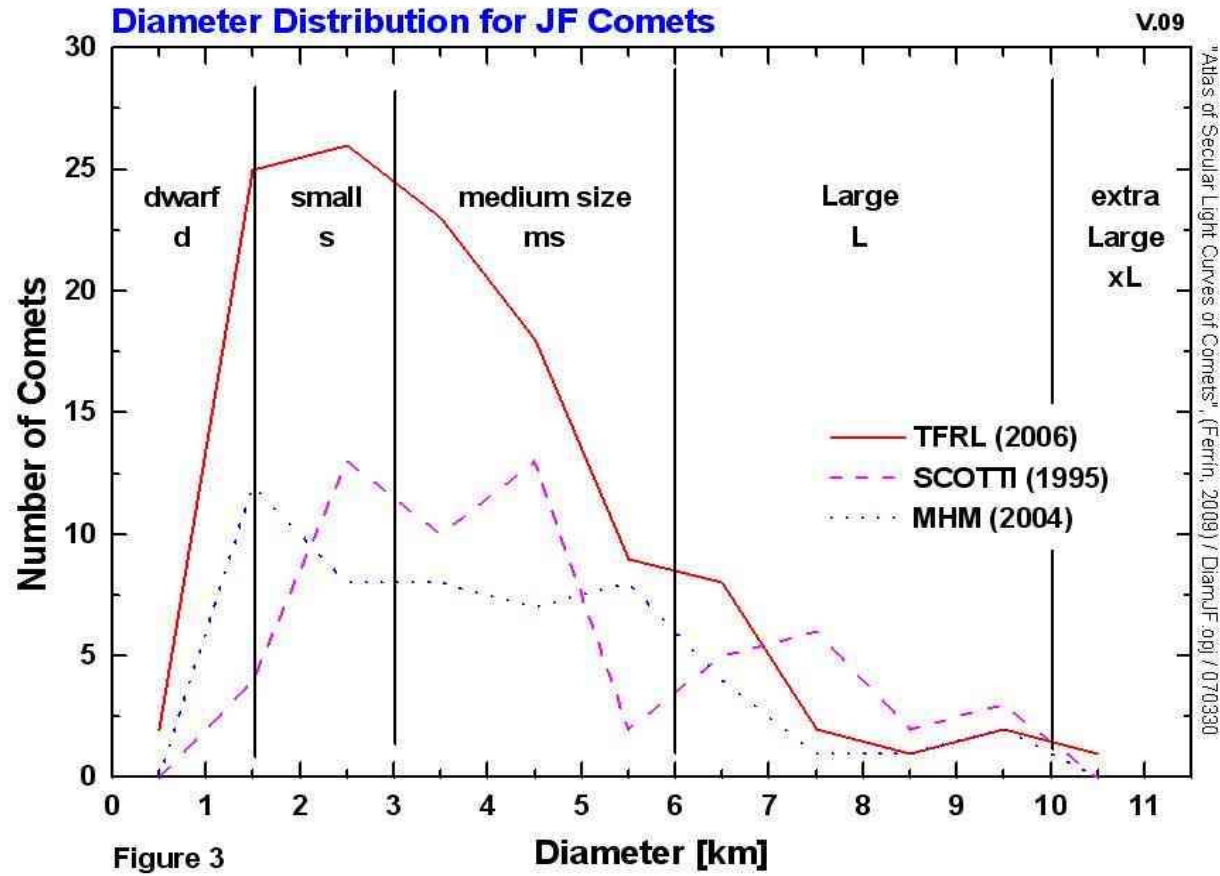

Figure 3. Histogram of diameters of Jupiter Family comets. The distribution has been divided into bins to define size classes. It can be ascertained that most Jupiter Family comets have diameters less than 10 km. TFRL = Tancredi et al. (2006). SCOTTI = Scotti (1995). MHM = Meech et al. (2004).

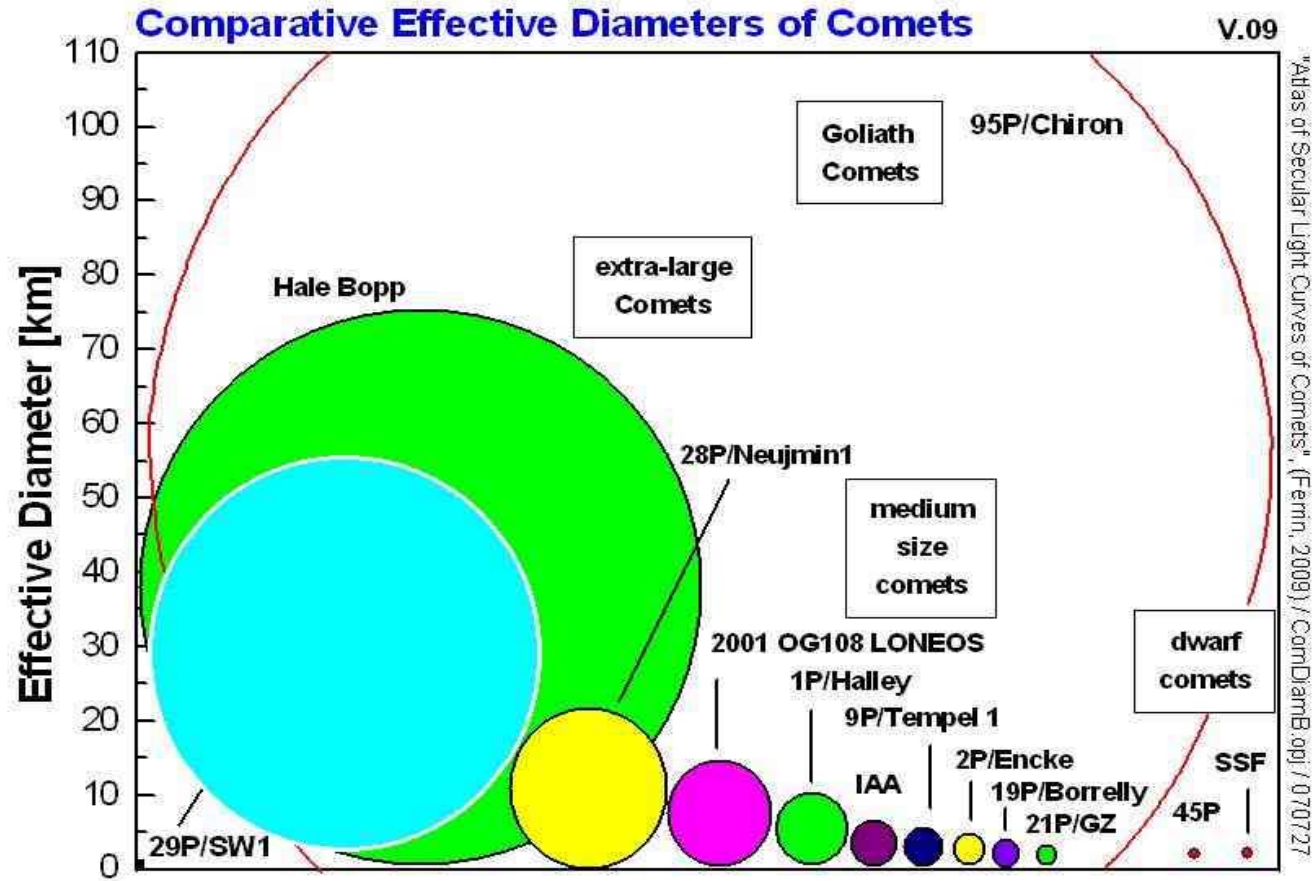

Figure 4. Comparative effective diameters of comets. There are goliath comets that if they were to break, they would give origin to thousands of small comets.

Let us define

0 < P-AGE < 4 cy, baby comet, b.
 4 < P-AGE < 30 cy, young comet, y.
 30 < P-AGE < 70 cy, middle age comet, ma.
 70 < P-AGE < 100 cy, old, O
100 cy < P-AGE, methuselah comet, M.

It is useful to think that 1P/Halley is a young large comet (classified yL) and that Hale-Bopp was a baby goliath comet (classified bG). It places the comets in perspective.

Using this classification we find also find that comet 29P/SW 1 is a baby goliath comet (bG). 28P/Neujmin 1 and C/2001 OG108 LONEOS are methuselah extra large comets (MXL). 107P/Wilson-Harrington and 133P/Elst-Pizarro are methuselah small comets (Ms). 67P/Churyumov-Gerasimenko and 85P/Boethin are middle age medium size comets (mams). Sugano-Saigusa-Fujikawa is a middle age dwarf (mad).

Some researchers do not like to place comets into boxes and would argue that there is a continuum of ages and sizes. This is correct. Then an alternative scheme of classification is to give the two numbers P-AGE and $D_{EFFE}$. Notice that 90% of comets lie inside 0 < P-AGE < 100 cy and 95% inside 0 < $D_{EFFE}$ < 10 km. *Any object above 100/10 or near 0/0 is exceptional (very old, very large, very young, very small).* For example 29P/SW1 with 0.7/54, is exceptional. In this scheme 1P/Halley has measurements 7/10, 28P/Neujmin 1 has 100/23, C/2001 OG108 has 102/16. Comets 107P/WH and 133P/E-P have respectively 1370/3 and 1640/5, while 85P/Boethin and SSF have 49/6 and 53/0.7.

A comparative diagram of the diameters of comets, can be found in Figure 4. It shows that there are a few very large comets that if they were to break they would give origin to thousands of minor comets.

## 6 OVERVIEW

Next we are going to describe several cometary phenomena deduced from the plots. It is beyond the scope of this Atlas to give a complete and coherent explanation of these phenomena. In a few cases we will only suggest a working hypotheses. A compilation of the parameters deduced from the plots is given in Table 4. Numerous positive correlations are found from this table. A statistical analysis of this data and a more in depth study of some of the physical phenomena, is reserved for a dedicated paper to be carried out elsewhere.

### 6.1 General

### 6.1.1 Phases of the SLC.

It can be seen that most comets have three phases in their secular light curves: a nuclear phase ($R < R_{ON}$), a coma phase ($R_{ON} < R < R_{OFF}$), and again a nuclear phase ($R_{OFF} < R$) (in the time domain $\Delta t < T_{ON}$, $T_{ON} < \Delta t < T_{OFF}$, and $T_{OFF} < \Delta t$ where $\Delta t$ is the time with respect to perihelion).

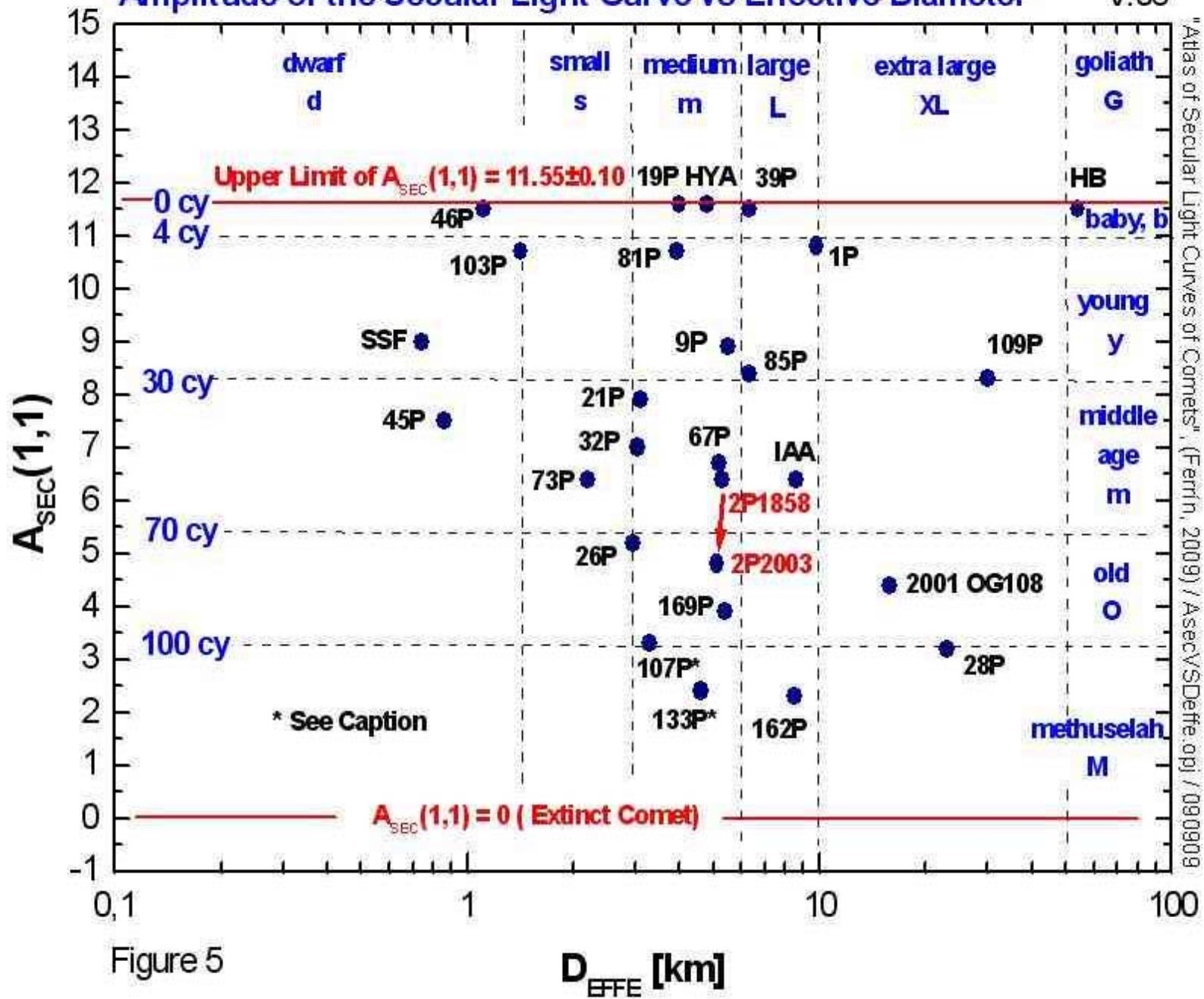
Figure 5

Figure 5. Amplitude of the secular light curve, $A_{SEC}(1,1)$ as a function of diameter. The amplitude of the secular light curve, $A_{SEC}$, is an indication of activity and decreases as a function of age. It can be seen that there is no correlation with diameter, implying that the activity of a comet depends on the composition, age and perhaps other parameters like the crust thickness but not the diameter. This is evident from the fact that dwarf comets like 46P/Wirtanen and 103P/Hartley 2 (Paper VI) are as active as goliath comets like C/1995 O1 Hale-Bopp. Notice that there is a maximum value of $A_{SEC}$ =11.55±0.10 magnitudes. We interpret this as the maximum amount of sublimation that a comet surface can output per unit area. The fact that there is a maximum value implies that we can set a minimum value for the diameter of a comet. This value might be too low, but nevertheless sets a minimum limit for the size. Notice also that comet 2P/Encke changed position from 1858 to 2003. This implies that this plot is an *evolutionary diagram*. Comet classes have been indicated above and on the right hand side. For comets 107P and 133P $A_{SEC}(1,LAG)$ has been plotted because $A_{SEC}(1,1)$ can not be defined. $A_{SEC}(1,1)=0$ implies a dead comet.

%%%%%%%%%%%%%%%%%%%%%%%%%%%

### 6.1.2 Shape of the SLC and P-AGE.

The secular light curves have been organized by increasing ID number to facilitate their access. However it is also interesting to order them by P-AGE. It is then possible to see that P-AGE orders the comets by shape of the SLC, a most interesting property. Young comets with tall and wide SLCs come first, while old comets with short and thin SLCs come last (Figure 6).

### 6.1.3 $H_2O$, CO and $CO_2$ ice sublimation.

Before P-AGE ~< 30 cy, some comets exhibit a linear law after turn on. This is an indication of CO or $CO_2$ ice sublimation, because the vapor pressure curve of those substances is linear. The existence of $CO_2$ and CO has been determined in a number of comets. After P-AGE ~30 cy, comets exhibit a SLC that shows curvature. Curvature is a clear indication of water ice sublimation, because the water vapor pressure vs temperature shows strong curvature and the SLC is a reflection of this law. Thus the shape after turn on might be an indicator of composition.

### 6.1.4 Threshold coma magnitude.

In Paper I it was shown that comet 28P/Neujmin 1 exhibited a coma with a secular amplitude $A_{SEC}(1,1)$ = 3.2 mag, while 133P does not exhibit a coma with $A_{SEC}(1,R)$ = 2.3 mag and 2P does not exhibit a coma at aphelion with $A_{SEC}(1,R)$ = 2.7 mag. Thus it seems that there is an intermediate value at which the coma becomes undetectable. We call that a threshold coma magnitude, TCM, and we estimate its value at TCM= ~3.0±0.3 mag *above the nucleus*. Below TCM the *comet is active but the coma is below detection level* and the nucleus looks stellar.

### 6.1.5 Maximum value of $A_{SEC}$ and $R_{OFF}/R_{ON}$.

There seems to be a maximum value of the amplitude of the secular light curve, $A_{SEC}(1,1)(MAX) = 11.55\pm0.10$ mag. This number measures the maximum sublimating capacity that a comet surface made of water ice can output, but up to now does not have yet a theoretical foundation (see Figure 5). There seems to be also a maximum value of the parameter $R_{OFF}/R_{ON} = 2.0\pm0.1$. This may have to do with the propagation of the thermal wave inside the nucleus, but up to now does not have a theoretical foundation either.

### 6.1.6 Validation of the concept of photometric age, P-AGE.

In Figure 6 the envelopes of comets are compared. It is apparent that older comets are nested inside the envelope of younger comets. On the left hand side of the plot the comet number is listed. On the right hand side of the plot, the photometric age is listed. The conclusion is that P-AGE classifies comets by shape of their secular light curve, a most interesting property. Figure 6 also shows that as comets age, the amplitude of the secular light curve diminishes, and $R_{SUM}$ decreases in value.

### 6.2 Classes of comets

Looking at the plots new classes of comets arise:

**1) Comets with power laws at turn on.** Several comets exhibit power laws at turn on. This is an indication that they are sublimating something more volatile than water ice, because water ice shows curvature. Comets that belong to this class are listed in Table 1. It can be seen that all are young comets.

Notice that only 6 comets out of 26 (23%) exhibit power laws at turn on, and the ones that do so, do so only partially. Table 1 gives evidence to conclude that the predictive power of the H10 system is restricted to young comets with power laws and even so the H10 system does not work for them all the way up to perihelion because of the break point. So this system should be discontinued.

**2) Comets with power law break.** Young comets have a tendency to exhibit a break in the incoming branch of the SLC, where the curve changes slope abruptly. This has been interpreted as a change from sublimating something more volatile than water ice (probably $CO_2$), to sublimating water ice. Comets that belong to this class are listed in Table 2. Once again we see that all comets in this category are young.

In most comets one single substance, water ice or $CO_2$ controls sublimation. If the interpretation of the break is a change of controlling substance, then all comets with power law breaks are controlled by two substances (Papers I, III). The best example of this is comet 9P. At turn on sublimation takes place causing a linear law, until at R= -2.08 AU water ice takes over and curvature is then apparent. Our preliminary interpretation is that CO2 is responsible for the linear law, while H2O is responsible for the law with curvature.

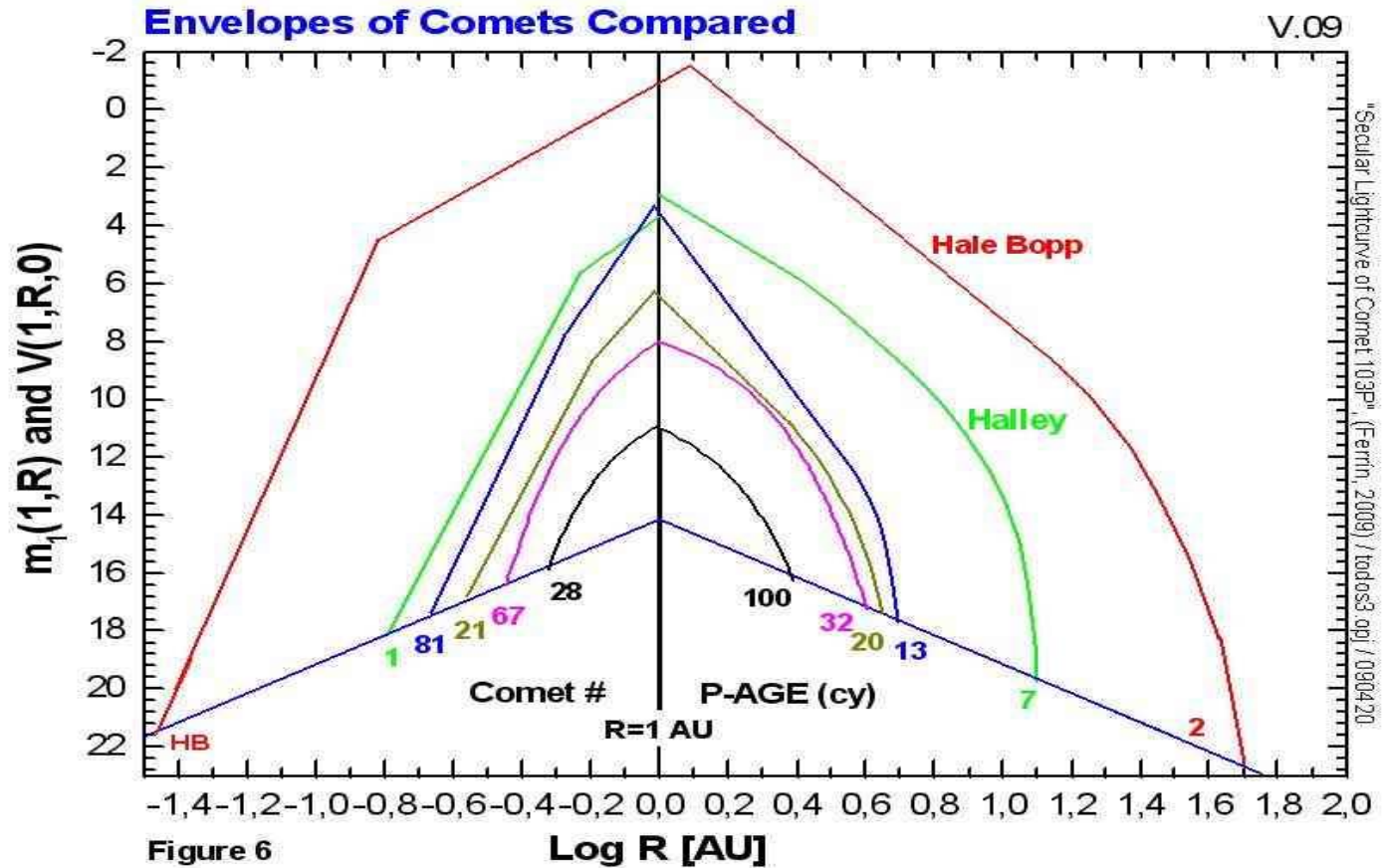

Figure 6. Validation of the concept of photometric age, P-AGE: Envelopes of comets compared. It is apparent that older comets are nested inside the envelope of younger comets. The conclusion is that P-AGE classifies comets by shape of their secular light curve, a most interesting property. On the left hand side of the plot the comet number is listed. On the right hand side of the plot, the photometric age is listed. This plot also shows that as a function of age, the amplitude of the secular light curve and $R_{SUM}$ diminish in value.

**3) Comets with belly.** A belly can be defined as an excessive and asymmetric extension of the secular light curve after perihelion. Young comets are the only ones that exhibit a belly: 1P (P-AGE= 7 cy), C/1995 O1 HB (1.8 cy), 19P (14 cy), 9P (21 cy), 21P, (20 cy), 6P (24 cy).

**4) Comets with steep slope at onset of sublimation.** Some comets exhibit a significant slope at turn on. Example, 6P with P-AGE = 24 cy has a slope at turn on of -0.17 mag/d (in the Log R plot, $R^{+70}$). There is no theoretical explanation of this fact.

**5) Comets with steep slope at turn off of sublimation.** Several comets exhibit a steep descent at turn off. Members of this class are 1P, 85P, C/1996 B2.

**6) Comets that turn on after perihelion.** Two methuselah comets exhibit activity only past perihelion (107P, 133P).

**7) Comets with outbursts after turn off.** 1P/Halley had a large outbursts *after turn off*, with $\Delta m$= -6.5 magnitudes ( - sign indicates brighter, + sign fainter). This new class is opened because there is evidence to believe that other comets to be published also have this kind of behavior.

**8) Comets with separated activity at aphelion.** A comet that clearly belong to this class is 2P/Encke (Paper V). The amplitude of the activity is $\Delta m$= -2.7 mag. Aphelion activity is suspected in comets 133P/Elst-Pizarro and 9P.

**9) Comets with an 'S'-shaped SLC.** Several comets exhibit an S-shape of the SLC after perihelion (19P and C/2001 OG108, 101P/Chernyk). There is no current explanation for this feature. This effect is different from the belly because the belly takes place after the comet exhibits a power law in the log plot. Or either the power law in these two comets is much larger than in normal comets.

**10) Comets with extreme LAG.** Several comets exhibit extreme value of LAG. The most likely explanation is that the pole of the nucleus lies near the orbital plane and the pole points to the Sun at time = LAG. Members of this class are 6P (LAG= +46 d), 65P (+242 d), 107P (+42 d), 133P (+155 d), and C/1995 O1 (+45 d). 101P is the only one with a significant negative LAG, LAG= -126 d. Positive LAGs are more common than negative LAGs (13 vs 6).

**11) Spill over comets.** Definition: *'A spill over comet is one that exhibits a continuous decrease in brightness as a function of solar distance, but it does not reach to its nuclear magnitude when it reaches aphelion '.* Thus the object spills over its activity from one orbit to the next. Members of this class are 29P, 39P, 65P, 73P/SW3C in 1990, and 81P. Probable members are 9P and 19P. Actually these comets exhibit activity at aphelion and thus are related to the class previously defined, but the activity *is not* separated like in comet 2P (Paper V).

**12) Comets with q-effect.** Definition: *'The q-effect or perihelion effect is a change in the maximum brightness exhibited by a comet at perihelion, as a consequence of a change in perihelion distance from one orbit to the next'.* Several comets exhibit this effect, one of the most significant being 101P/Chernykh which shows a change of perihelion distance of -0.212 AU causing a change of -4.9 mag/AU.

**13) Comets with photometric anomalies.** Several comets exhibit a photometric anomaly defined *as dip or a bump in the envelope of the SLC that repeats at different apparitions*. This definition excludes outbursts that are transient in nature and do not repeat from one apparition to the next. Members of this class are 1P, 2P, 67P.

**- 1P/Halley.** Exhibits a small photometric anomaly post-perihelion, of $\Delta m$= +0.5 mag amplitude. The confirmation that this is a real feature is that the same dip appears at the same place in infrared $J_o$ and visual $C_2$ observations (Morris and Hanner 1993).

**- 2P/Encke** exhibits a dip in the SLC *at aphelion*, of $\Delta m$= +1.7 mag. The magnitude decays linearly from $\Delta t$ = +312 to +393 d until it reaches the nucleus (confirmation the plot in Additional Plots).

**- 67P/Churyumov-Gerasimenko** exhibits a decrease in brightness just before perihelion of $\Delta m$=+1.9 mag. This could be the result of a topographic effect or the turn off of an active region.

**- Comet C/1995 O1 Hale-Bopp** exhibits an increase in brightness *above the power law*, centered at perihelion of $\Delta m$= -3.4 mag. This is a *perihelion surge* since it is symmetric with respect to q.

**- Comet C/1996 B2 Hyakutake** shows a most peculiar increase in brightness at minimum approach to Earth, of $\Delta m$= -1.1 mag. We would expect a decrease in brightness due to the Delta-effect (Kamel 1997). However what is observed is *an increase* in brightness. The explanation of this discrepancy is that comet C/1996 B2 had many water production outbursts along the orbit (Schleicher & Osip 2002) and by chance one of them took place near R= 1 AU, masking any Delta-effect.

While the photometric anomalies of 1P, 2P and 67P might be due to a topographic feature, the real reason for the perihelion surge of Hale-Bopp is still a matter of debate.

### 6.3 Comets with peculiar secular light curves.

Definition of a normal comet: *A normal comet is one that exhibits sustained activity quasi-symmetric with respect to perihelion, within a well defined turn on point before perihelion, and a well defined turn off point after perihelion.* Based on this definition, some comets listed below have peculiar secular light curves.

**6P/D'Arrest.** Due to a very asymmetric light curve with LAG= +48 d, it is difficult to measure its absolute magnitude. The nucleus may have a very odd shape, or the distribution of volatiles on the surface may be very odd, or both.

**9P/Tempel 1.** While the sublimation of most comets is controlled by one substance, the sublimation of 9P is controlled by two substances. In the Log R plot it can be seen how, at turn on, the slope is linear (clear indication of $CO_2$ or CO sublimation) and at R= -2.08 AU it converts to a law with curvature (clear indication of $H_2O$ sublimation).

**19P/Borrelly.** Shows an odd shape of the SLC after perihelion in the form of an 'S'. At the present time there is no explanation for this feature.

**29P/SW 1.** This comet is permanently beyond Jupiter. It does not exhibit a regular SLC. It shows an irregular activity. It is active all around the orbit at 5.9 AU from the Sun.

**67P/Churyumov-Gerasimenko.** The photometric anomaly pre-perihelion of this comet is very intense and unique.

**101P/Chernykh.** The SLC is asymmetric exhibiting a huge pre-perihelion LAG. The nucleus has a very odd shape, or the distribution of volatiles is very odd, or both.

**107P/Wilson-Harrington.** The activity was detected photographically on only one night in 1949. This is one of the oldest comets measured by P-AGE in this Atlas, and the SLC is like no other.

**133P/Elst-Pizarro.** This methuselah comet, besides presenting a brief activity after perihelion, exhibits a very thin tail that does not widen with distance.

**C/2001 OG108.** Shows an odd shape of the SLC after perihelion in the form of an 'S'. At the present time there is no explanation for this feature.

### 6.4  Look alike comets

Although there are not two comets that are identical, some have look alike SLCs. Examples are listed in Table 3. In most cases the pair has about the same P-AGE, but not always.

## 7  COMMENTS ON SPECIFIC COMETS

### 7.1  Log R Plots

**1P/Halley, Log R plot, Figure 7.** Since this young large comet has q < 1 AU, Log 0.587 = 0.231 has been subtracted before perihelion and added after perihelion to the horizontal axis to make room for observations inside the Earth's orbit. We can observe the three phases of the secular light curve, the nuclear phase, the coma phase and the nuclear phase again. On Log R plots, power laws plot as straight lines, therefore the nucleus makes a pyramidal line at the bottom, since it follows a $R^{+2}$ law. Most nuclear magnitudes lie inside the amplitude of the *rotational* light curve range ( $A_{ROT}$ ). The turn-on and turn-off points are very sudden affairs, so it is easy to decide what is a nuclear magnitude and what observations are coma contaminated. The coma is described by two power laws before and one after perihelion. There is a noteworthy break in the slope at R = -1.70 AU for no apparent reason although we suspect it is due to the onset of water sublimation. The coma reaches a pointed sharp maximum after perihelion, and turns off with a steep descent. There is an outburst of 6.7 mag after the nucleus turns off. The secular light curve is very asymmetric with $R_{OFF} / R_{ON}$ = 2.02. SEK = Sekanina (1985). V1+V2+G = Keller et al., (1987), based on Vega 1, Vega 2 and Giotto observations. BWWS = Belton et al. (1986). MJR = Meech et al. (1986). W = West (1990). WJ = West and Jorgensen (1989). See Paper I for additional information.

**2P/Encke, Log R plot, 1858, Figure 8.** The SLC from 1858 compared with the SLC from 2003. The data has been taken from Kamel (1991). The SLC has been narrowing with time, a result that is in accord with the time evolution of the comets presented in this Atlas. It is possible to derive a photometric age for this Epoch. Seven physical parameters show evolution with respect to 2003 (Paper V).

**2P/Encke, Log R plot, 2003, Figure 9.** The x-axis has been increased in Log q = ±0.47965 to make room for observations inside the Earth's orbit of this old medium size comet. Notice the sharp turn on point, and the activity at aphelion evidenced by observations well above the nuclear line. Although the SLC exhibits almost the same turn on and turn off distances, the SLC is asymmetric at 1 AU from the Sun, as evidenced by the fact that the absolute magnitude before perihelion, $m_{1B}(1,1)$= 9.8±0.1, while after perihelion is $m_{1A}(1,1)$= 11.5±0.1, an asymmetry of 1.7±0.15 magnitudes at 1 AU. The photometric age derived from this plot is P-AGE= 98±8 cy (comet years), corresponding to an old comet. The mean value <V-R> and $p_R$ are taken from the literature. The vertical distribution of nuclear observations are from Fernandez et al. (2005). The nuclear line is defined from Paper V where there is additional information.

**6P/D'Arrest Log R plot, Figure 10.** The SLC of this young dwarf comet is very peculiar. Because the comet has a very large LAG= +42 d (confirmation the time plot), the Log R plot is distorted. It is then very difficult to determine the absolute magnitude. We know from the Log R plot that $A_{SEC}$ > 10.0 mag. Since the maximum observed for comets in this Atlas is $A_{SEC}$ ~ 11, then it is a good guess that $A_{SEC}$ for 6P is ~10 to ~11. This makes the comet very active, and this is an indication of youth. The estimated values are P-AGE < 24 cy and T-AGE < 26 cy. Because the SLC exhibits a large LAG and a sudden turn on with a very steep power law, this comet must be very peculiar, either with an *odd shape or with an odd distribution of volatiles,* and a pole laying near the orbital plane, or both. It is very probable that one pole points to the Sun at LAG causing the maximum brightness. The nuclear line can only be defined using the pre-turn on observations.

**9P/Tempel 1, Log R plot, Figure 11.** This young medium size comet has a most peculiar light curve with a break in the power law at turn on. The slope changes abruptly at R = -2.08, and then exhibits curvature, clear indication of water ice sublimation. The comet presents activity pre-turn on that has to be studied more carefully. The post-perihelion branch is also anomalous because it exhibits a linear decay over the whole range. CJ = Chen & Jewitt (1994), FMLAPB = Fernandez et al. (2003), LTAWW = Lamy et al. (2001), LFCW = Lowry et al. (2003), M et al. = Meech et al. (2000), B et al. = Belton et al. (2005), LF= Lowry & Fitzsimmons (2001), FMLAPB = Fernandez et al. (2003). HMCSM = Hergenrother et al. (2006). See Paper IV for additional information.

**19P/Borrelly, Log R plot, Figure 12.** Only one power law is needed to describe the envelope before perihelion of this young medium size comet. After perihelion the secular light curve shows an S-shape characterized by an upturn at perihelion, shown also by comet C/2001 G108 LONEOS. The reason for this odd shape is unknown. The turn-off point is uncertain. The nuclear magnitudes are reasonably well determined and observations lie inside the rotational light curve amplitude. LTW = Lamy et al. (1998), BHSBOH = Buratti et al. (2003), MS = Mueller & Samarasinha (2002). See Paper I for additional information.

**21P/Giacobinni-Zinner, Log R plot, Figure 13.** Three power laws are needed to describe the light curve of this young medium size comet. Notice the sharp change in slope at R = - 1.58 AU. M = Mueller (1992), CJ = Chen & Jewitt (1994), H = Hergenrother (Scotti, 2001). See Paper I for additional information.

**26P/Grigg-Skejellerup, Log R plot, Figure 14.** The size and P-AGE of this comet indicate that it is an old small nucleus, and that can be confirmed from the fact that the secular light curve is dominated by the nuclear phase, while the coma phase is of short duration. BRBS = Boehnhardt et al. (1998), HL = Hergenrother & Larson (Scotti 2001), LTLRH = Licandro et al. (2000), S = Scotti (1995). See Paper I for additional information.

**28P/Neujmin 1, Log R plot, Figure 15.** The nuclear phase of this methuselah giant comet dominates the light curve. This is a textbook example of how the nucleus follows a $R^{+2}$ law and the nuclear observations lie inside the *rotational* light curve amplitude. Notice, for example, the two clusters of vertical points post-perihelion. The comet is active for about 9 months, and the amplitude $A_{SEC}$ of the secular light curve has diminished significantly in comparison with 1P/Halley. This is a Saturn family comet, as deduced from the fact that the aphelion distance Q= 12.07 AU, is beyond Saturn, and that the orbital inclination is i = 14°, indicating that it is neither a nearly isotropic comet nor a Jupiter family comet in the classification of Levison (1999). In the 2002 apparition the comet was in conjunction with the Sun at perihelion and the coma phase could not be observed. Campins et al. (1995) have recalculated the diameter and albedo of this comet, changing the parameters assumed for the standard thermal model, and the resulting albedo is twice the former value (Campins et al ., 1987). However notice that the diameters quoted in their Table 2 correspond to a sphere of the same projected area at the maximum of the rotational light curve, thus they are not effective diameters as defined in equations (11) and (12). The Epoch is 1913, which means a) that the coma photometry is poor, and b) that there is an urgent need for observations of the coma phase of this comet to test for evolution. Thus the next opportunity will be in the next perihelion passage, in 2021.2, with observations starting around 2020.6 . DMHD = Delahodde et al. (2001), JM = Jewitt & Meech (1988), CAM= Campins et al. (1987). See Paper I for additional information.

**29P/Schwassmann-Wachmann 1, Log R plot, Figure 16.** This baby goliath comet exhibits an irregular light curve and must be classified as odd. There is no clear trend with solar distance. The comet is in a state of 'boiling' with irregular outburst spaced along the orbit. The extrapolation of these magnitudes to perihelion result in the determination of a photometric age P-AGE= 0.3±0.2 cy, clearly a baby comet. The nucleus dimensions are due to Stansberry et al. (2004). MBMDL = Meech et al. (1993). S = Scotti (1995). S et al. = Stansberry et al. (2004).

**32P/Comas-Sola, Log R plot, Figure 17.** This middle age small comet has a well defined SLC. The apparitions of 1927-1969 require a correction of +0.82 mag, while that of 1978 requires +0.31 mag to place them at the level of 1996. Thus this comet has been decreasing significantly in brightness. The nuclear line is due to Tancredi et al. (2006).

**39P/Oterma, Log R plot, Figure 18.** This young medium size comet had a q= 3.39 AU and thus never got near to the Sun. Its activity is restricted from q to Q and it is active all around the orbit. There is no nuclear phase. The comet suffered an encounter with Jupiter in 1963 that changed the orbit. The new q is q= 5.47 AU and the new Q= 9.01 AU. Thus the comet went from being a JF comet to a Centaur. Consequently the activity will be reduced further and it is expected that it will only raise ~3 magnitudes above the nucleus. This value is identical to the threshold coma magnitude defined above, and thus it is probable that the comet will not exhibit a coma in this new orbit. This comet is visible all around the orbit. The comet resembles 65P/Gunn. It has not been ascertained if this comet is in the sustained regime or in the outburst regime. The nuclear magnitude is that deduced from Fernandez's observations in IAUC7689. Most of the observations are photographic so these results are tentative.

**45P/Honda-Mrkos-Pajdusakova, Log R plot, Figure 19.** The light curve of this middle age dwarf comet of only 0.9 km diameter is well defined. Notice the asymmetry of the SLC, and the steep descent at turn off, which marks the initiation of the nuclear phase. LTAW = Lamy et al. (1999).

**65P/Gunn, Log R plot, Figure 20.** This young medium size comet resembles 39P/Oterma. In fact both had orbits inside Jupiter's orbit, before 39P was moved to an external orbit. 65P/Gunn does not have a nuclear phase since it is active all around the orbit. This is an annual comet. Notice the asymmetry around perihelion caused by a strong perihelion LAG observed in the time plot. The most probable explanation is that the pole of this comet lies near the orbital plane and that it points to the Sun at time = LAG. The nuclear magnitude is due to Tancredi et al. (2006) = TFRL. S = Scotti (1995).

**67P/Churyumov-Gerasimenko, Log R plot, Figure 21.** By this P-AGE the secular light curve of this middle age medium size comet adopts a rounded shape that can not be described by a power law. Notice the existence of a photometric anomaly in the secular light curve from Log R = - 0.27 to Log R = -0.12. This piece of the secular light curve is enlarged and flattened in additional plots. This comet will be visited in 2014 by the Rosetta spacecraft. The location of the encounter, orbiter, landing and $R_{ON}$ have been indicated. HM = Hainaut & Martinez (2004), LTWJK = Lamy et al. (2003), M = Mueller (1992). TBDB= Tubiana et al. (2008). See Paper I for additional information.

**73P/Schwassmann-Wachmann 3C, Log R plot, 1990 and 1995, Figure 22.** This is a split middle age small comet for which the SLC from 1990 and 1995 are quite different. Notice the huge increase in brightness in 1995. Consequently the photometric age also changed. In 1990 we find P-AGE= 55 cy while in 1995 we find 21 cy. The comet rejuvenated due to the outburst. This comet is an excellent example of a splitting event producing an outburst. However small outbursts may be due to surface activity and not to an splitting. The nucleus observations are from Boehnhardt et al. (1998) = BRBS. G = Green (2007). LJ = Luu & Jewitt, cited by Scotti (2001).

**73P/Scwassmann-Wachmann 3C, Log R plot, 2001, Figure 23.** The SLC in 2001 still contained a lot of activity and the photometric age is comparable to that of 1995, P-AGE= 18 cy. Notice that the activity seems to reach to aphelion both pre and post perihelion. S = Scotti (1995). HL = Hergenrother & Larson, cited in Scotti (2001).

**81P/Wild 2, Log R plot, Figure 24.** Different symbols correspond to different apparitions identified in the time plot. This young medium size comet exhibits a photometric behavior requiring three power laws, as for comet 1P/Halley. Notice the change in slope at R = -1.88 AU. Pittichova & Meech (2001) found the comet active at R=4.98 AU post-perihelion (Log R=0.70), while aphelion takes place at R=5.3 AU (Log R=0.72). Thus it is likely that this comet is active at aphelion. It is even plausible that this comet may be active *beyond* aphelion, in which case it would be a member of the *spill-over comets*, comets whose activity spills over from one orbit to the next. The 2003 perihelion could not be observed because the comet was in conjunction with the Sun. Notice the secular increase in brightness with time: The 1990 observations had to be raised by 1.25 mag, and the 2003 apparition seems to have had a brighter maximum magnitude. LTLRH = Licandro et al. (2000), CJ = Chen & Jewitt (1994), LFC = Lowry et al. (2003), SEK = Sekanina (2003) , MN = Meech & Newburn (1998), B et al. = Brownlee et al. (2004). See Paper I for additional information.

**85P/Boethin, Log R plot, Figure 25.** This is a middle age medium size comet with P-AGE= 49 cy. Notice the large slope at turn on, with n= 28. The comet changed orbits from 1975 to 1986. Although q was larger in 1986, the comet was brighter which does not make much sense. The nuclear magnitude of this comet can be derived from the two post perihelion, post turn off observations of Roemer = R, listed by Kamel (1992). This comet is lost. It did not return in the 2008 apparition. GK = Gilmore and Kilmartin (IAUC 4121).

**101P/Chernykh, Log R plot, Figure 26.** This young medium size comet belongs to the class of odd comets because of the very strong LAG before perihelion. The plot had to be folded at q, because the folding at R= 1 AU did not make much sense. Notice also that the observations in 1992 with smaller q are brighter than the 1978 observations, as expected. This comet split on 1991 April 15 at R= -3.32 AU, thus this SLC must be anomalous. The nuclear lines are due to Scotti (1995) = S, and TFRL = Tancredi et al. (2006). CJ = Chen & Jewitt (1994).

**107P/Wilson-Harrington, Log R plot, Figure 27.** This methuselah small comet exhibited very feeble activity in 1949 so there is the question if this is a comet or not. If a solar system object exhibits a tail, it must be a comet. Applying this definition to 107P, and considering other physical properties, it is concluded that this is a comet, albeit a rather old one. The definition of the photometric age is not valid for this comet because the turn on point if past perihelion. Accordingly a photometric age can not be calculated. However the definition of T-AGE is still valid and we find T-AGE(1,q)= 760 cy, a methuselah comet. The nuclear magnitude is derived from data compiled in this work.

**109P/Swift-Tuttle, Log R plot, Figure 28.** This middle age giant comet has a very pointed light curve with a very large slope at R= 1 AU. Notice the smoothness of the envelope pre-perihelion. Ó'Ceallaigh et al. (1995) = OFW, Boehnhardt, et al. (1996) = BBO, Fomenkova et al. (1995) = FJPPSGJ. Green et al. (1997) = GMOFWI.

**133P/Elst-Pizarro, Log R plot, Figure 29.** Most of the data is taken after perihelion, where this methuselah medium size comet shows brief activity. The definition of photometric age fails for this comet because the turn on point is after perihelion. However the definition of T-AGE is still valid and we find T-AGE(1,q) = 280 cy, a methuselah comet. The Asteroid Dynamic Site data ( http://hamilton.dm.unipi.it/astdys/) is plotted as open squares, and is used to corroborate the photometric data (black circles and squares). The magnitude increase at Log R = +0.516 has not been corroborated with independent measurements. The main outburst is not well sampled. Two observations near aphelion and 0.6 magnitudes above the nucleus may suggest activity at aphelion. The turn off is determined with great accuracy thanks to our own observations (downward open triangles) and to serendipitous observations made by Lowry & Fitzsimmons (2004) = L&F (dark upward triangles), HJF= Hsieh et al. (2004). The nuclear magnitude is by Hsieh, Jewitt & Fernandez (2004) = HJF. See Paper III for additional information.

**C/1983 H1 IRAS-Araki-Alcock, Log R plot, Figure 30.** This old medium size comet approached Earth within a very small distance. The SLC allows the determination of P-AGE= 84 cy, an old comet. The nuclear line is due to Sekanina (1988) = SEK.

**C/1983 J1 Sugano-Saigusa-Fujikawa, Log R plot, Figure 31.** Since observations of this middle age dwarf comet were carried out only after perihelion, we had to assume a light curve before perihelion. It is clear from looking at other SLCs that comets past middle age have almost symmetrical light curves with $R_{OFF}/R_{ON}$ ~1.0. Thus we find P-AGE= 53 cy, a middle age comet. Unfortunately due to the eccentricity being e=1, the comet will not return in the near future. Notice that this is a very small comet of diameter D= 0.7 km. This value and the nuclear line come from Hanner et al. (1987).

**C/1985 XII Shoemaker, Log R plot, Figure 32.** The nucleus of this comet has not been observed, so we only have an upper limit. The turn off point can be determined with some precision, but the turn on point is unknown. To determine P-AGE we follow an iteration. We first assume $R_{ON} = R_{OFF}$ and calculate P-AGE with Asec = 9.6. We find that it has a very small P-AGE and that the SLC resembles that of comet C/1995 O1 Hale Bopp or that of comet 1P/Halley. It is then reasonable to assume that $R_{OFF}/R_{ON}$ = 2.04 similar to those two comets. Then $R_{ON}$ = -11.8 AU, and P-AGE can be iterated to a second more precise value of P-AGE ~4.3 cy, corresponding to a young giant object. MHM = Meech, Hainaut, Marsden (2004).

**C/1995 O1 Hale-Bopp, Log R plot, Figure 33.** Notice the sharp turn on and turn off points of this baby goliath comet. Compare this SLC with that of comet 1P/Halley. They look very similar. Notice the perihelion surge at q. The diameter of this comet comes from Paper V and was obtained scaling the SLC to that of comet 1P/Halley to whom it most resembles. Notice the break point at R=-6.3 AU pre-perihelion. RB = Rivkin & Binzel (IAUC 8479), TS= Tsumura (2006). SKS = Szabo et al. (2008).

**C/1996 B2 Hyakutake, Log R plot, Figure 34.** This was a bright young medium size comet that approached Earth, but it can be seen that its greatness was more due to the approach than to its absolute magnitude of m(1,1)= 4.3. The approach was at $\Delta$= 0.10 AU but the SLC does not exhibit any decrease or increase, thus no evidence of a Delta-effect. What is apparent at minimum approach is an increase in brightness, a kind of inverse Delta-effect.

This is most apparent in the time plot. The comet had an outburst at this time masking any Delta-effect (Schleicher and Osip, 2002). The nuclear line is due to Lisse et al. (1999).

**C/2001 OG108 LONEOS, Log R plot, Figure 35.** This methuselah goliath comet exhibits an odd shaped light curve, with a prominent S-Shape that up to now does not have a physical explanation. It is a methuselah comet with P-AGE= 102 cy. Although far into the future the comet will return in 2050. This is a large comet with D= 15.8 km. The nuclear line is due to Abell et al. (2005).

## 7.2 Time plots

**1P/Halley, time plot, Figure 36.** This secular light curve is unusual in that the comet exhibits a prominent belly, in fact looking as a pregnant comet. The turn-off of activity is a very sudden event. Additionally the nucleus had a significant outburst at $\Delta t$ = + 1772 d. The thermal wave penetration into the nucleus is clearly seen, thus sublimation must be taking place *in depth.*

**2P/Encke, time plot, Source 1, 1858, Figure 37.** This figure confirms the findings of the Log R plot. There is a significant difference between the two apparitions. The shift of the SLC from 1858 to 2003 is what is expected from cometary evolution: a narrowing of the SLC and a decrease in absolute magnitude. Since this conclusion is derived from a SLC all past criticisms about fading of comets not been realistic do not apply to this case. Other comets in this Atlas exhibit a narrowing of the SLC with P-AGE. Source of data, Kamel (1991).

**2P/Encke, time plot, Source 1, 2003, Figure 38.** Notice the time delay between perihelion and maximum brightness, LAG. The time-age derived from this plot is T-AGE= 103±9 cy, corresponding to a methuselah comet.

**6P/D'Arrest, time plot, Figure 39.** This comet has one of the largest values of LAG = +42 d. This causes a distortion of the SLC making difficult the determination of some parameters. The turn on point exhibits a very large slope of 0.17 mag/d. The turn off is indeterminate.

**9P/Tempel 1, time plot, Figure 40.** Notice the rounded shape of the secular light curve, and possible activity before turn on. CJ = Chen & Jewitt (1994).

**19P/Borrelly, time Plot, Figure 41.** The secular light curve exhibits a prominent belly after perihelion. The turn-off is beyond 320 days. BHSBOH = Buratti et al. (2003). LTW = Lamy et al. (1998).

**21P/Giacobinni-Zinner, time plot, Figure 42.** The break in the SLC can be detected in this plot before perihelion. The turn off is not well determined. CJ = Chen & Jewitt (1994). M = Mueller (1992), CJ = Chen & Jewitt (1994), H = Hergenrother (Scotti 2001).

**26P/Grigg-Skejellerup, time plot, Figure 43.** Notice the round shape of the secular light curve characteristic of old comets and water ice sublimation, and the short time of activity that does not reach 7 months. LTLRH = Licandro et al. (2000).

**28P/Neujmin 1, time plot, Figure 44.** Notice the round shape of the secular light curve and the short interval of activity. The Epoch of this plot is 1913 which implies that the photometry must be poor and that there is an urgent need to observed this comet in the coma phase. In 2003 perihelion coincided with conjunction with the Sun, so the coma phase went unobserved. The next opportunity will be in October of 2020.

**29P/Schwassmann-Wachmann 1, time plot, Figure 45.** The SLC shows that the comet is active all around the orbit although there seems to be a minimum of activity ~-2000 d before perihelion. So observers should look for periods of minimum activity to observe the nucleus. These are infrequent.

**32P/Comas Sola, time plot, Figure 46.** This SLC seems to be well defined, but due to the intense decay in brightness with time it would be worthwhile to observer every apparition carefully. If the trend continues the comet will turn off in a few more revolutions.

**39P/Oterma, time plot, Figure 47.** This comet is active all around the orbit. The comet is annual. The turn on and turn off points are beyond aphelion. LAG= ~0 for this comet. The approach to Jupiter took place in 1963. F = Fernandez (IAUC7689).

**45P/Honda-Mrkos-Pajdusakova, time plot, Figure 48.** The light curve is well defined. Notice the steep descent at turn off. The comet has almost no lag. Notice that this is a dwarf comet of only 0.9 km diameter.

**65P/Gunn, time plot, Figure 49.** This comet is very similar to 39P but with a perihelion LAG= +242 d, a huge value. A probable interpretation is that the nucleus' pole points to the Sun at LAG.

**67P/Churyumov-Gerasimenko, time plot, Figure 50.** Notice the rounded shape of the secular light curve and the uncertainty in the nucleus magnitude. The photometric anomaly has been expanded in an additional plot. LTWJK = Lamy et al. (2003).

**73P/Schwassmann-Wachmann 3C, 1990 and 1995, time plot, Figure 51.** This comet experienced an outburst, changing the SLC from 1990 to 1995. Due to the outburst the comet rejuvenated, its P-AGE going from 55 cy in 1990 to 21 cy in 1995. There is no way to calculate a T-AGE because we do not know the turn off date. This plot is an excellent example of a splitting event producing an outburst. The two outbursts had about the same amplitude, 6.4 vs 6.3 magnitudes. G = Green (2007). S = Scotti (1995). CJ = Chen & Jewitt (1994).

**73P/Schwassmann-Wachmann 3C, 2001, time plot, Figure 52.** The comet seems to be even more rejuvenated. S = Scotti (1995). HL = Hergenrother and Larson, cited in Scotti (2001).

**81P/Wild 2, time plot, Figure 53.** Notice the secular increase in maximum brightness. CJ = Chen & Jewitt (1994). LTLRH = Licandro et al. (2000). LFC = Lowry et al. (2003)

**85P/Boethin, time plot, Figure 54.** This middle age comet exhibits a round SLC typical of water ice dominated comets. R= Roemer. GK = Gilmore & Kilmartin (IAUC 4121).

**107P/Wilson-Harrington, time plot, Figure 55.** The activity of this comet took place over a few nights, thus it is not surprising that its T-AGE is 720 cy, a methuselah comet.

**109P/Swift-Tuttle, time plot, Figure 56.** This young giant comet has a very smooth light curve. Unfortunately this comet will not be seen for another century.

**133P/Elst-Pizarro, time plot, Figure 57.** The shape of the outburst is clearly seen, as well as the possible activity at aphelion. The outburst was initiated at $T_{ON}$ = 42±4 d (Sekanina, IAUC 6473), reached a maximum amplitude of $A_{SEC}(1,R)$ = 2.3±0.2 magnitudes at LAG=155±10 d, and turned off at $T_{OFF}$= 233±10 d. A Time-Age, T-AGE = 80 (+30, -20) cy can be calculated from this plot. By good luck our own observations (open down triangles) and those of Lowry and Fitzsimmons help define the extent of activity. The date of the next perihelion is indicated. L&F= Lowry & Fitzsimmons (2005).

**C/1983 H1 IRAS-Araki-Alcock, time plot, Figure 58.** This famous comet approached Earth at very short distance. The SLC is somewhat uncertain, but there is evidence that this comet is old.

**C/1983 J1 Sugano-Saigusa-Fujikawa, time plot, Figure 59.** The SLC of this comet is somewhat uncertain, but there is evidence to conclude that this is an old comet.

**C/1995 O1 Hale-Bopp, time plot, Figure 60.** Compare this plot with that of comet 1P/Halley. They are very similar, both with an extensive belly. Notice that the surge is not apparent in the time plot. RB = Rivkin & Binzel (IAUC 8479), TS= Tsumura (2006). SKS = Szabo et al. (2008).

**C/1996 B2 Hyakutake, time plot, Figure 61.** The comet exhibits a normal SLC with a very small negative LAG. There is a photometric anomaly at t~ -39 d, seen expanded in a view in additional plots due to a water outburst (Schleicher & Osip 2002).

**C/2001 OG 108 LONEOS, time plot, Figure 62.** This very old Halley type comet exhibits the same S-shaped anomaly as 19P/Borrelly. The origin of this feature is unknown. The reason for having observations well below the nucleus is not understood either.

**7.3  Additional plots**

**1P/Halley, photometric anomaly, Figure 63.** The break point is clearly seen. The brightness changes slope abruptly at perihelion. The gaps at perihelion and at Log R= +0.75 are due to conjunction with the Sun. There is a dip in the SLC from R= +1.11 AU to R= +1.24 AU, of amplitude m= 0.5±0.1 mag, or T= +45 to +56 d. The reason we believe this photometric anomaly is real is because there was a dip in the $C_2$ production rate and the $J_o$ infrared value, at R= +1.21 AU (Morris & Hanner 1993). Most probably this feature is due to a topographic effect.

**2P/Encke, Epoch 2003, time plot, photometric anomaly at aphelion, Source 2, Figure 64.** Source 1 is active at perihelion. Source 2 is active at aphelion (Paper V). The dashed line tries to describe an envelope of the observations but in no way implies that the activity is sustained. Between $\Delta t$ = +310 and +392 d there is a well observed linear decay in brightness that might be an indication of topography. The maximum activity (2.7±0.2 mag) is quite significant. The nuclear magnitude derived from the phase plot is in agreement with the lowest observations at aphelion by Fernandez et al. (2005) and Meech et al. (2001) at $\Delta t$= +390 days.

**2P/Encke, Epoch 2003, photometric anomaly at aphelion, expanded, Figure 65.** The anomaly has been expanded from the previous Figure to see its extent and shape. The brightness decreases linearly from $\Delta t$ = +310 to +392 d after perihelion, an interval of 82 days. There seems to be an absolute minimum of the secular light curve at $\Delta t$= +393 to 413. The photometric anomaly can be due to a topographic effect or the turn off of an active region. Additional observations of this region are highly desirable because this might be the absolute minimum brightness of the whole orbit.

**67P/Churyumov-Gerasimenko, photometric anomaly, Figure 66.** This comet exhibits a prominent photometric anomaly from -119 d to -6 d before perihelion. It could be due to an odd shape of the nucleus, an odd distribution of volatiles, or to the turn off of an active region. The anomaly has an amplitude of 1.9 magnitudes, and observations from the apparitions of 1982, 1996 and 2002 show it, thus it must be a permanent feature of the SLC.

**C/1995 O1 Hale-Bopp, photometric anomaly, Figure 67.** This feature is placed exactly at perihelion, being approximately symmetric with respect to it. An increase in brightness *above the power law* is apparent. This phenomenon does not have as of now a satisfactory explanation.

**C/1996 B2 Hyakutake, photometric anomaly, Figure 68.** The anomaly appears at the time of minimum approach, when we should expect the action of the Delta-effect. This effect is a decrease in brightness, but what is seen is *an increase* in brightness. A comparison with the water production rate shows that the brightness increase correlates with a water outburst.

Plots and tables in ASCII format can be downloaded from the site: http://webdelprofesor.ula.ve/ciencias/ferrin.

**ACKNOWLEDGMENTS**

To Matthew M. Knight and to Carl Hergenrother for their many suggestions to improve the scientific value of this paper. Some secular light curves contain observations carried out at the National Observatory of Venezuela (ONV), managed by the Center for Research in Astronomy (CIDA), for the Ministry of Science and Technology (MinCyT). We thank the TAC of CIDA for assigning telescope time to observe these faint objects. To the Council for Scientific, Technologic and Humanistic Development of the University of the Andes for their support through grant number C-1281-04-05-B.

Table 1. Comets with power laws at turn on.

| Comet | P-AGE (cy) | Slope $n_{ON}$ |
|---|---|---|
| C/1995 O1 | 1.8 | 10.70 |
| 1P | 7 | 8.92 |
| 81P | 13 | 9.31 |
| 19P | 14 | 11.70 |
| 21P | 20 | 9.09 |
| 9P | 21 | 7.70 |
| C/2001 OG108 | 102 | 10.90 |

Table 2. Comets with a break point in the power law.

| Comet | P-AGE[cy] | $R_{BP}$ [AU] | $m_{BP}$ |
|---|---|---|---|
| C/1995 O1 | 1.8 | -6.3 | 4.6 |
| 1P | 7.2 | -1.70 | 5.6 |
| 81P | 13 | -1.9 | 10.0 |
| 21P | 20 | -1.58 | 10.6 |
| 9P | 21 | - 2.08 | 13.8 |

Table 3. Look alike comets

| Comet #1 | P-AGE #1 | P-AGE #2 | Comet #2 |
|---|---|---|---|
| 1P/Halley | 7 | 1.8 | C/1995 O1 Hale Bopp |
| 19P/Borrelly | 14 | 102 | C/2001 OG108 LONEOS |
| 21P/Giacobini-Zinner | 20 | 13 | 81P/Wild 2 |
| 26P/Grigg-Skjellerup | 89 | 100 | 28P/Neujmin 1 |
| 28P/Neujmin 1 | 100 | 76 | C/1983 H1 IRAS-Araki-Alcock |
| C/1983 H1 IRAS-Araki-Alcock | 76 | 89 | 26P/Grigg-Skjellerup |
| 39P/Oterma | 8 | 10 | 65P/Gunn |
| 107P/Wilson-Harrington | 1370 | 1460 | 133P/Elst-Pizarro |

Table 4. Basic photometric parameters deduced from the secular light curves in order or increasing P-AGE.

| Comet | Rank | Figure | $R_{ON}$ [AU] | $R_{OFF}$ [AU] | $T_{ON}$ [d] | $T_{OFF}$ [d] | P-AGE [cy] | T-AGE [cy] | $m_1$ | $V_N$ (1,1,0) | $A_{SEC}$ (1,1) | LAG [d] | q [AU] | Q [AU] | $D_{EFFE}$ [km] |
|---|---|---|---|---|---|---|---|---|---|---|---|---|---|---|---|
| 29P/SW 1 | 1 | 16,45 | ~80 | ~300 | ~7800 | ~6700 | ~0.5 | ~0.5 | -3.6 | 10.9 | ~14.7 | ----- | 5.44 | 7.28 | 54 |
| C/1995O1 Hale-Bopp | 2 | 33,60 | -17.3 | 33.9 | -2550 | ~7650 | 2.4 | ~0.75 | -1.2 | 10.6 | 11.5 | +45 | 0.91 | 370 | 54 |
| C/1984 K1 Shoemaker | 3 | 32 | -7.3 | 24.0 | ---- | ---- | <4.8 | ---- | 3.8 | <13.4 | >9.6 | ---- | 2.70 | e=1 | <14 |
| 1P/Halley | 4 | 7,36 | -6.15 | 12.6 | -497 | 1495 | 7.1 | 4.2 | 3.4 | 14.13 | 10.8 | +11.2 | 0.59 | 35.3 | 9.8 |
| 39P/Oterma | 5 | 18,47 | -7.7 | 6.6 | -2747 | 3000 | 9 | 1.4 | 3.0 | 14.52 | 11.5 | 0 | 3.39 | 4.53 | 6.3 |
| 65P/Gunn | 6 | 20,49 | -8.3 | 6.1 | -1417 | 1550 | 9 | 2.8 | 1.0 | 14.8 | 10.8 | +242 | 2.46 | 4.74 | 7.4 |
| 81P/Wild 2 | 7 | 24,53 | -5.1 | 4.5 | -648 | 690 | 13 | 6 | 5.8 | 16.53 | 11.4 | -13 | 1.58 | 5.30 | 3.9 |
| 19P/Borrelly | 8 | 12,41 | -2.9 | 5.8 | -266 | >300 | 14 | ---- | 4.2 | 15.9 | 11.6 | +6 | 1.37 | 5.87 | 4.5 |
| 73P/SW3C (2001) | 9 | 23,52 | -5.2 | 4.2 | -950 | 478 | 18 | 7 | 8.3 | 17.1 | 8.7 | ---- | 0.94 | 5.18 | 2.2 |
| C/1996B2 Hyakutake | 10 | 34,61 | -3.2 | 3.52 | -167 | 189 | 18 | 22 | 4.3 | 15.7 | 11.6 | -6 | 0.23 | 1901 | 4.8 |
| 101P/Chernykh | 11 | 26 | -3.1 | 4.9 | ---- | ---- | 22 | ---- | 10.2 | 16.1 | ----- | ---- | 2.36 | 9.24 | 5.0 |
| 21P/Giacobini-Zinner | 12 | 13,42 | -3.6 | 4.4 | -280 | 470 | 22 | 15 | 8.2 | 16.1 | 7.9 | -10 | 1.03 | 6.0 | 3.1 |
| 9P/Tempel 1 | 13 | 11,40 | -3.5 | 4.2 | -410 | 659 | 22 | 9 | 6.4 | 15.3 | 9.0 | -10 | 1.50 | 4.74 | 5.5 |
| 109P/Swift-Tuttle | 15 | 28,56 | -2.8 | 3.4 | -123 | 156 | 28 | 40 | 3.8 | 12.0 | 8.2 | +1 | 0.96 | 52 | 27 |
| 32P/Comas-Sola | 17 | 17,46 | -3.3 | 3.6 | -324 | 372 | 30 | 19 | 8.6 | 15.6 | 7.0 | 0 | 1.85 | 6.67 | 3.0 |
| 6P/D'Arrest | 14 | 10,39 | -1.6 | <3.2 | -70 | >250 | 32 | <26 | 7.2 | 18.3 | 9.5 | +42 | 1.35 | 5.62 | 1.6 |
| 67P/Churyumov-G | 16 | 21,50 | -2.8 | 4.0 | -232 | 408 | 37 | 21 | 9.6 | 15.4 | 5.8 | +33 | 1.30 | 5.73 | 5.5 |
| 85P/Boethin | 18 | 25,54 | -1.8 | 1.7 | -107 | 91 | 49 | 54 | 5.9 | 14.3 | 8.4 | -4 | 1.11 | 8.91 | 6.3 |
| C/1983 J1 SSF | 19 | 31,59 | ~1.5 | 1.5 | ~-60 | 60 | 55 | 84 | 10.5 | 19.8 | 8.8 | ---- | 0.47 | e=1 | 0.7 |
| 45P/Honda-Mrkos-P | 20 | 19,48 | -1.5 | 2.0 | -70 | 118 | 55 | 64 | 12.0 | 19.5 | 7.5 | +2 | 0.53 | 5.5 | 0.9 |
| 2P/Encke 1858 | 21 | 8,37 | -2.2 | 1.7 | -126 | 105 | 58 | 61 | 9.1 | 15.5 | 6.4 | -13 | 0.34 | 4.09 | 5.1 |
| C/1983H1 IRAS-A-A | 22 | 30,58 | -1.2 | 1.5 | -45 | 80 | 84 | 113 | 7.8 | 14.6 | 6.4 | +10 | 0.99 | 199 | 8.6 |
| 26P/G-Skjellerup | 23 | 14,43 | -1.4 | 1.8 | -73 | 132 | 89 | 85 | 11.4 | 16.7 | 5.2 | +14 | 0.99 | 4.93 | 3.0 |
| 2P/Encke 2003 | 24 | 9,36 | -1.6 | 1.5 | -87 | 94 | 98 | 103 | 10.8 | 15.5 | 4.8 | +6 | 0.34 | 4.09 | 5.1 |
| 28P/Neujmin 1 | 25 | 15,44 | -2.1 | 2.4 | -115 | 167 | 100 | 100 | 9.6 | 12.8 | 3.2 | +7 | 1.53 | 12.1 | 23 |
| C/2001 OG108 LON | 26 | 35,62 | -1.6 | 1.6 | -78 | 96 | 102 | 117 | 8.7 | 13.1 | 4.4 | -5 | 0.99 | 25.6 | 15.8 |
| 107P/Wilson-Harrington | 27 | 27,55 | +1.1 | 1.4 | +38 | 74 | ---- | 760* | 12.9 | 16.3 | 3.3* | +42 | 1.00 | 4.28 | 3.3 |
| 133P/Elst-Pizarro | 28 | 29,57 | +2.6 | 2.8 | +42 | 233 | ---- | 280* | --- | 16.0 | 1.4* | +155 | 2.63 | 3.68 | 4.6 |

**\*** T-AGE(1,LAG) is given because P-AGE(1,1) can not be calculated.

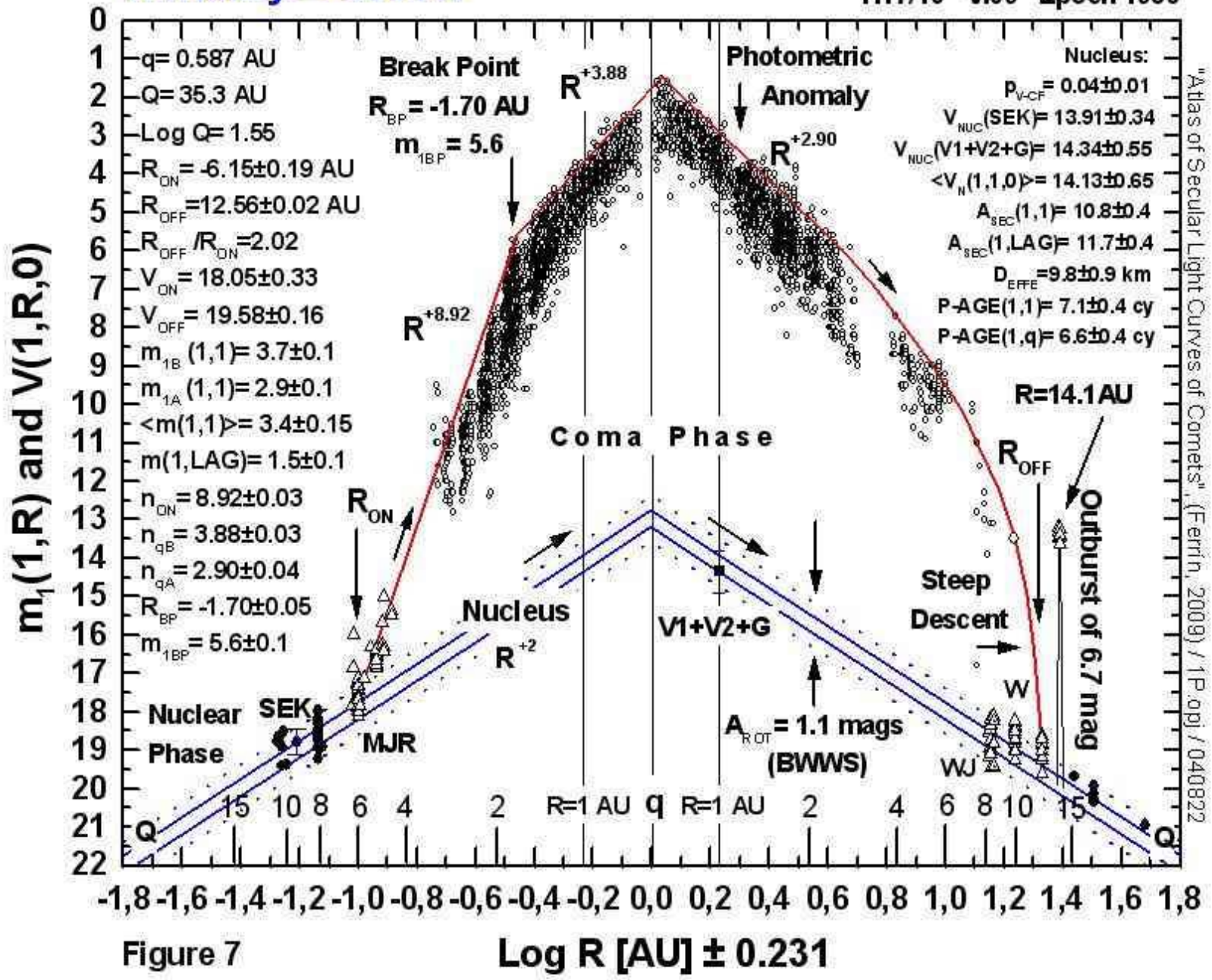

Figure 7

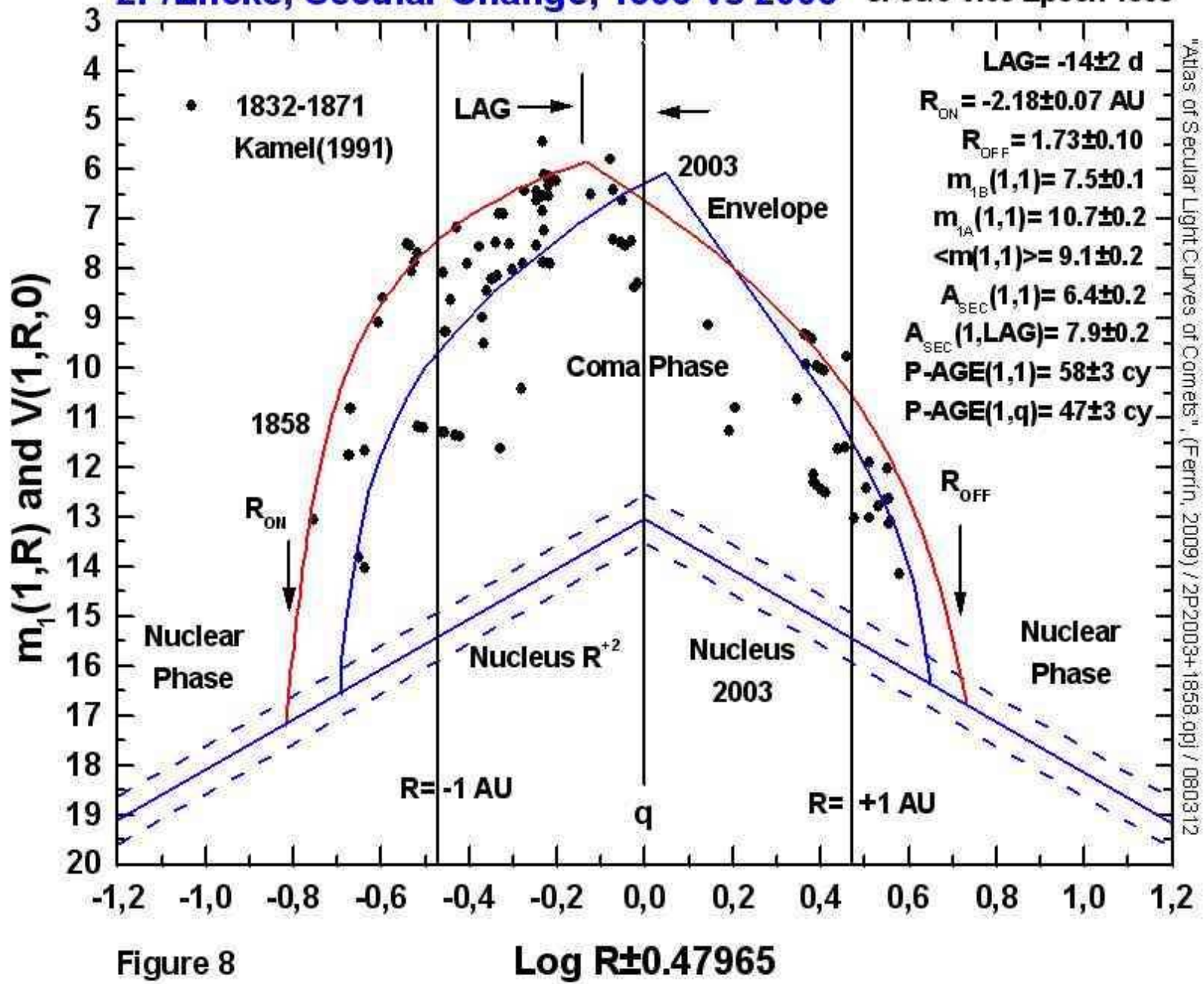

Figure 8

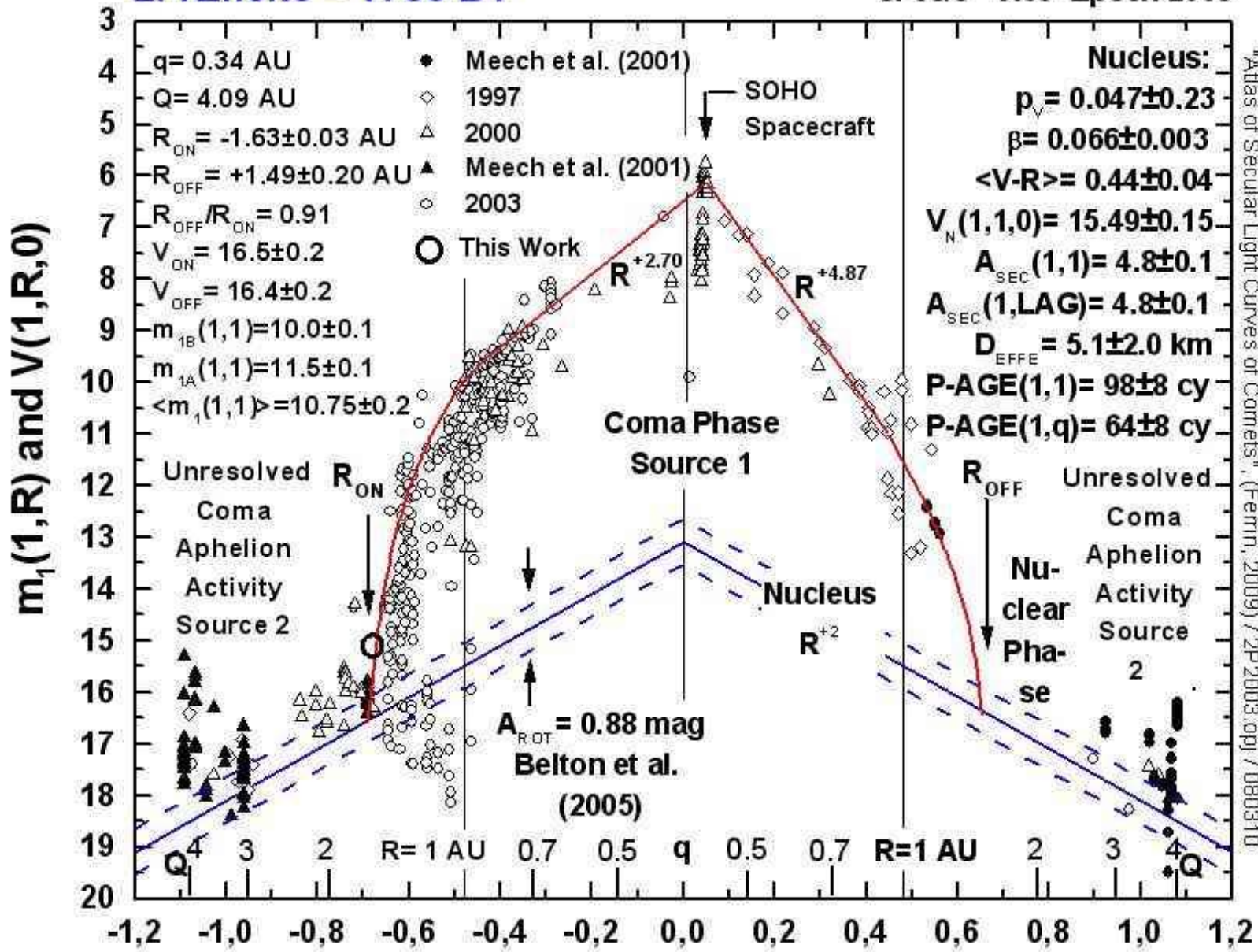

Figure 9

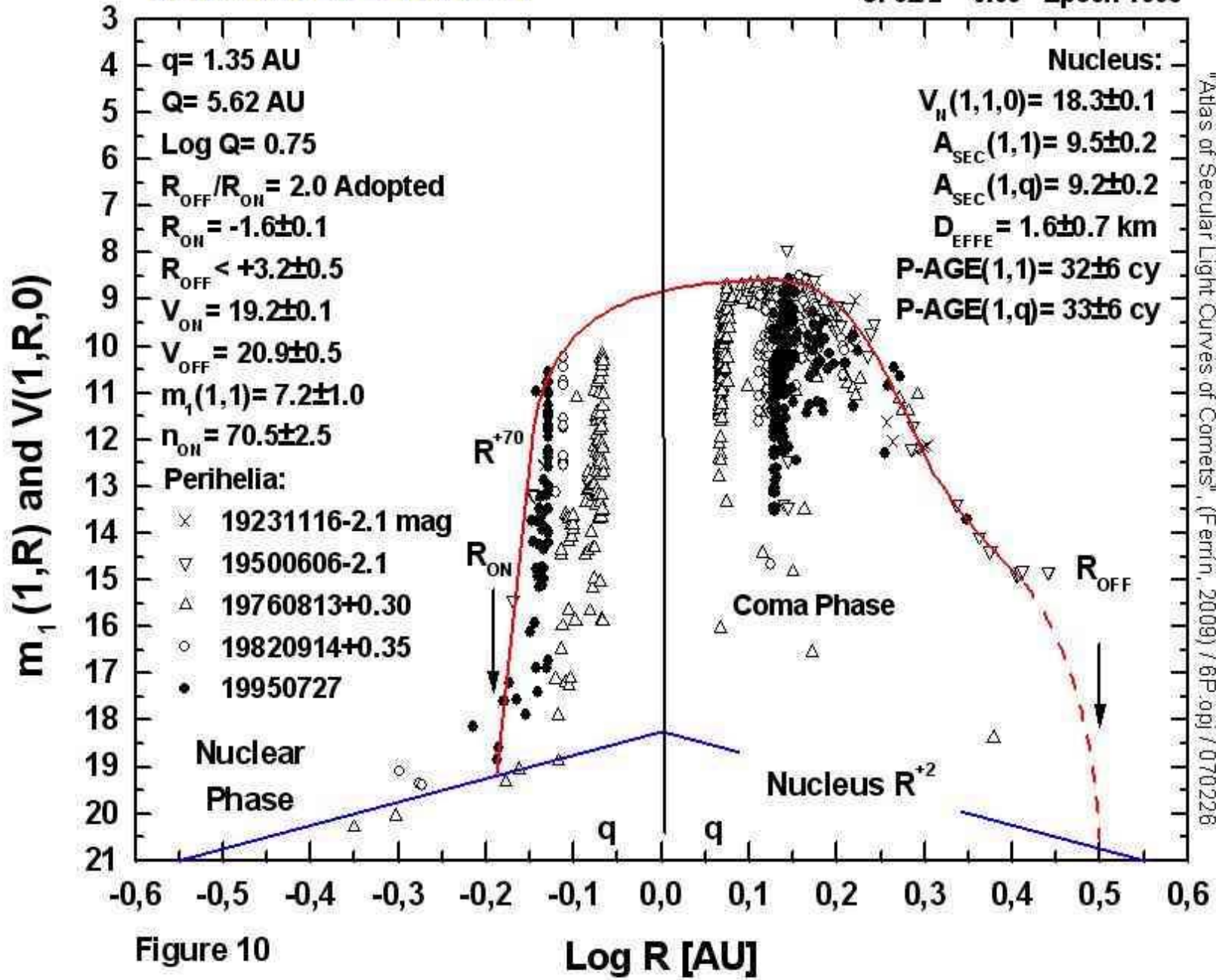

Figure 10

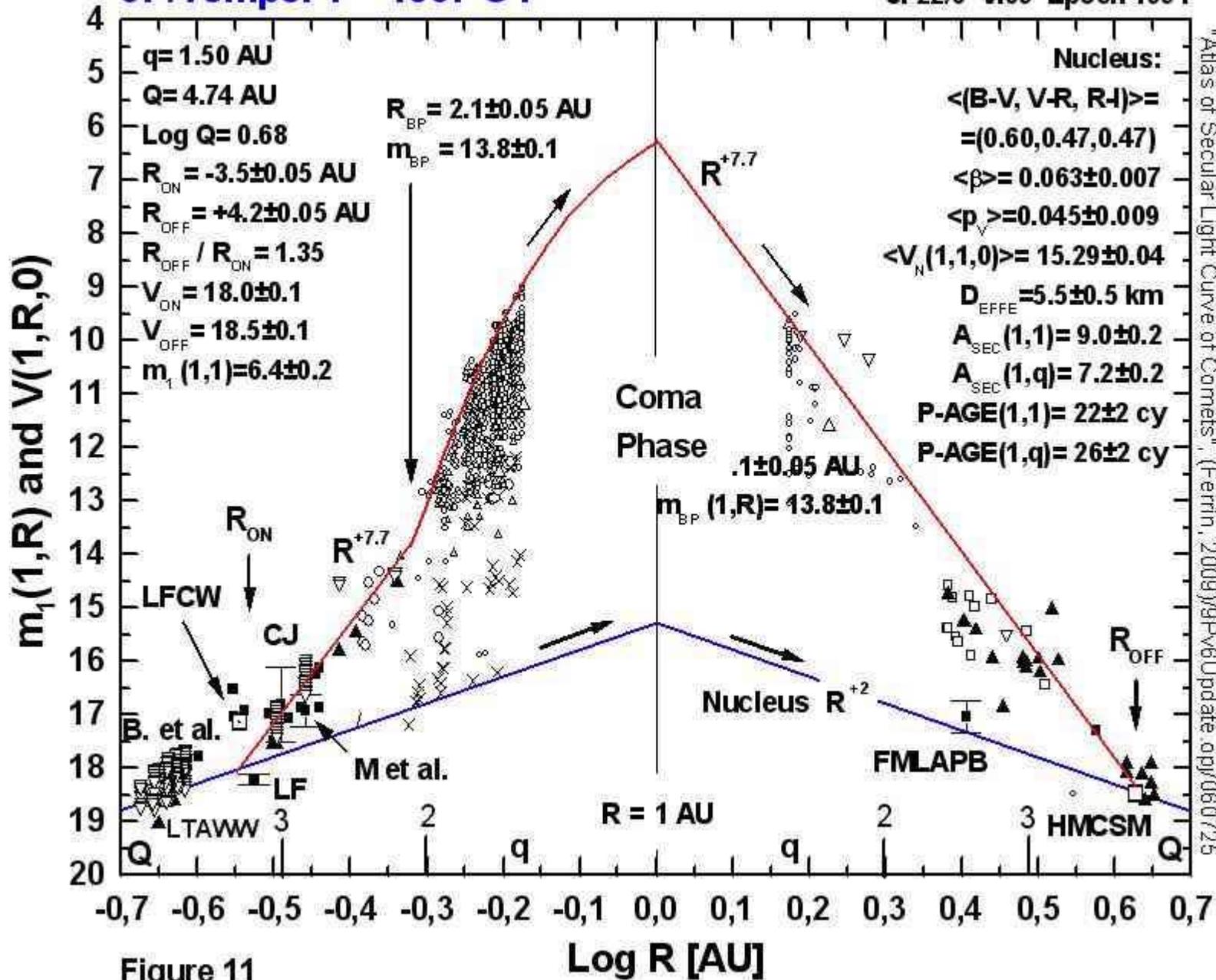

Figure 11

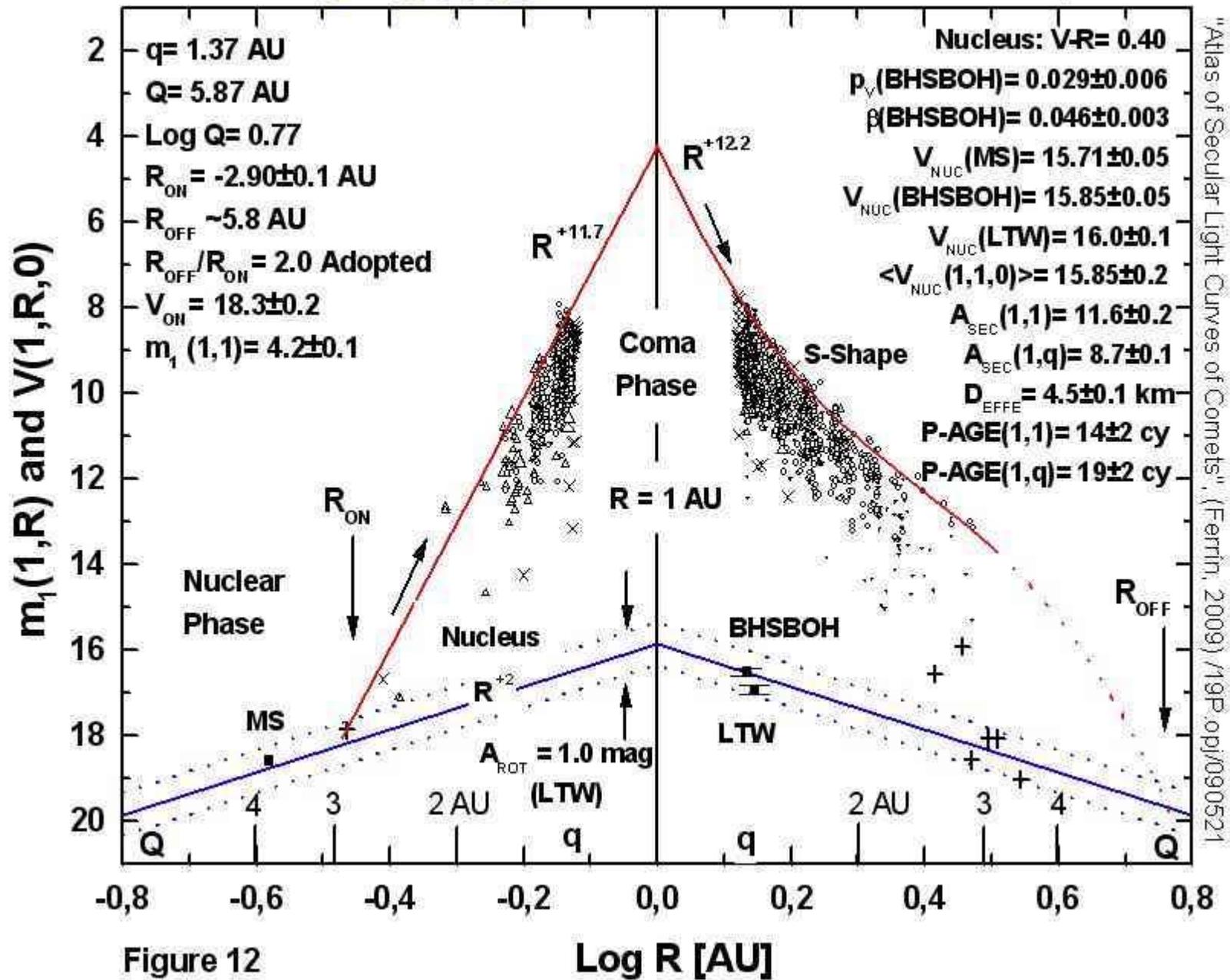

Figure 12

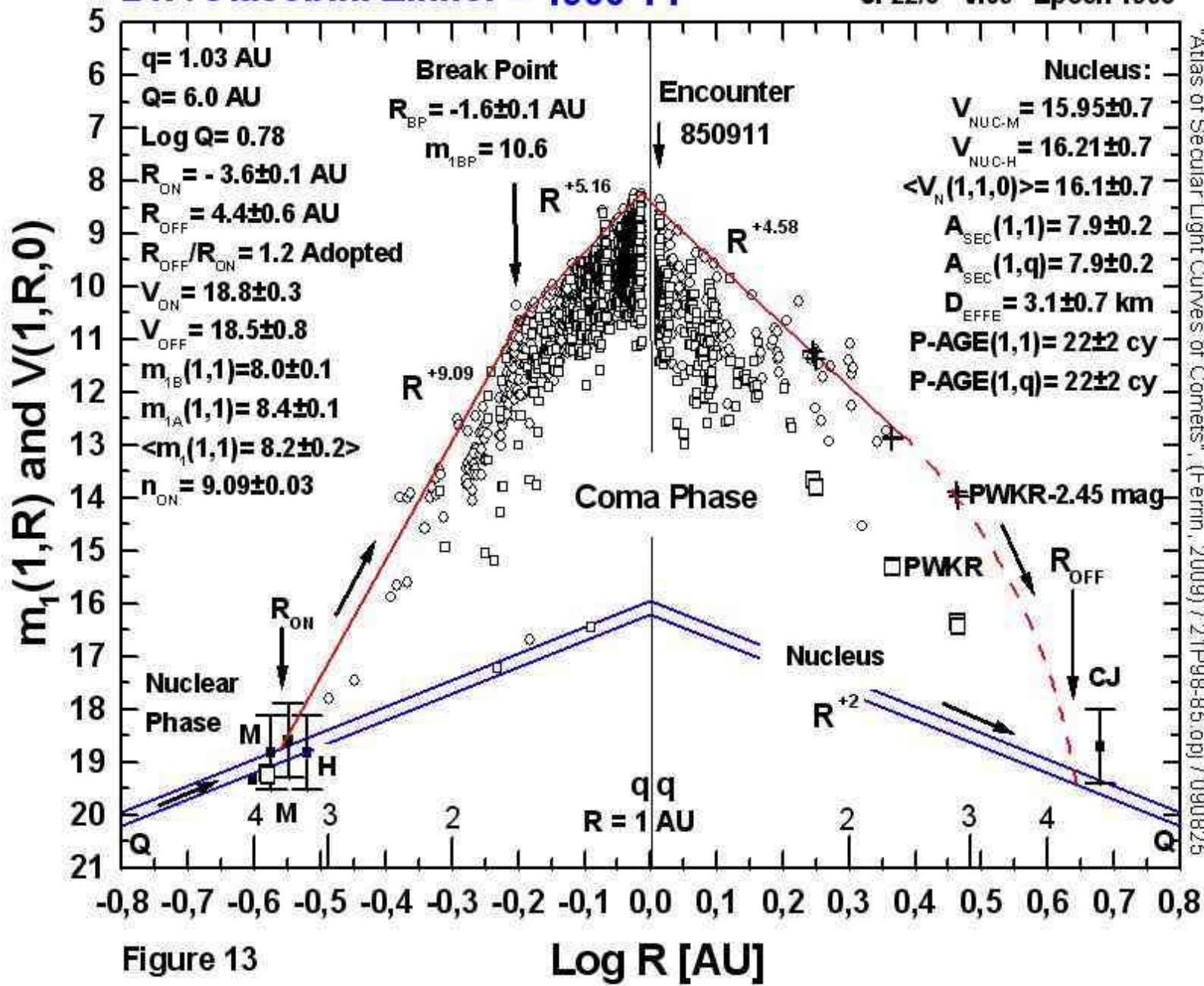

Figure 13

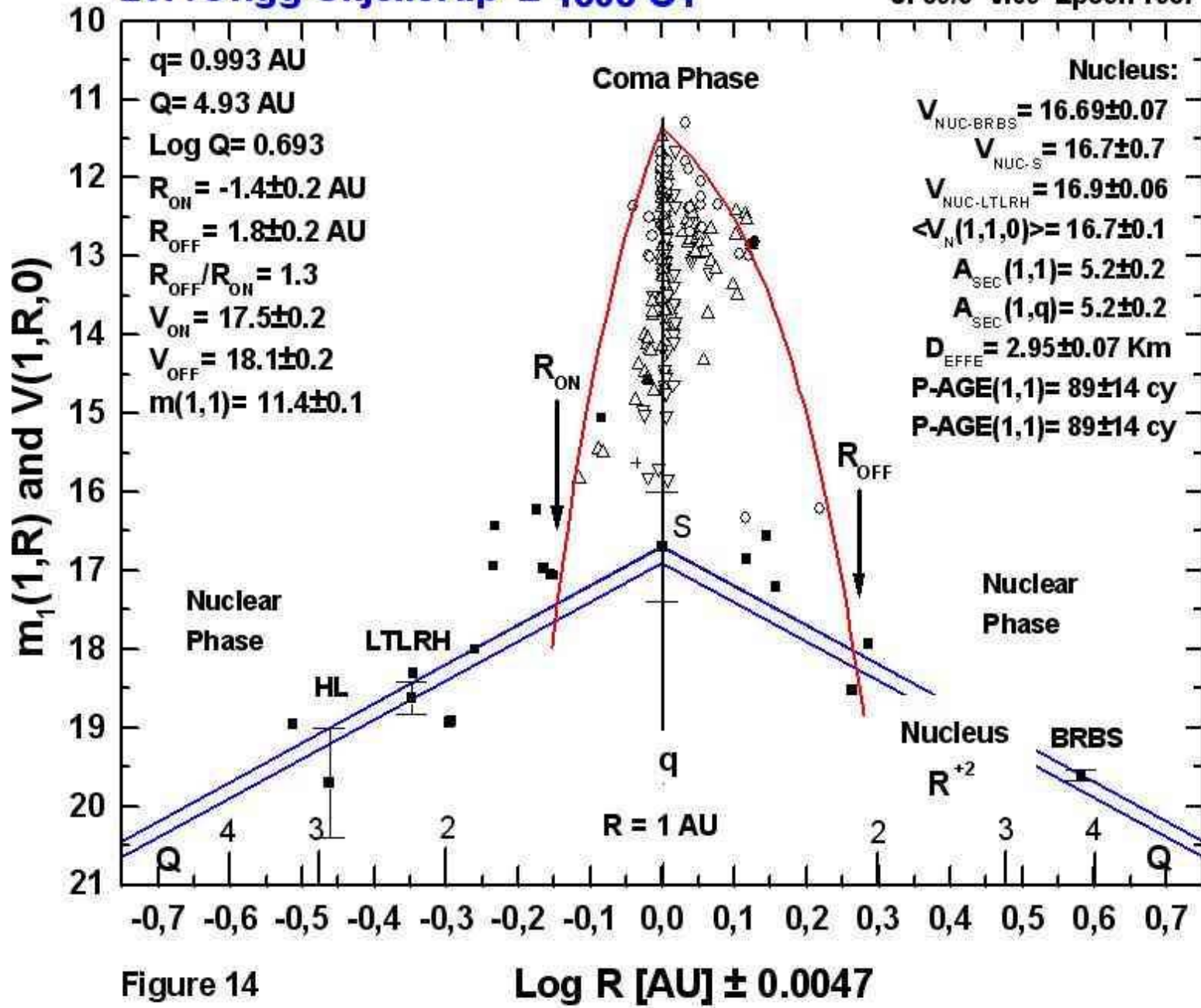

Figure 14

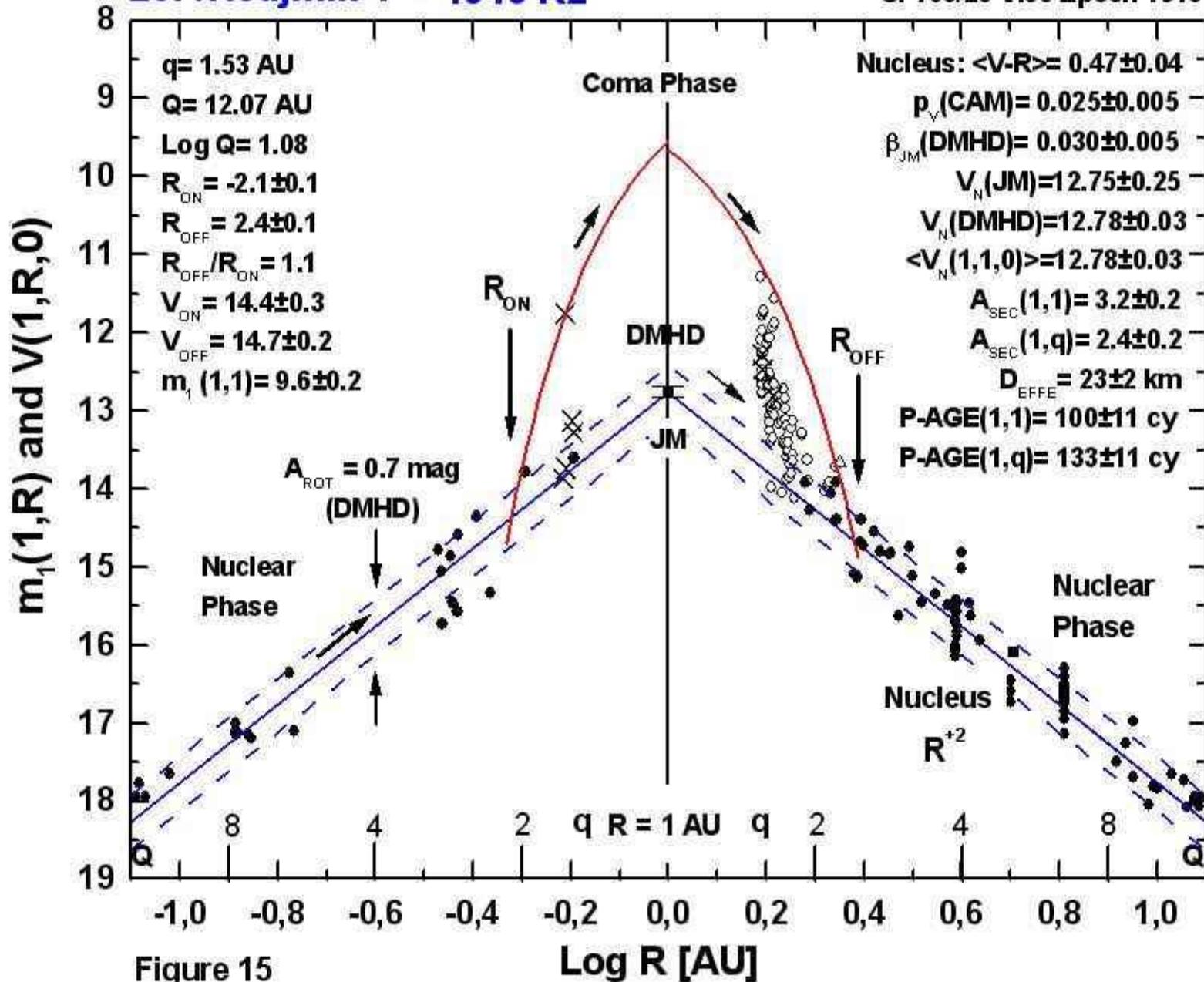

Figure 15

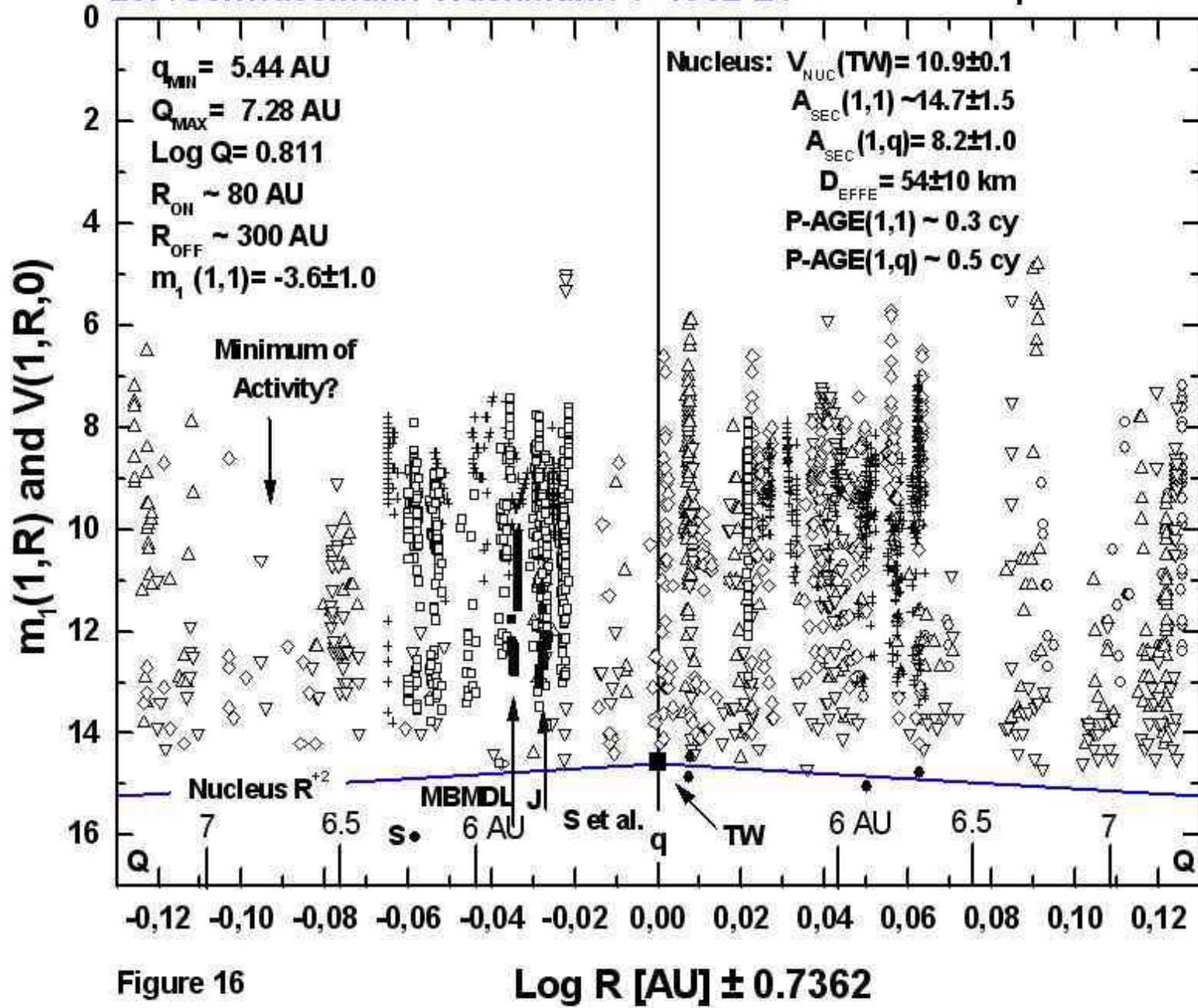

Figure 16

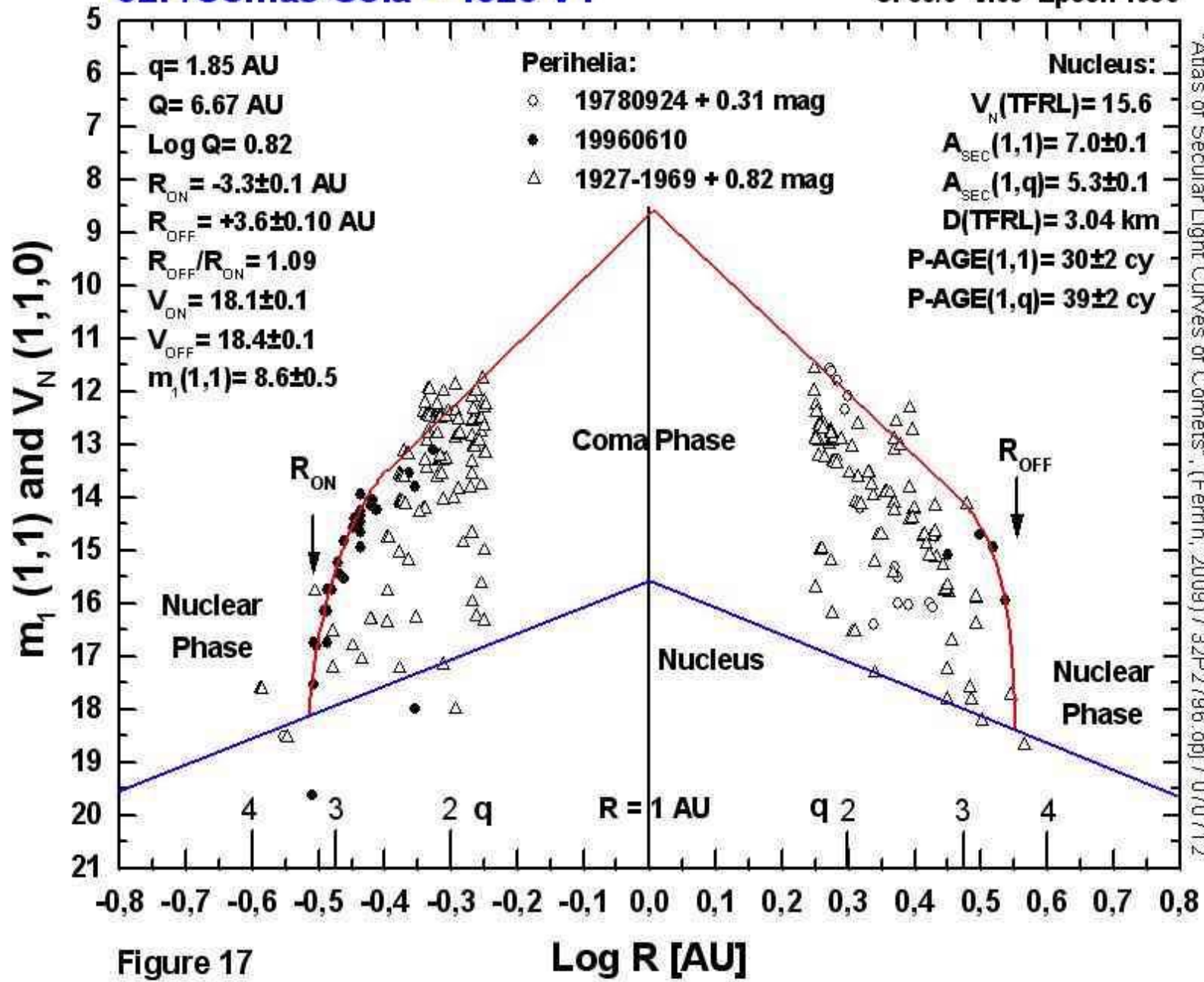

Figure 17

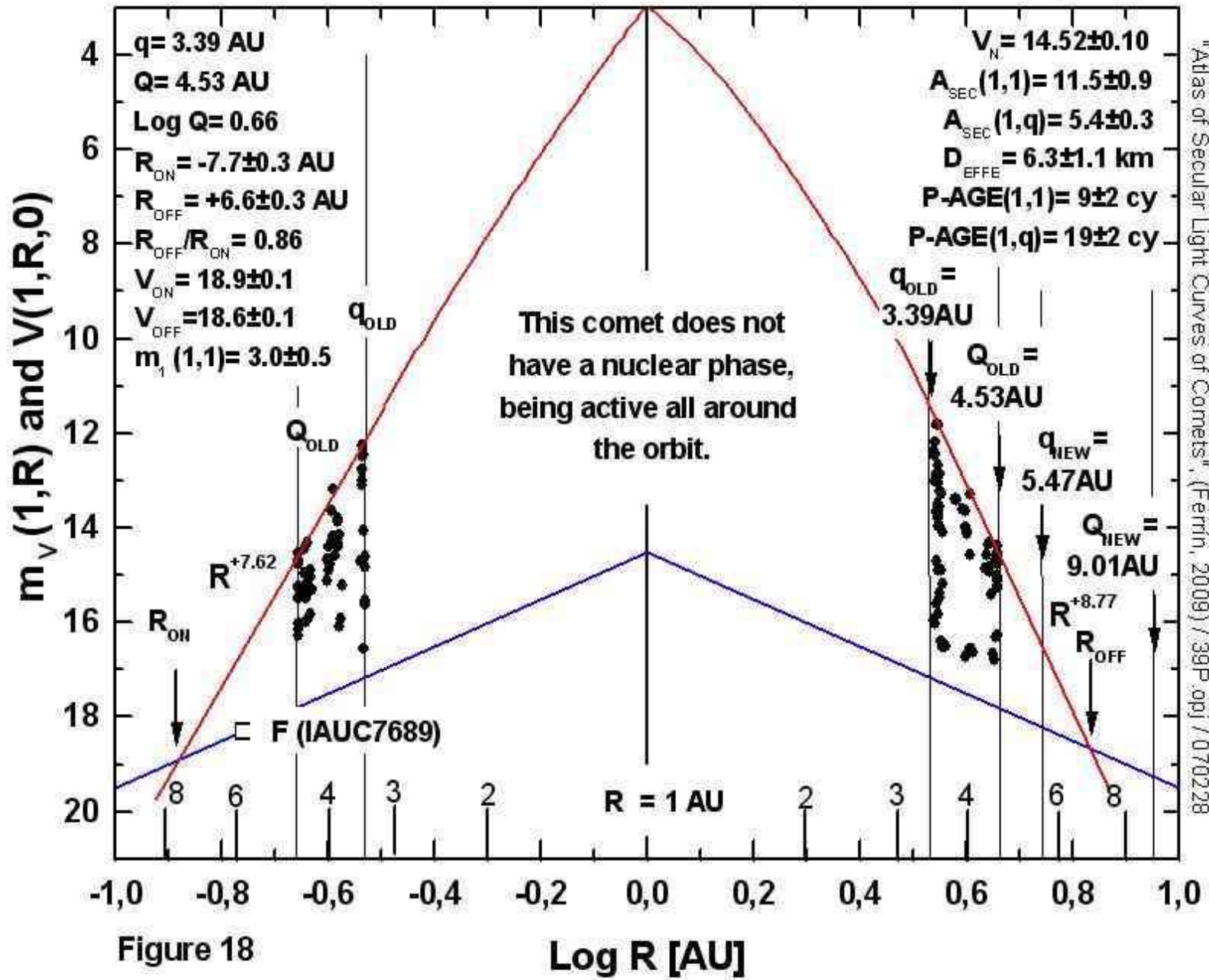

Figure 18

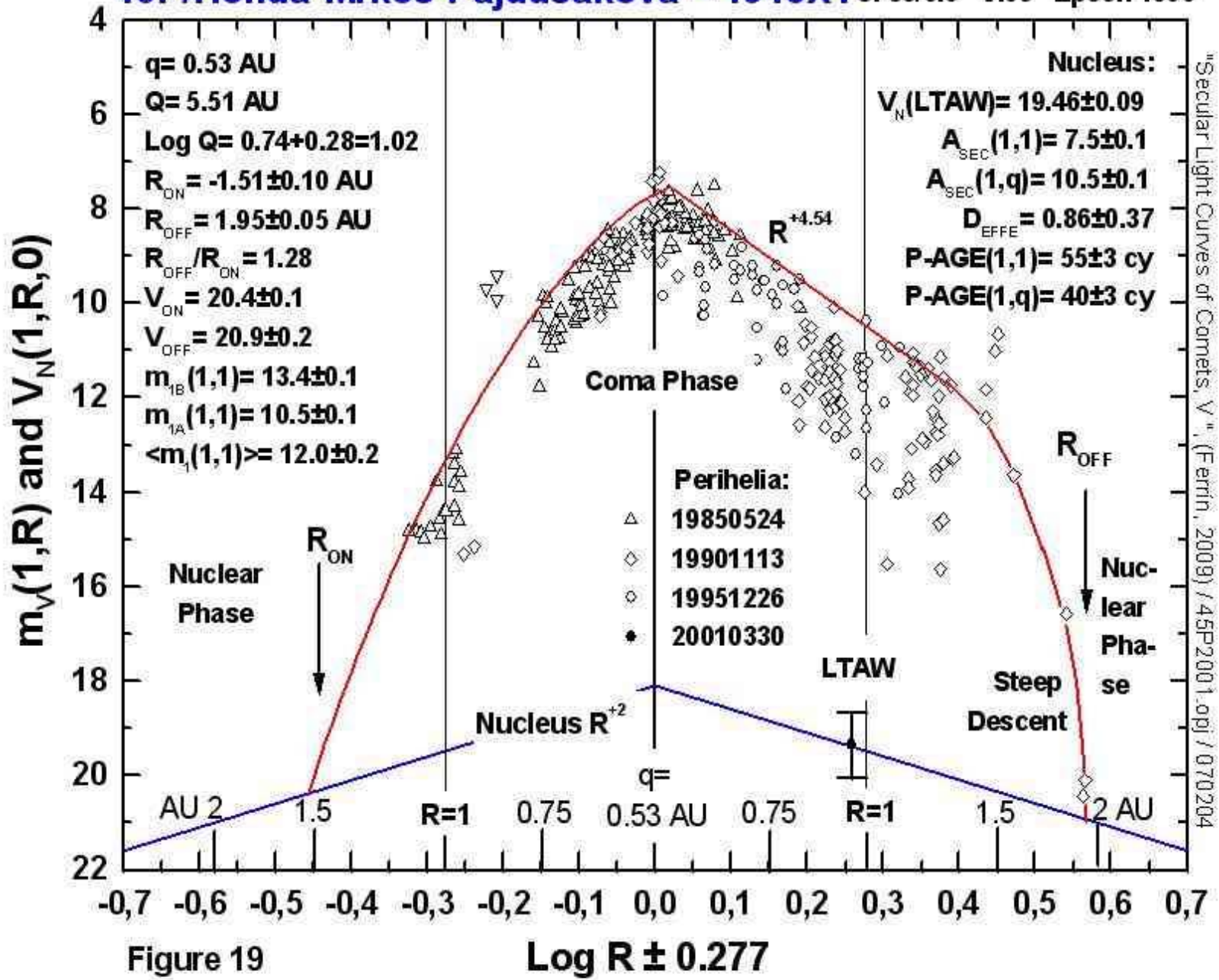

Figure 19

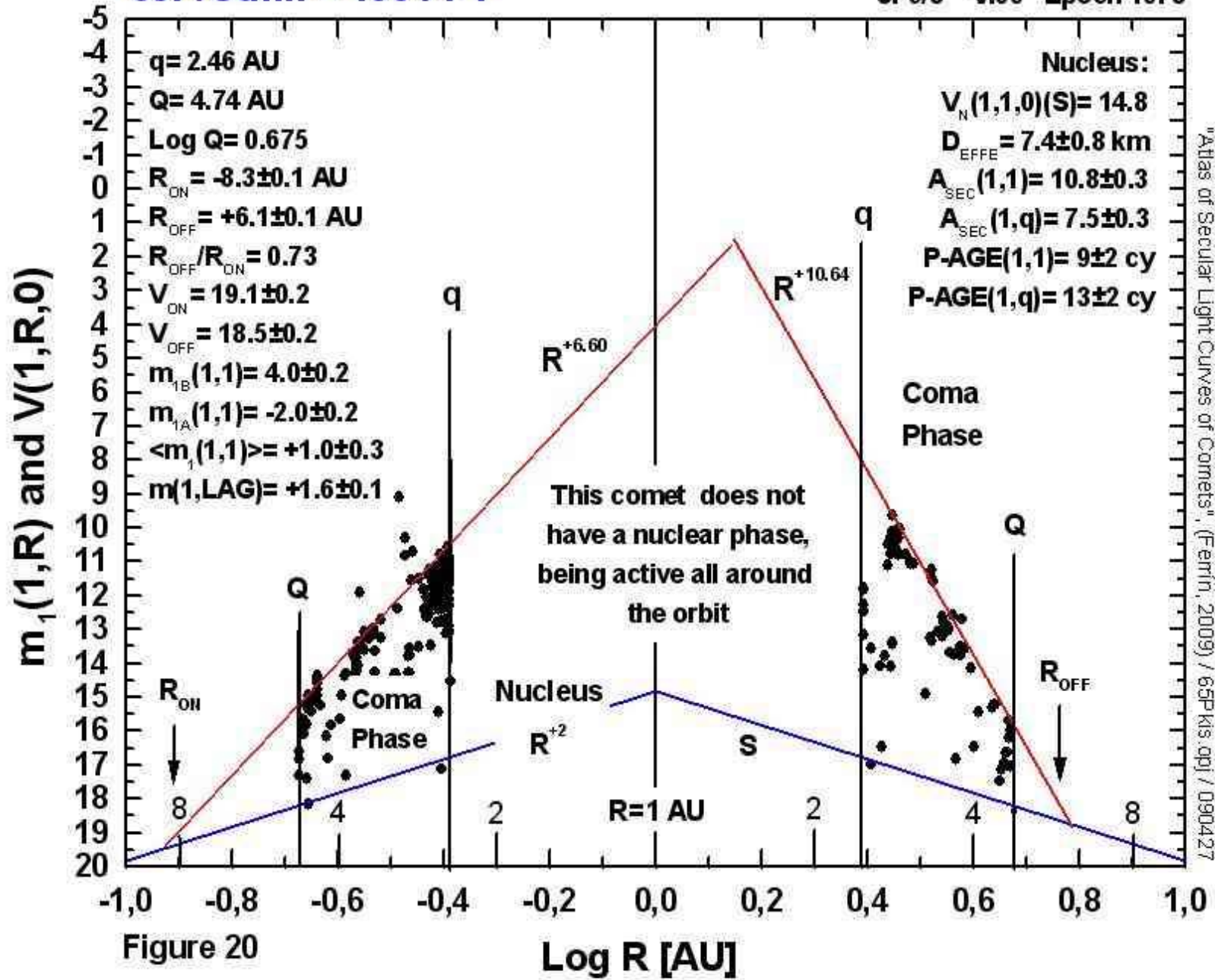

Figure 20

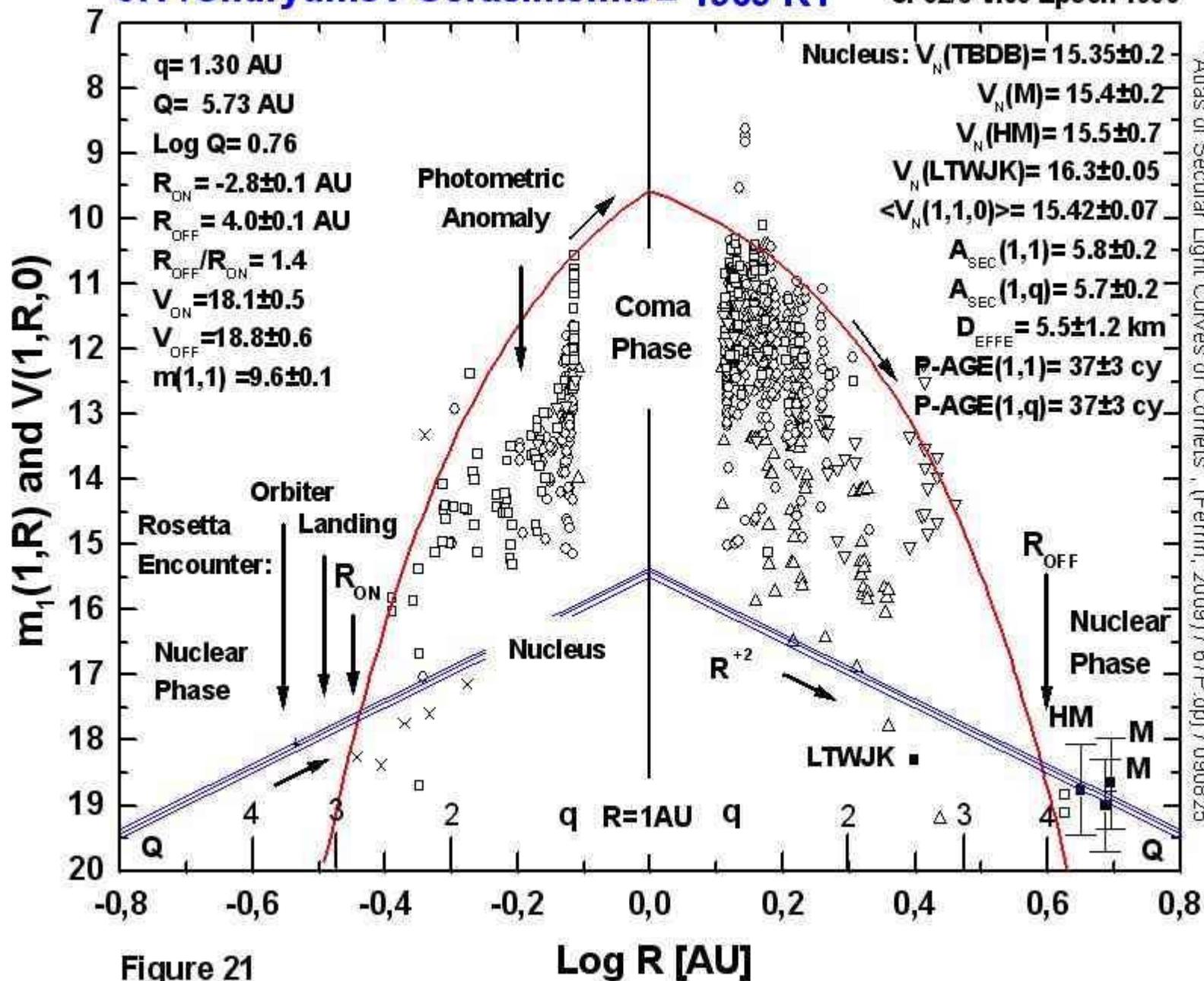

Figure 21

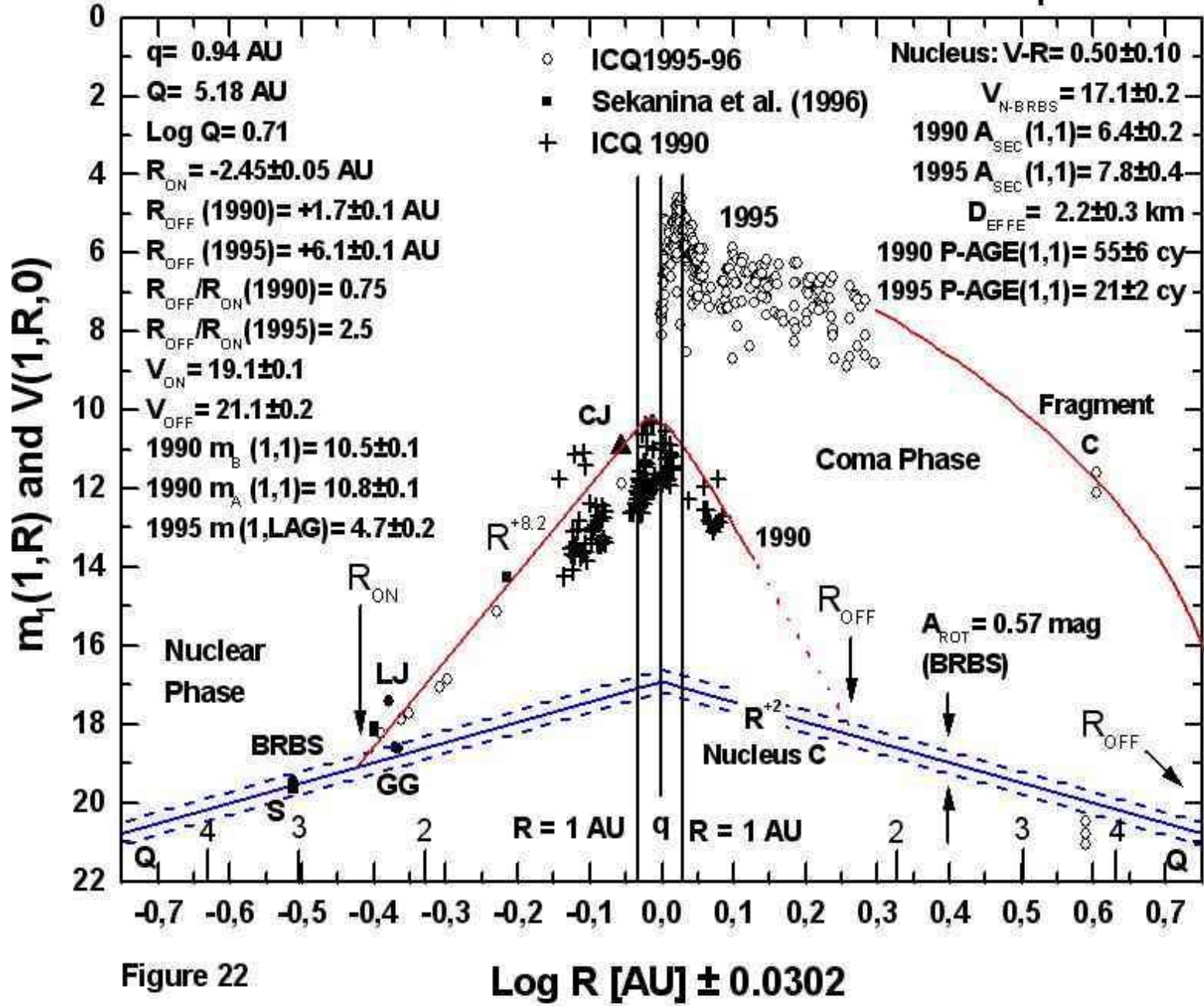

Figure 22

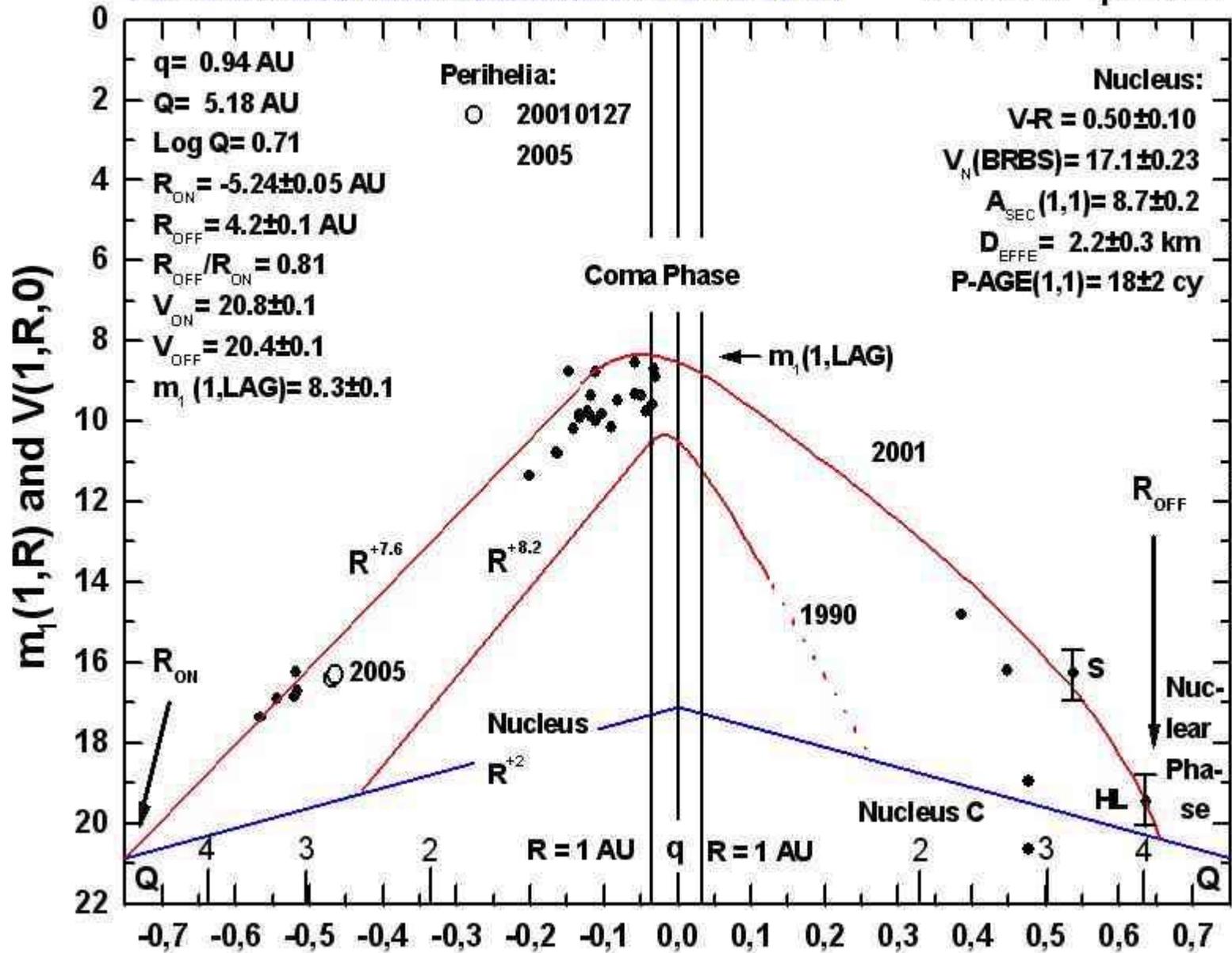

Figure 23

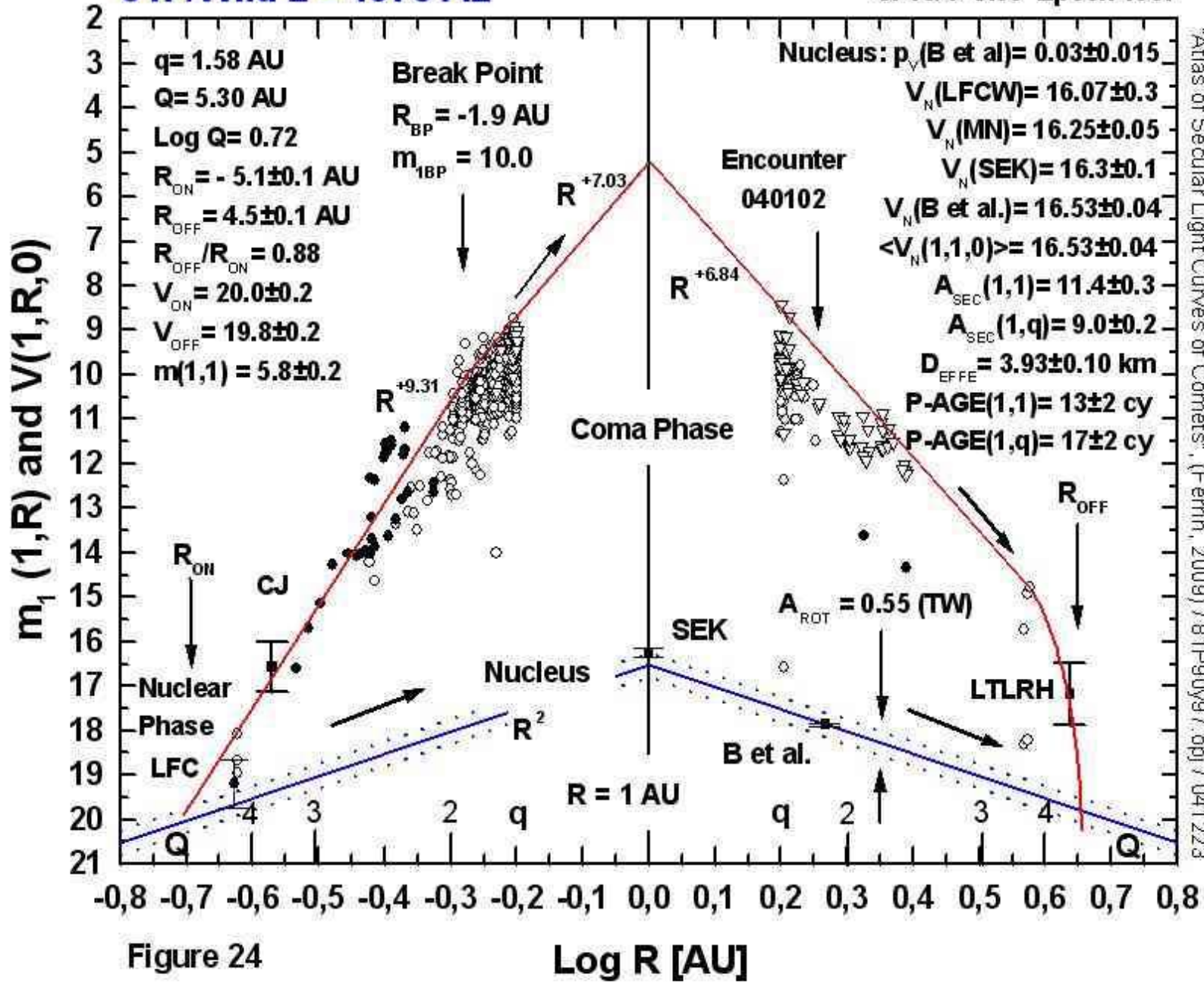
Figure 24

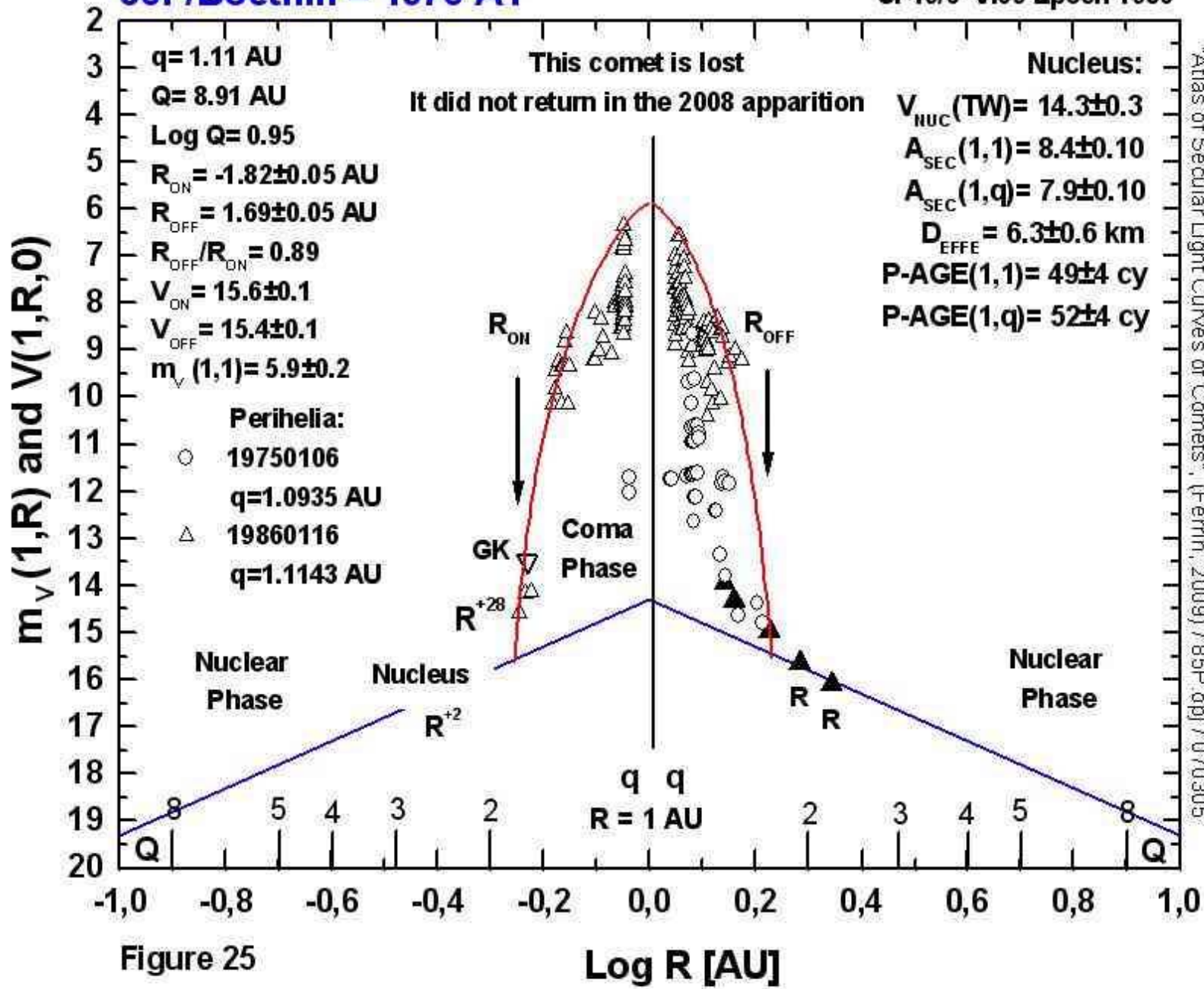

Figure 25

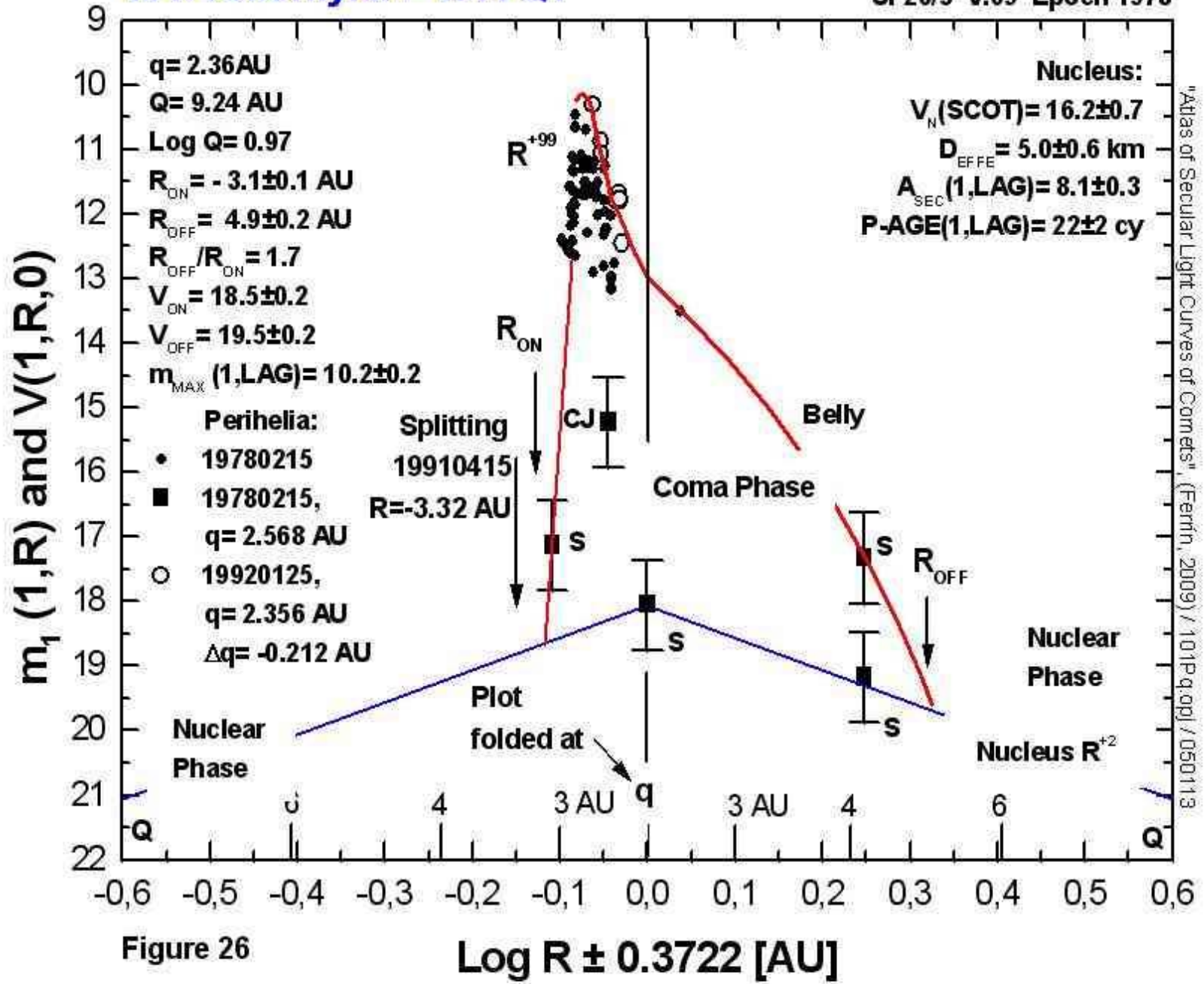

Figure 26

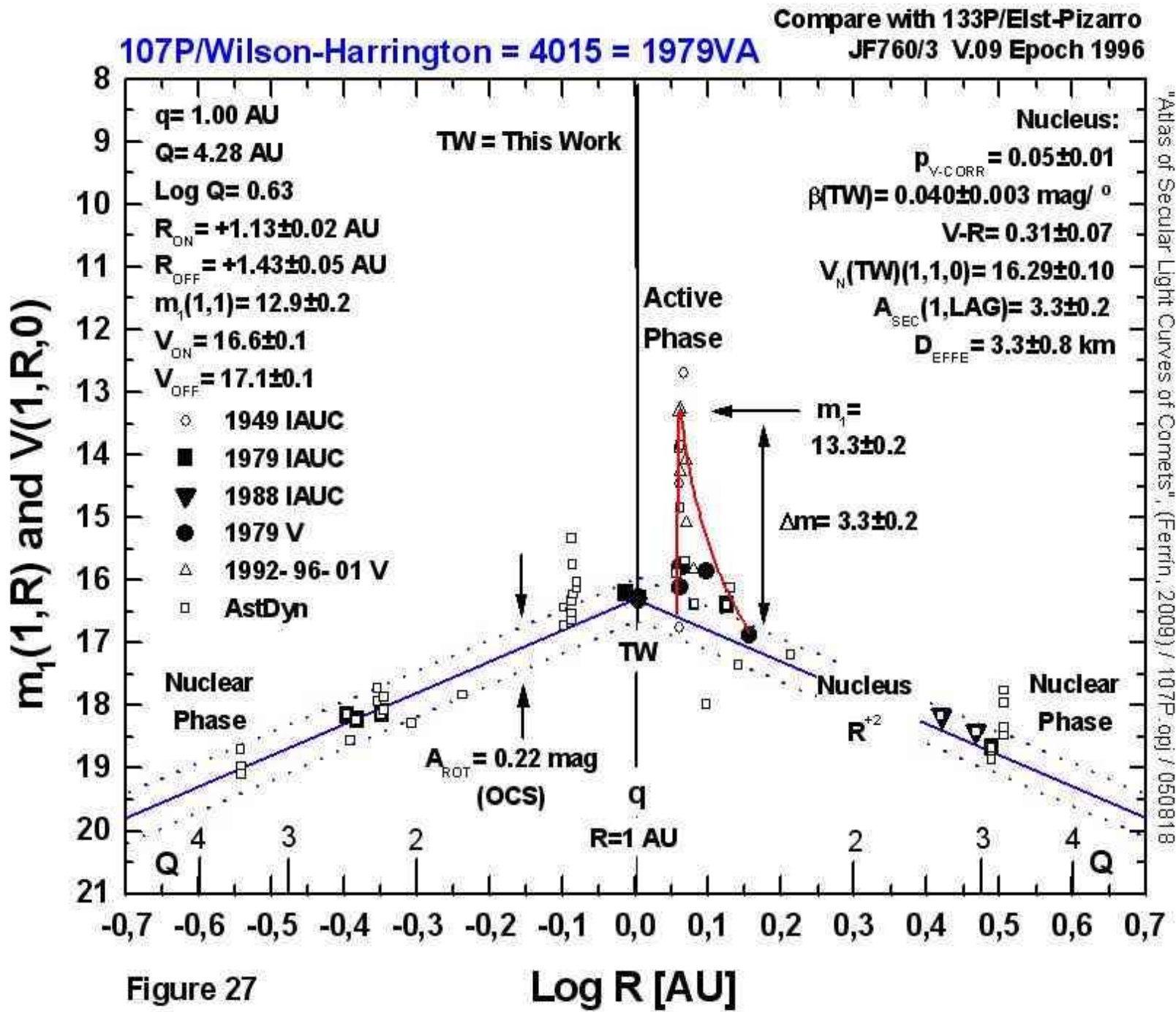

Figure 27

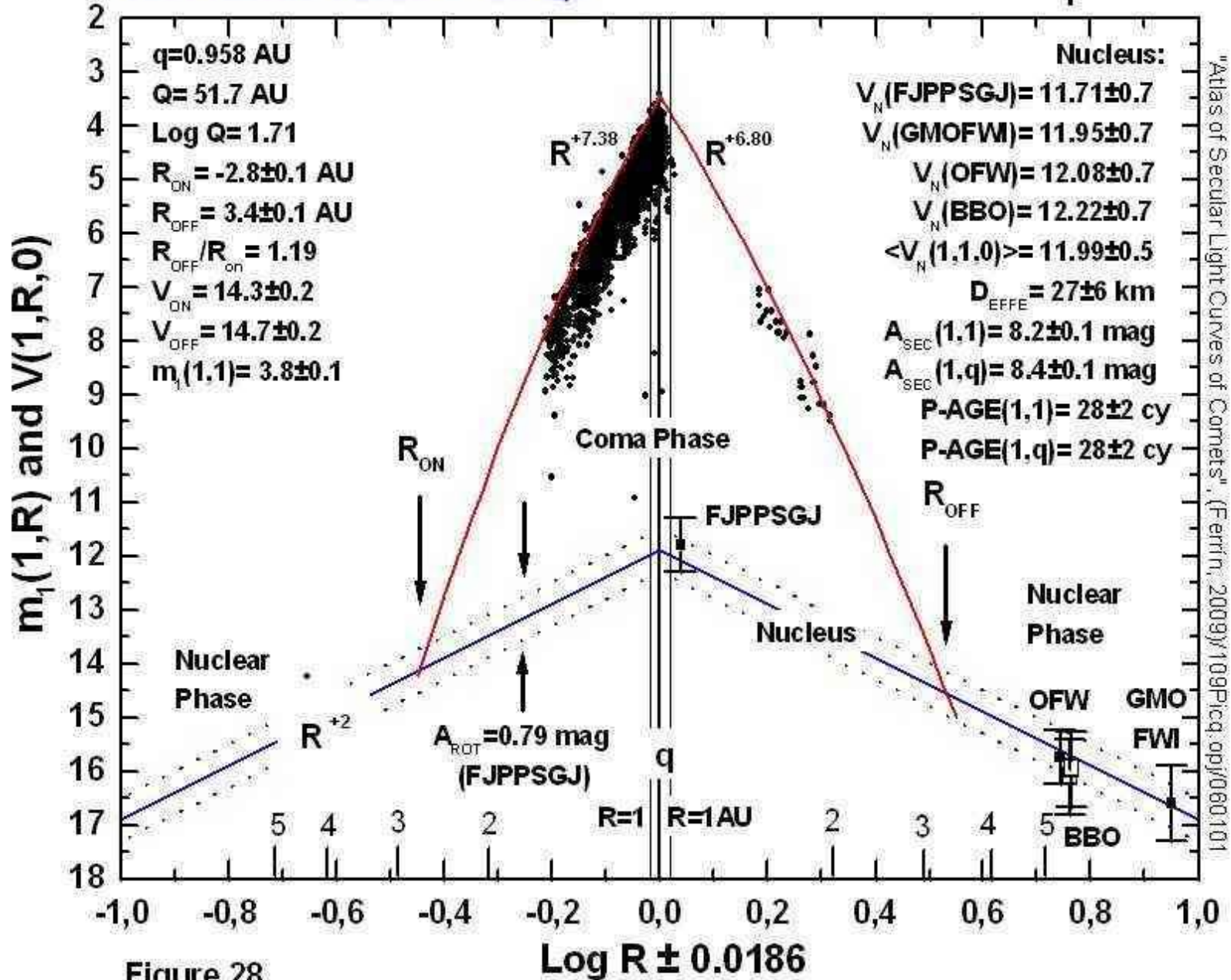

Figure 28

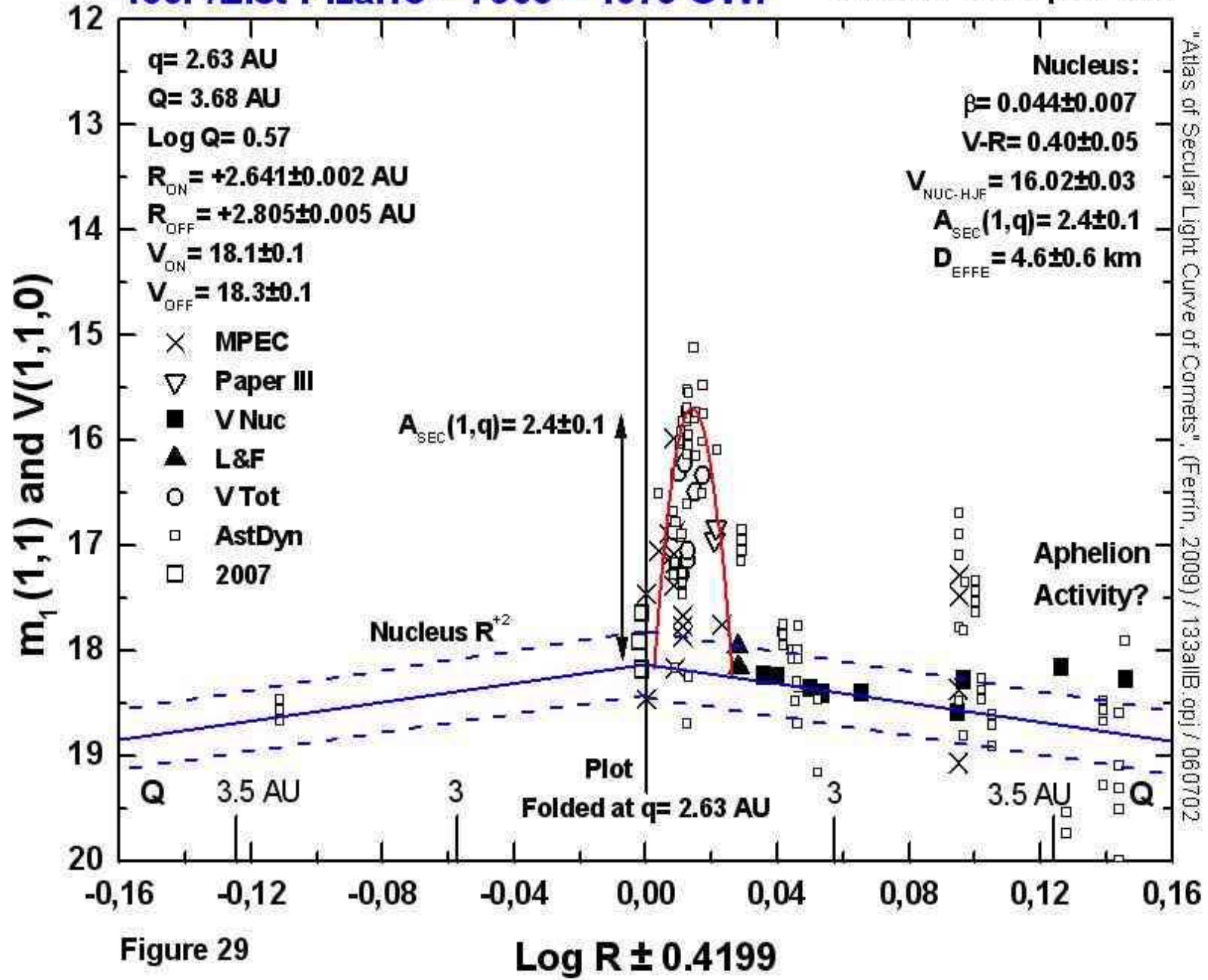

Figure 29

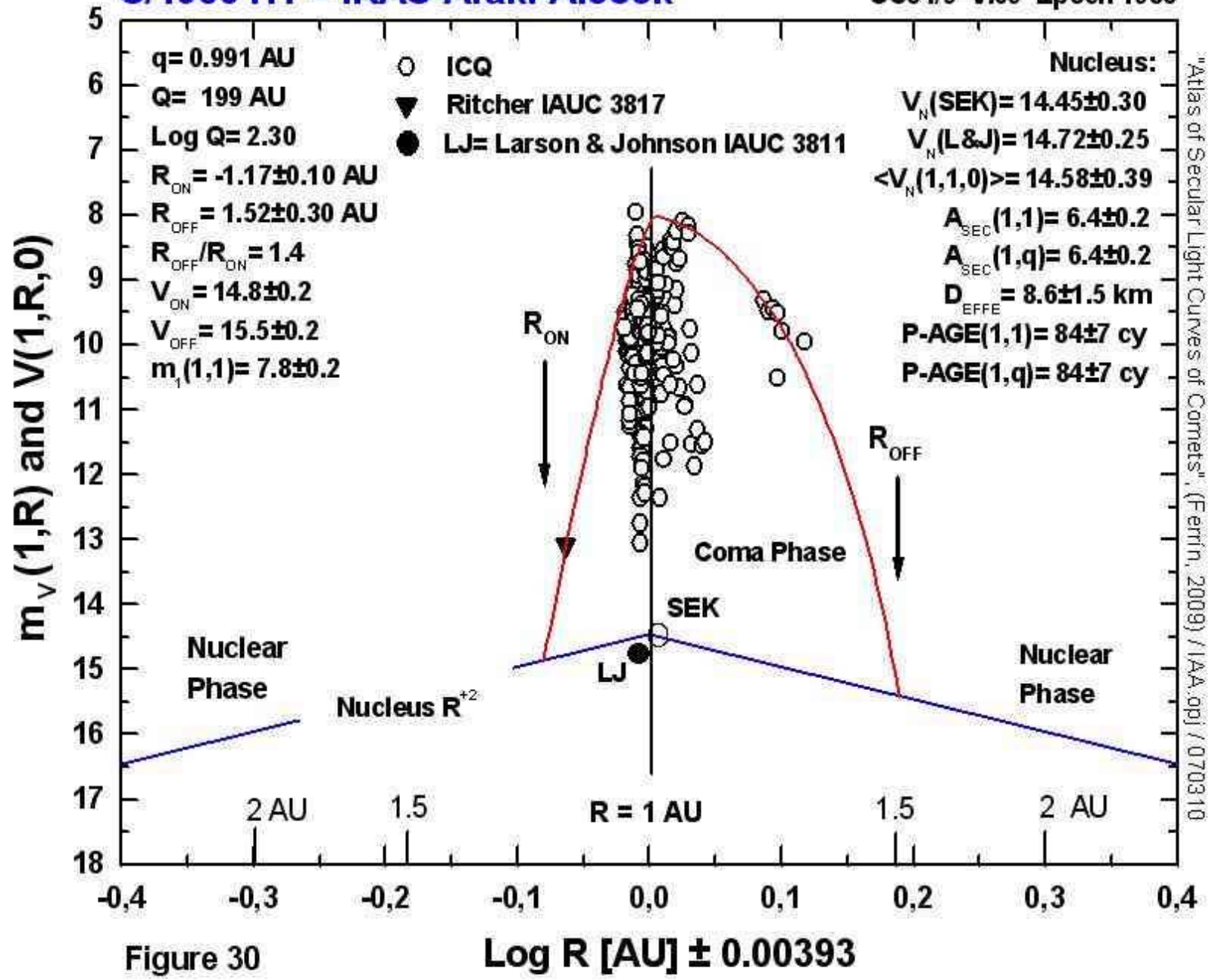

Figure 30

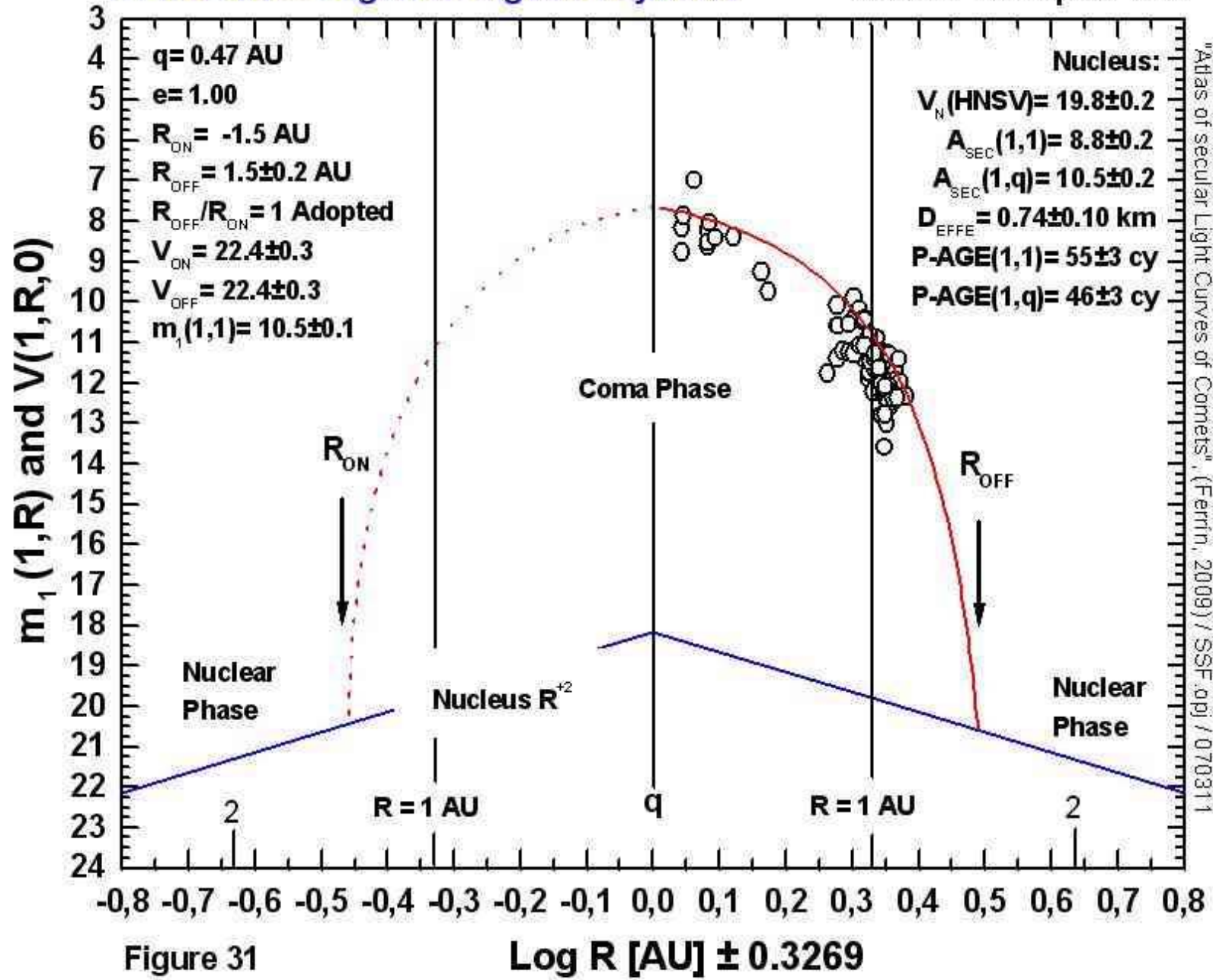

Figure 31

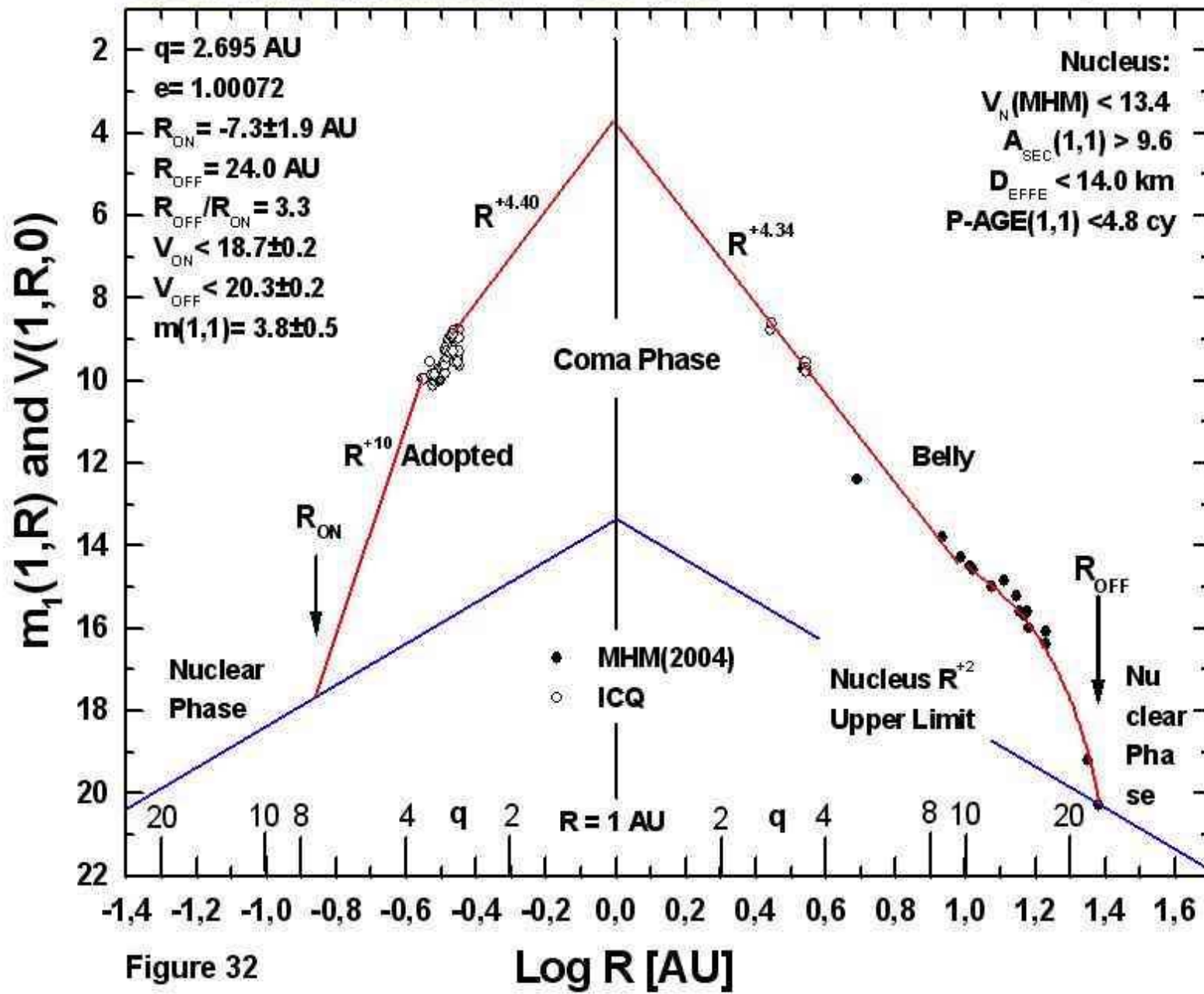

Figure 32

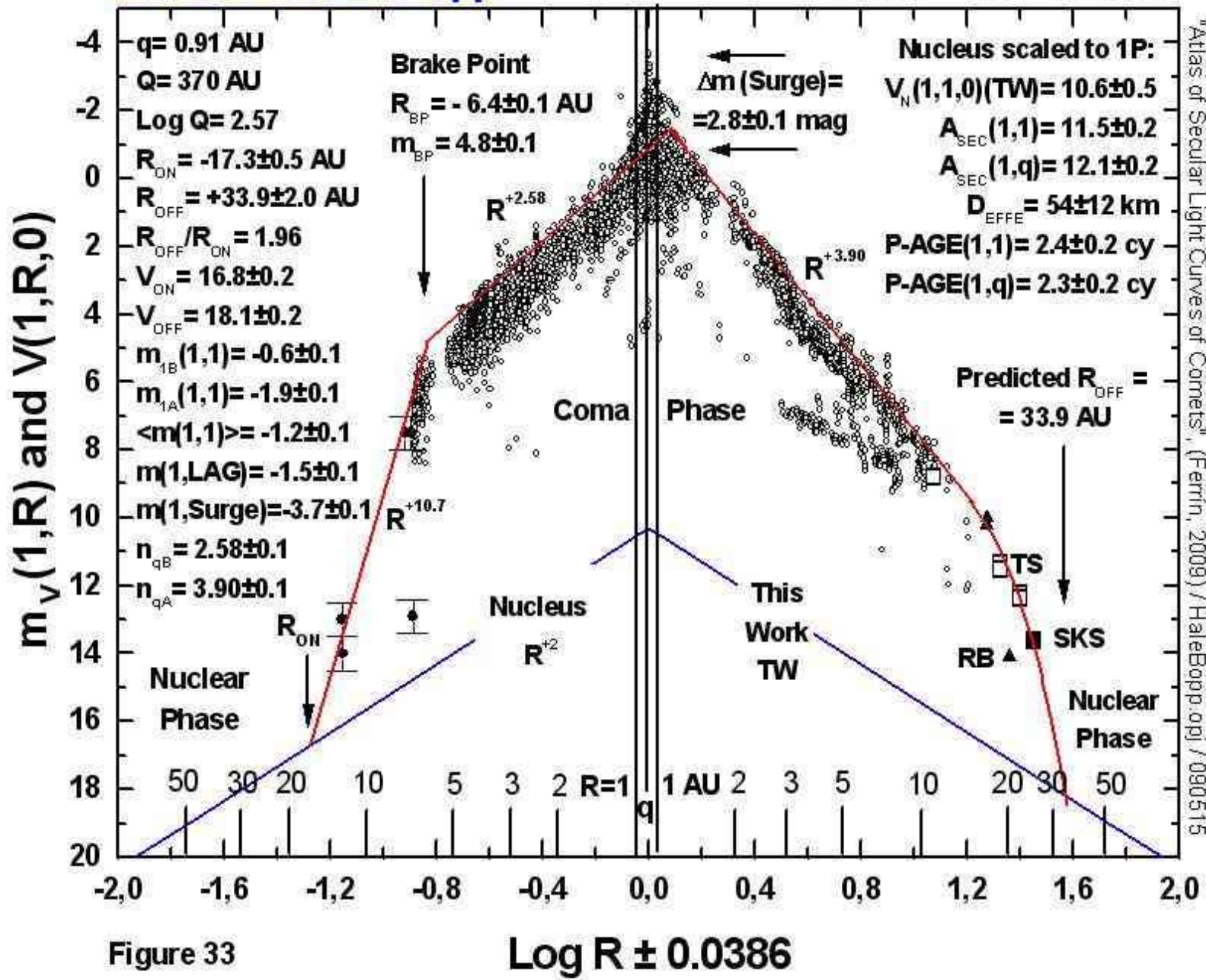

Figure 33

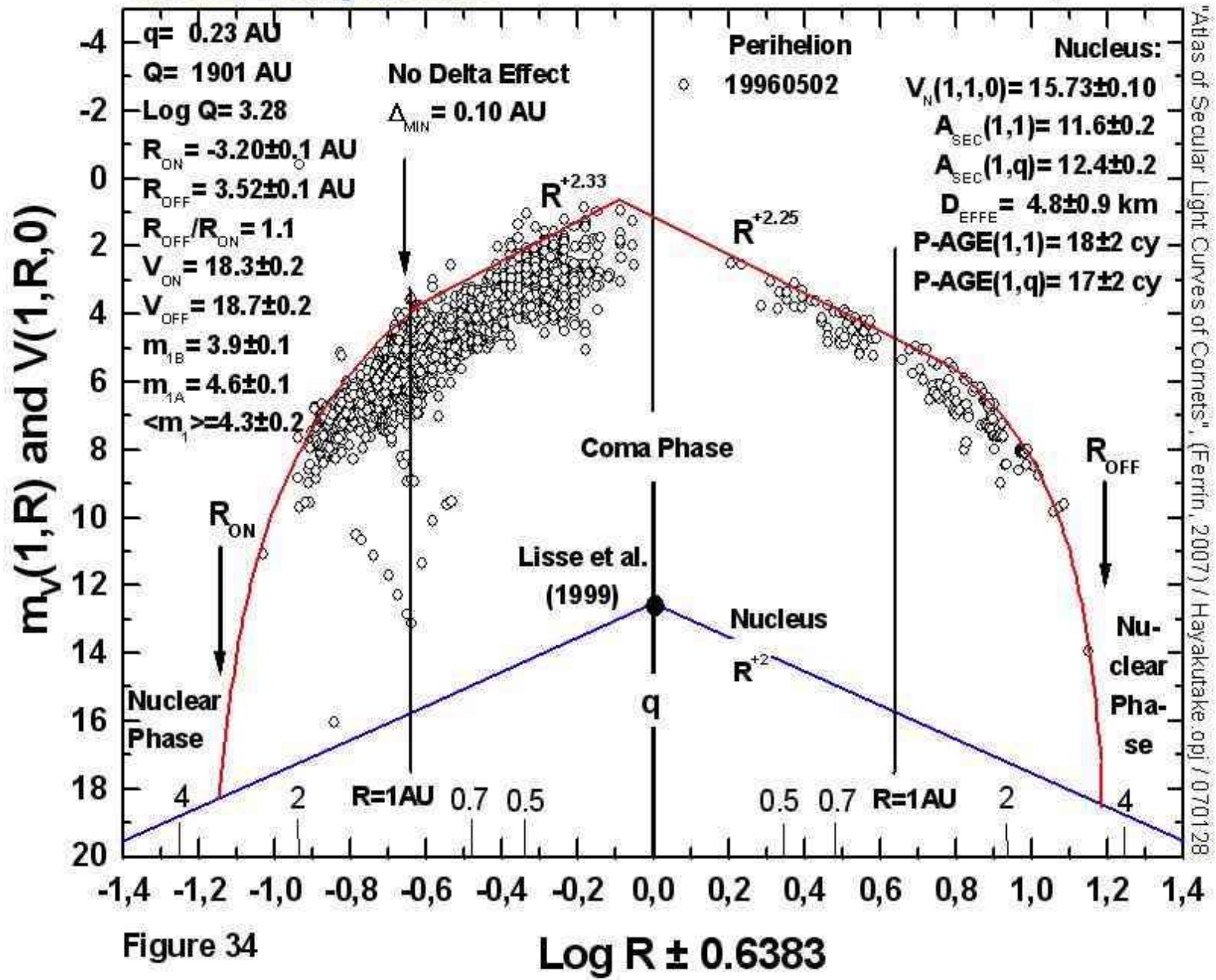

Figure 34

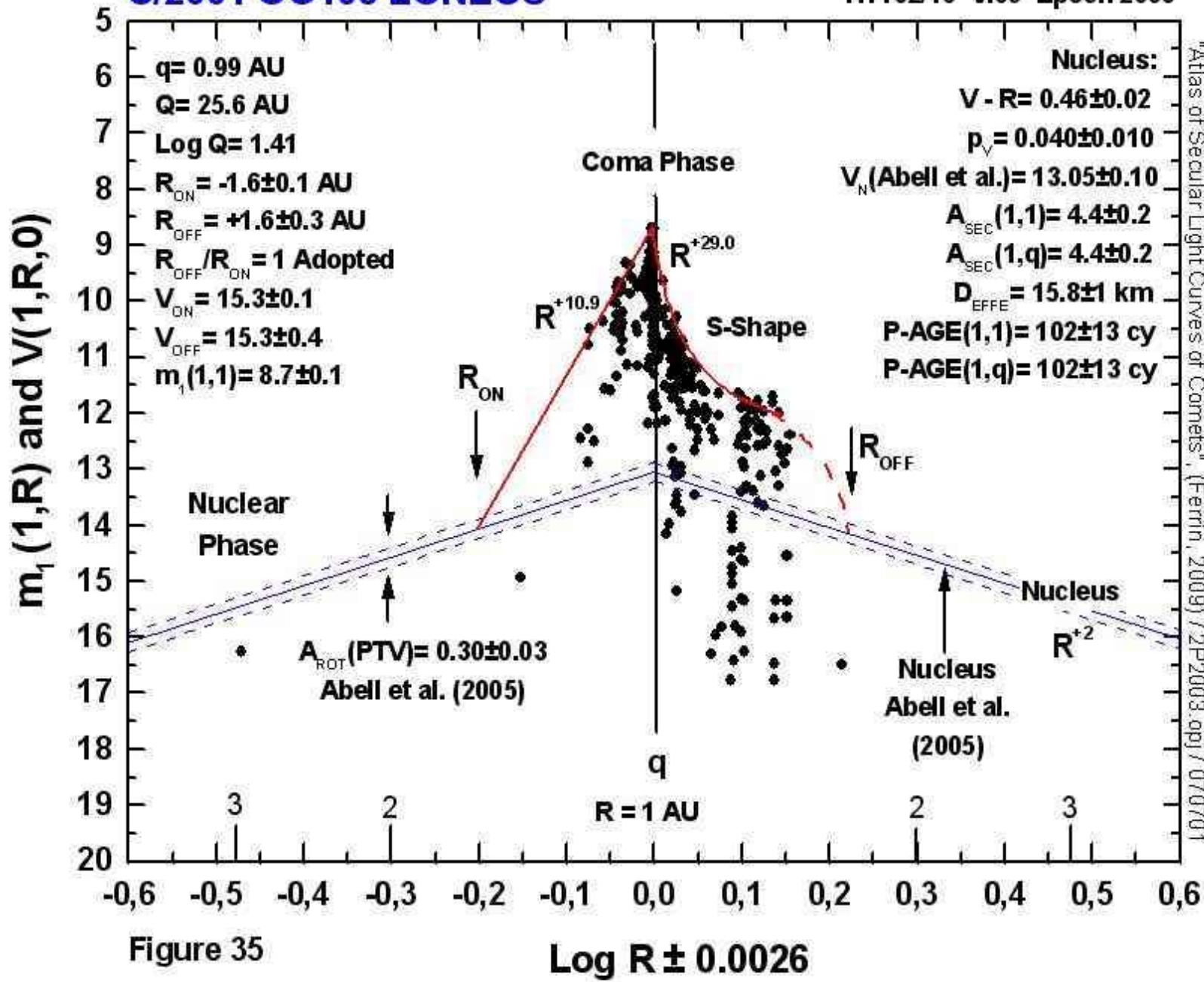

Figure 35

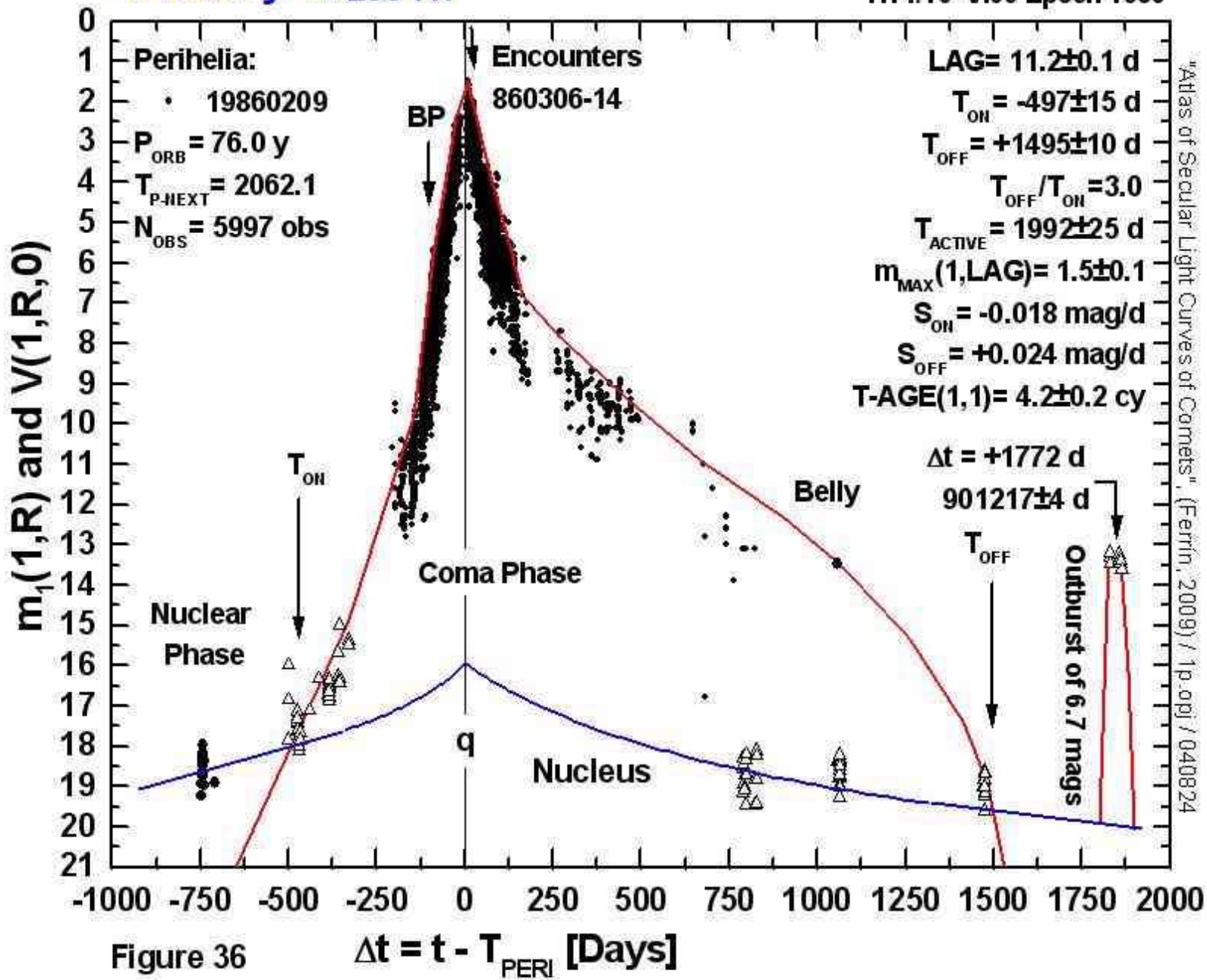

Figure 36

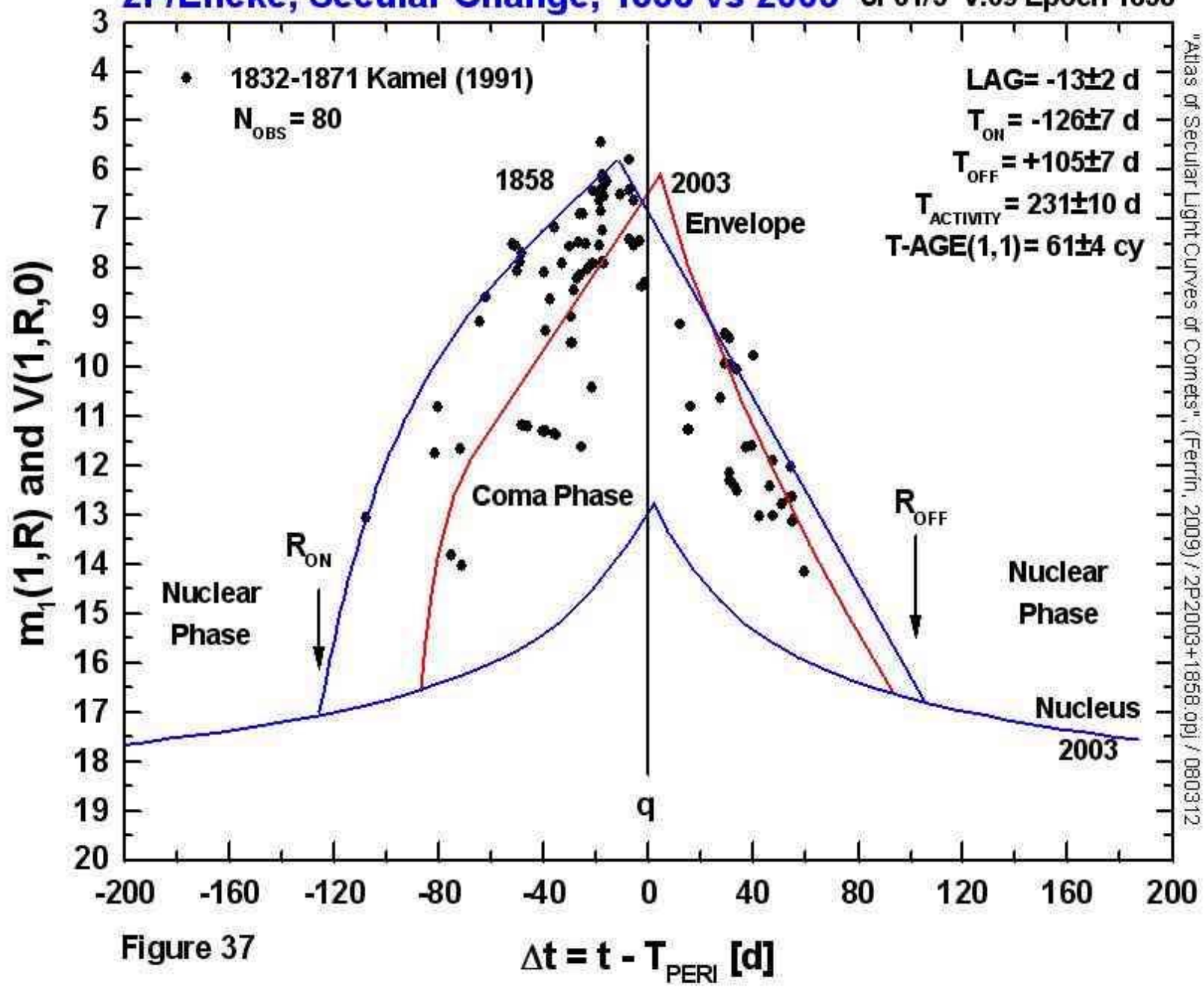

Figure 37

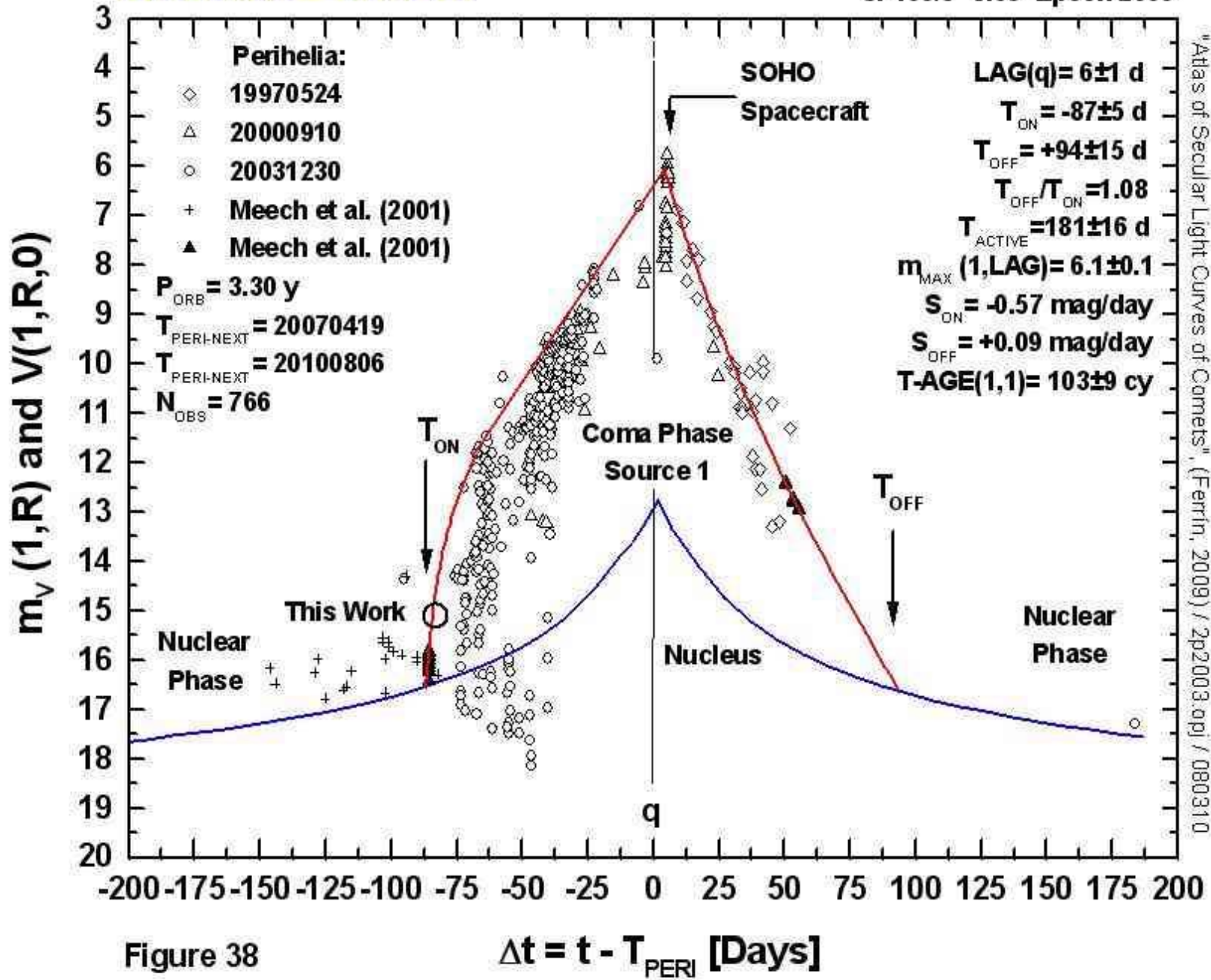

Figure 38

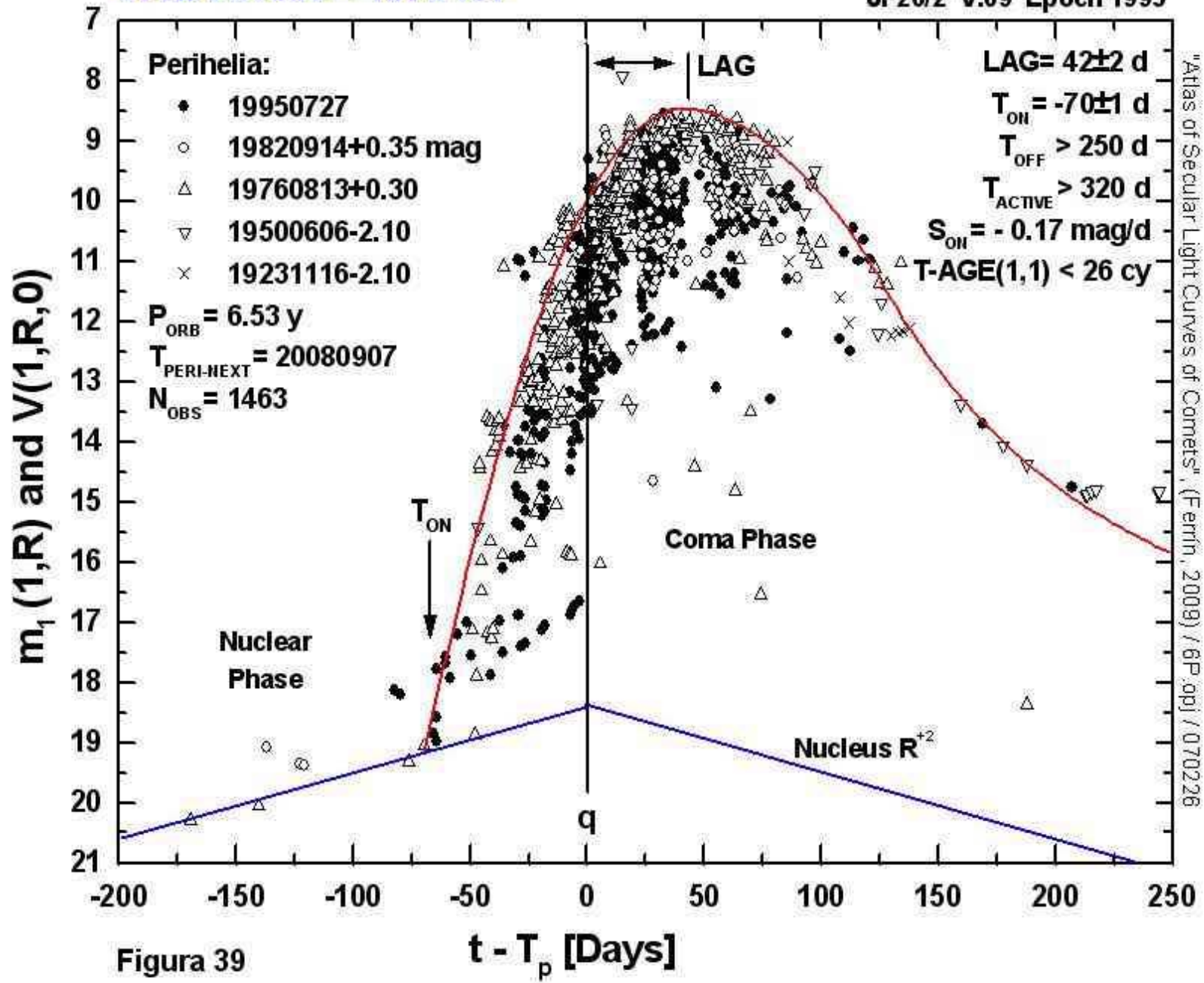

Figura 39

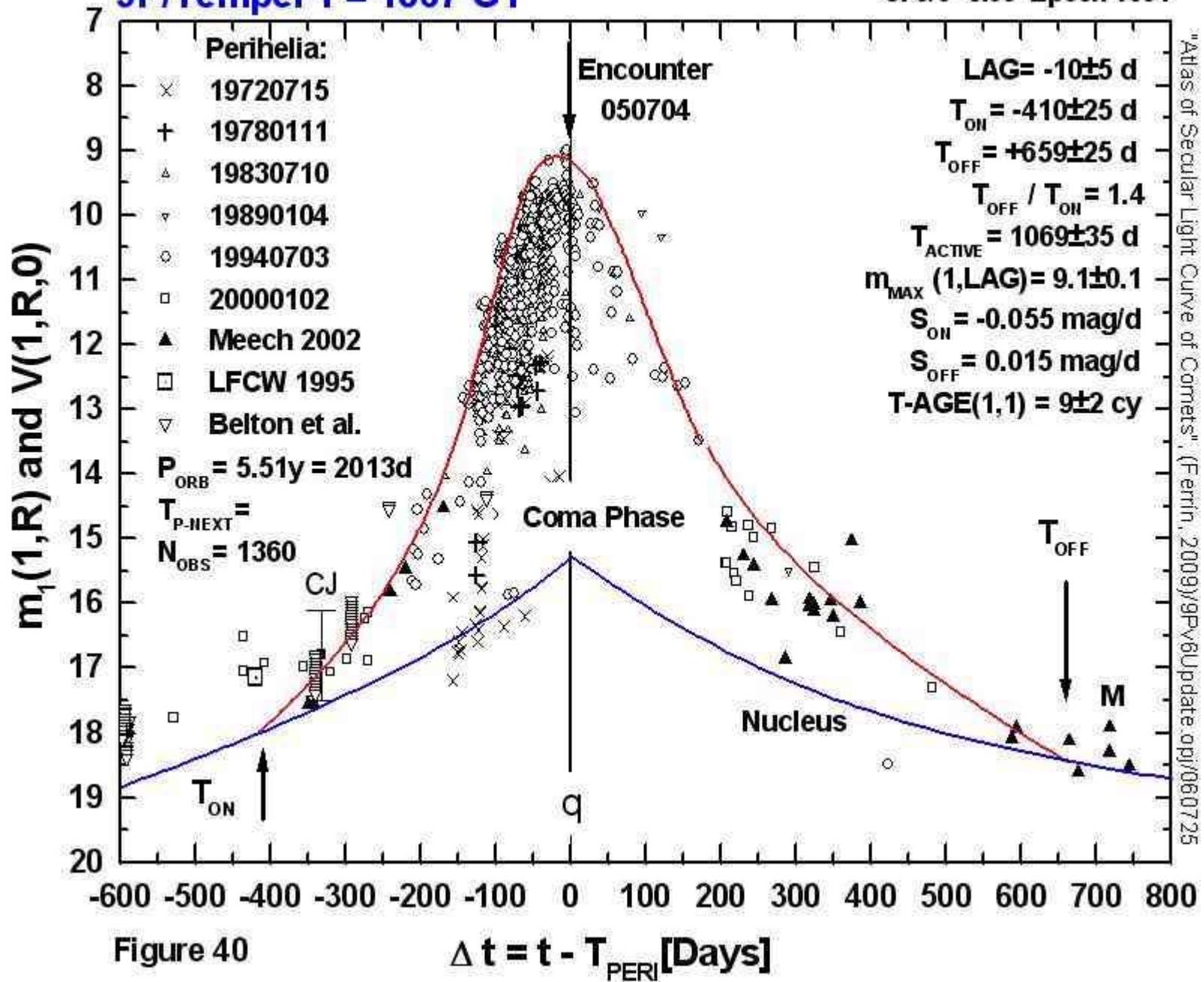

Figure 40

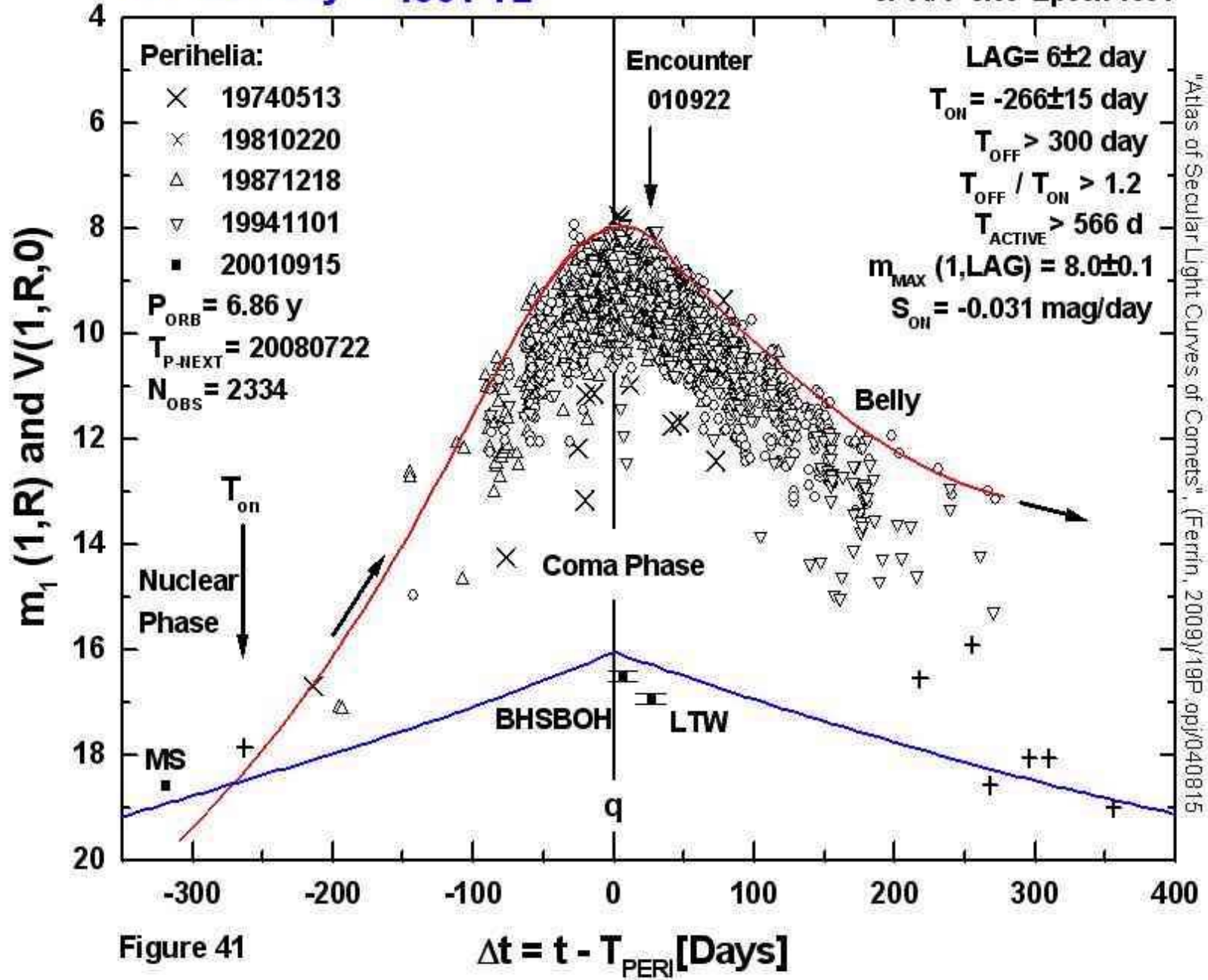

Figure 41

## Figure 42

**21P/Giacobini-Zinner = 1900 Y1**

Compare with 81P/Wild 2
JF11/3 V.09 Epoch 1998

Perihelia
- ○ 19981121
- □ 19850905 -0.3 mag

$P_{ORB}$ = 6.61 y
$T_{P-NEXT}$ = 20050703
$N_{OBS}$ = 1828

Encounter 850911
Break Point
$T_{ON}$
Nuclear Phase
M M H
Coma Phase
PWKR
PWKR − 2.45 mag
$T_{OFF}$
q
CJ

LAG = −10±1 d
$T_{ON}$ = −280±20 d
$T_{OFF}$ = +470±30 d
$T_{OFF}/T_{ON}$ = 1.7
$T_{ACTIVE}$ = 750±40 d
m(1,LAG) = 8.0±0.1
$S_{ON}$ = −0.041 mag/d
$S_{OFF}$ = 0.023 mag/d
T-AGE = 15±1 cy

Y-axis: $m_1(1,R)$ and $V(1,R,0)$
X-axis: $\Delta t = t - T_{PERI}$ [Days]

"Atlas of Secular Light Curves of Comets", (Ferrín, 2009) / 21P98-85.opj / 090825

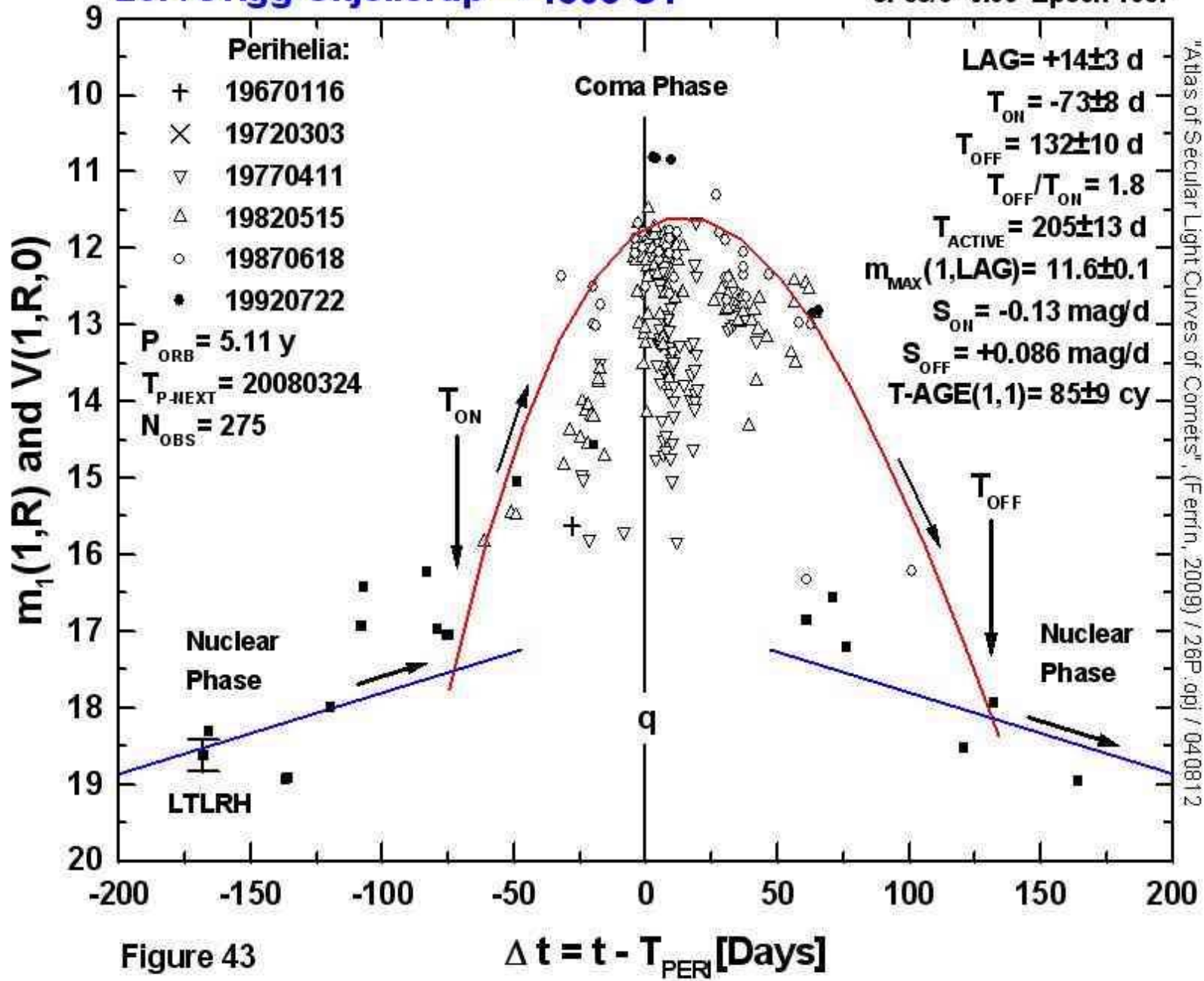

Figure 43

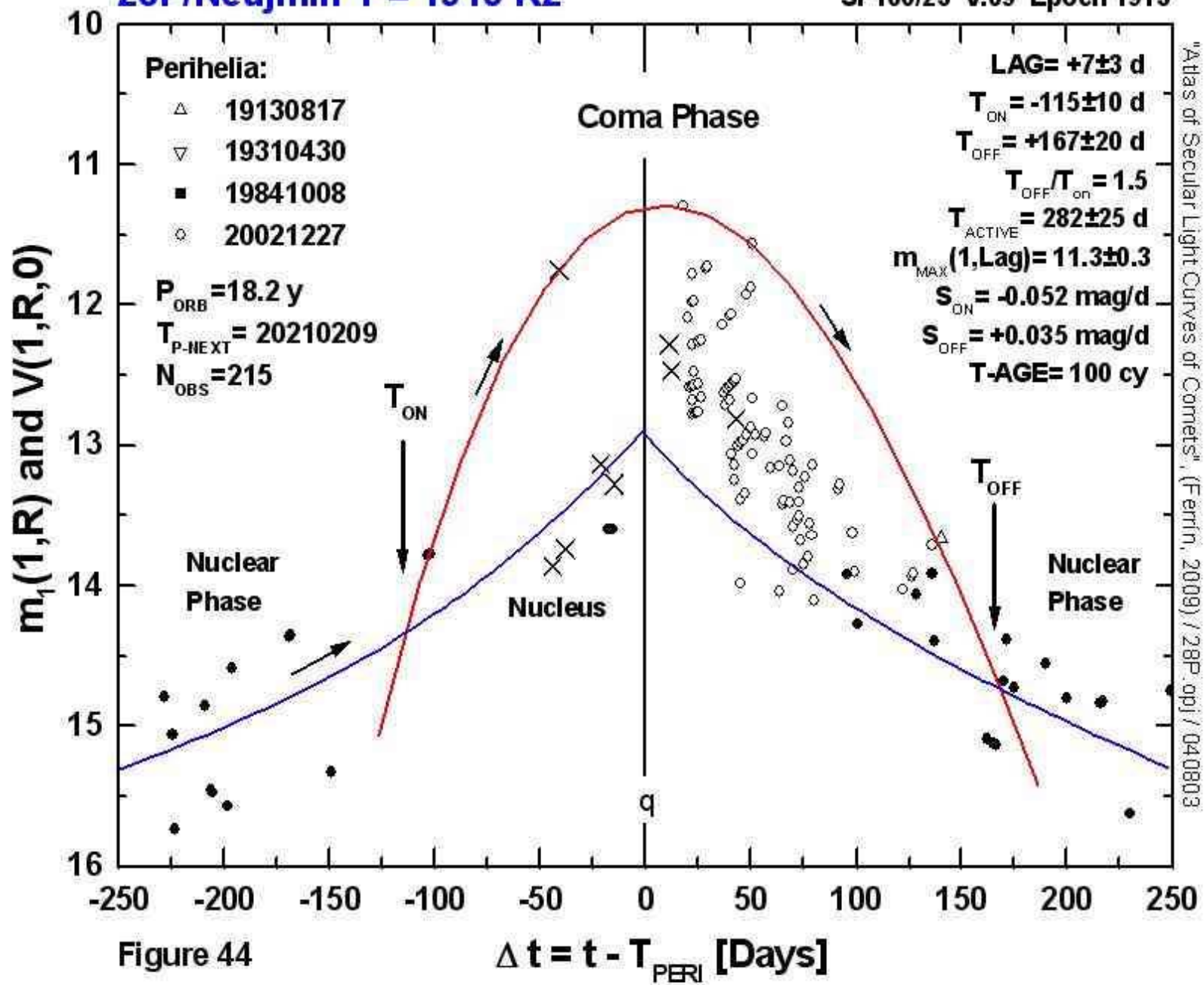

Figure 44

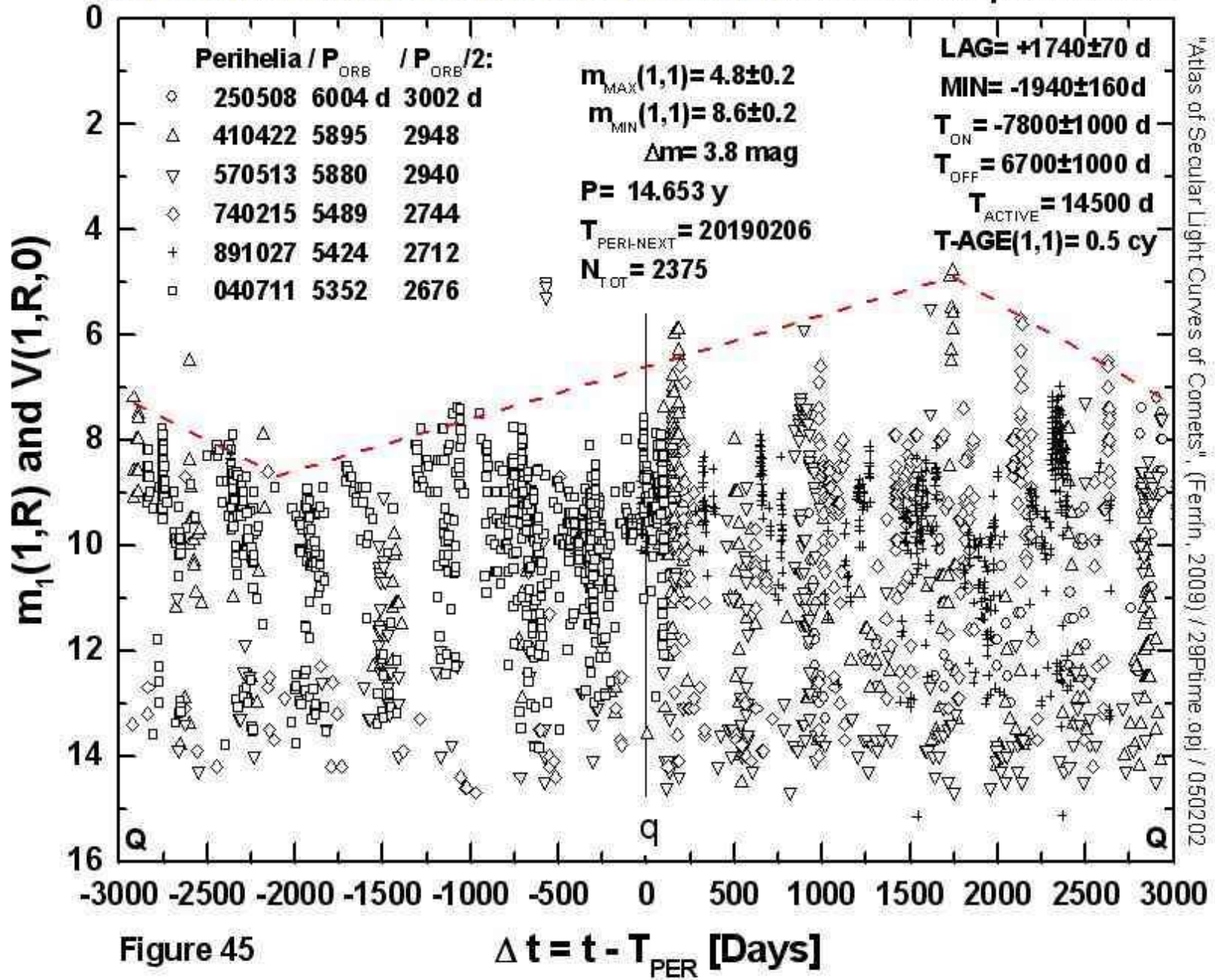

Figure 45

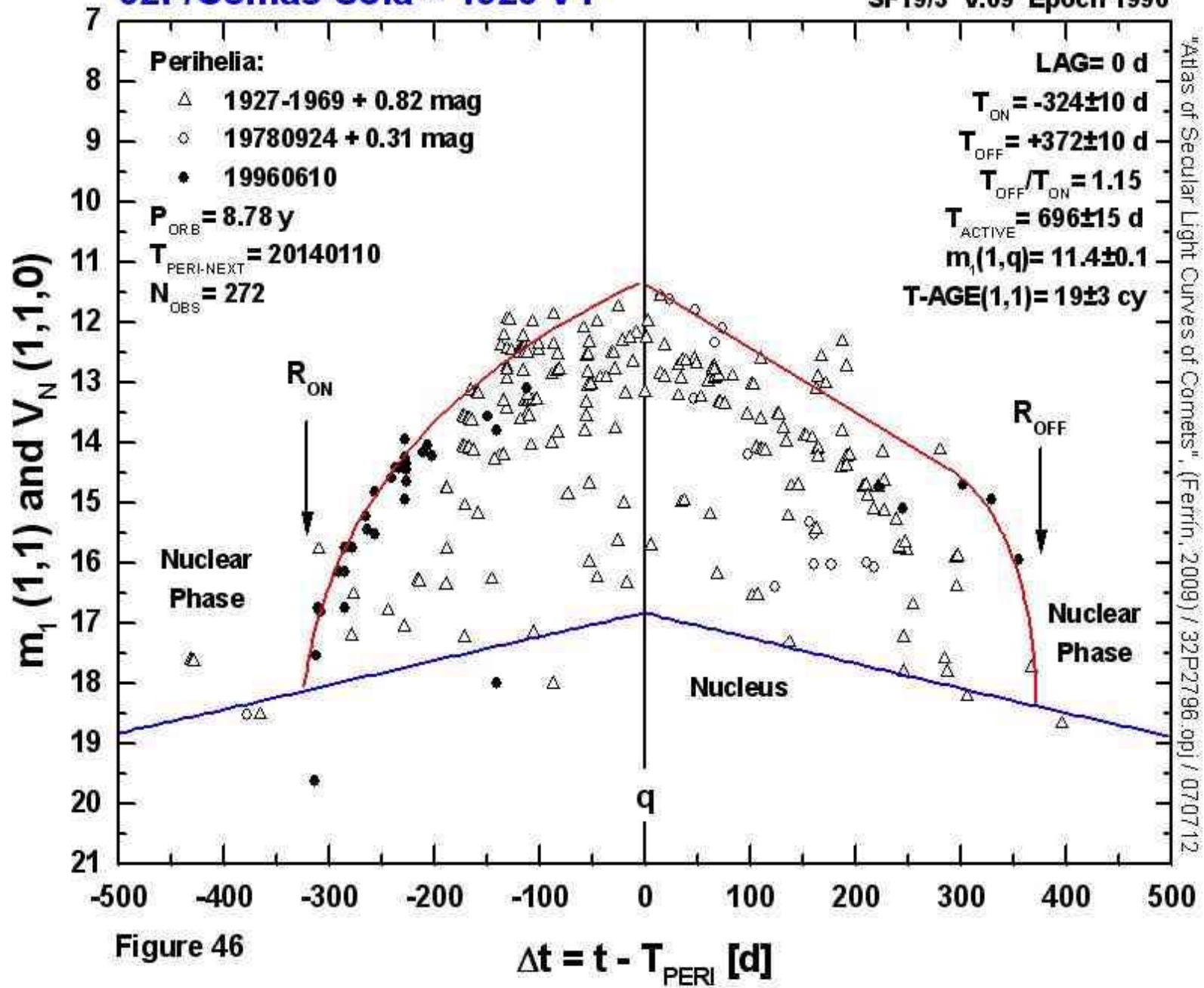

Figure 46

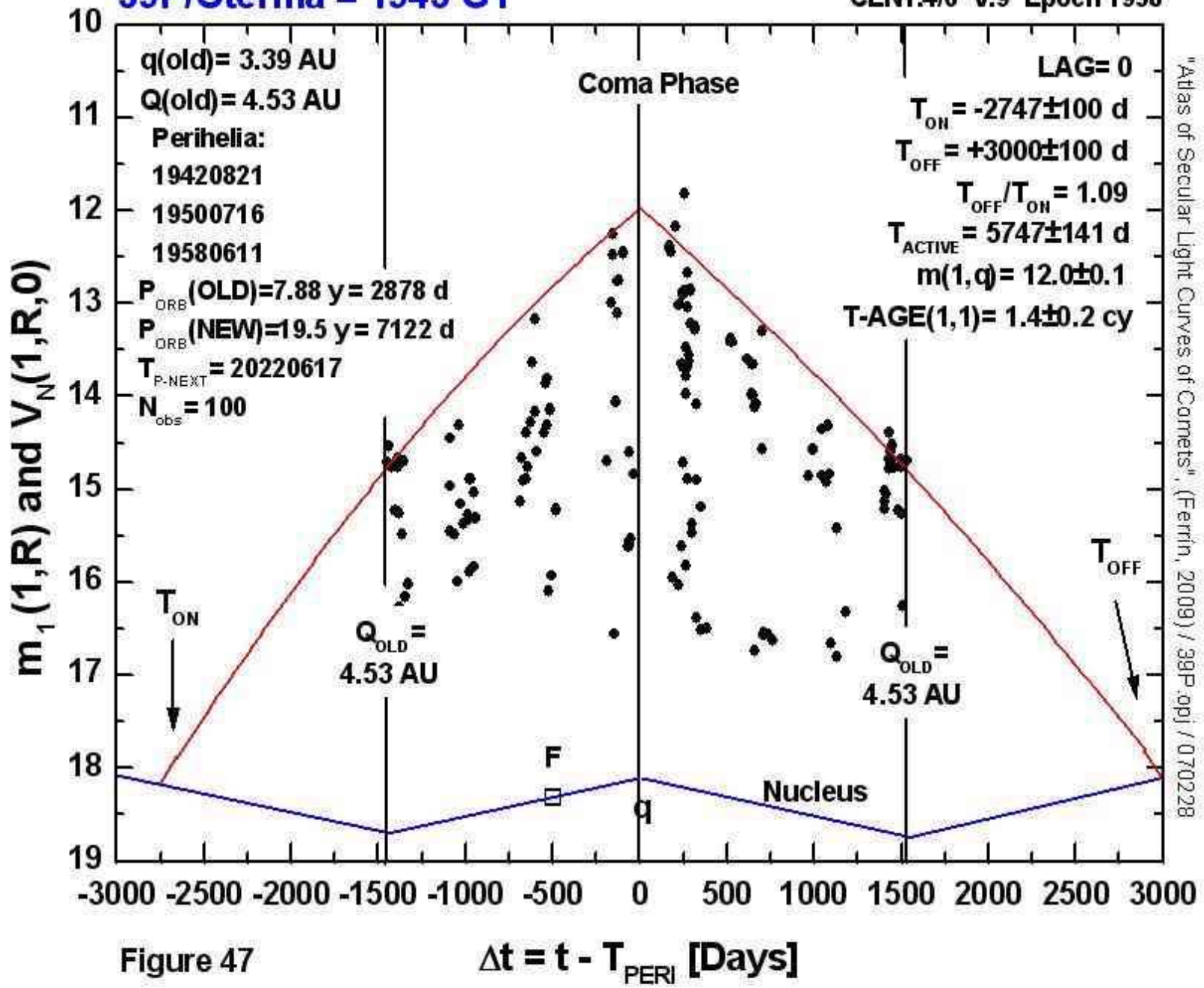

Figure 47

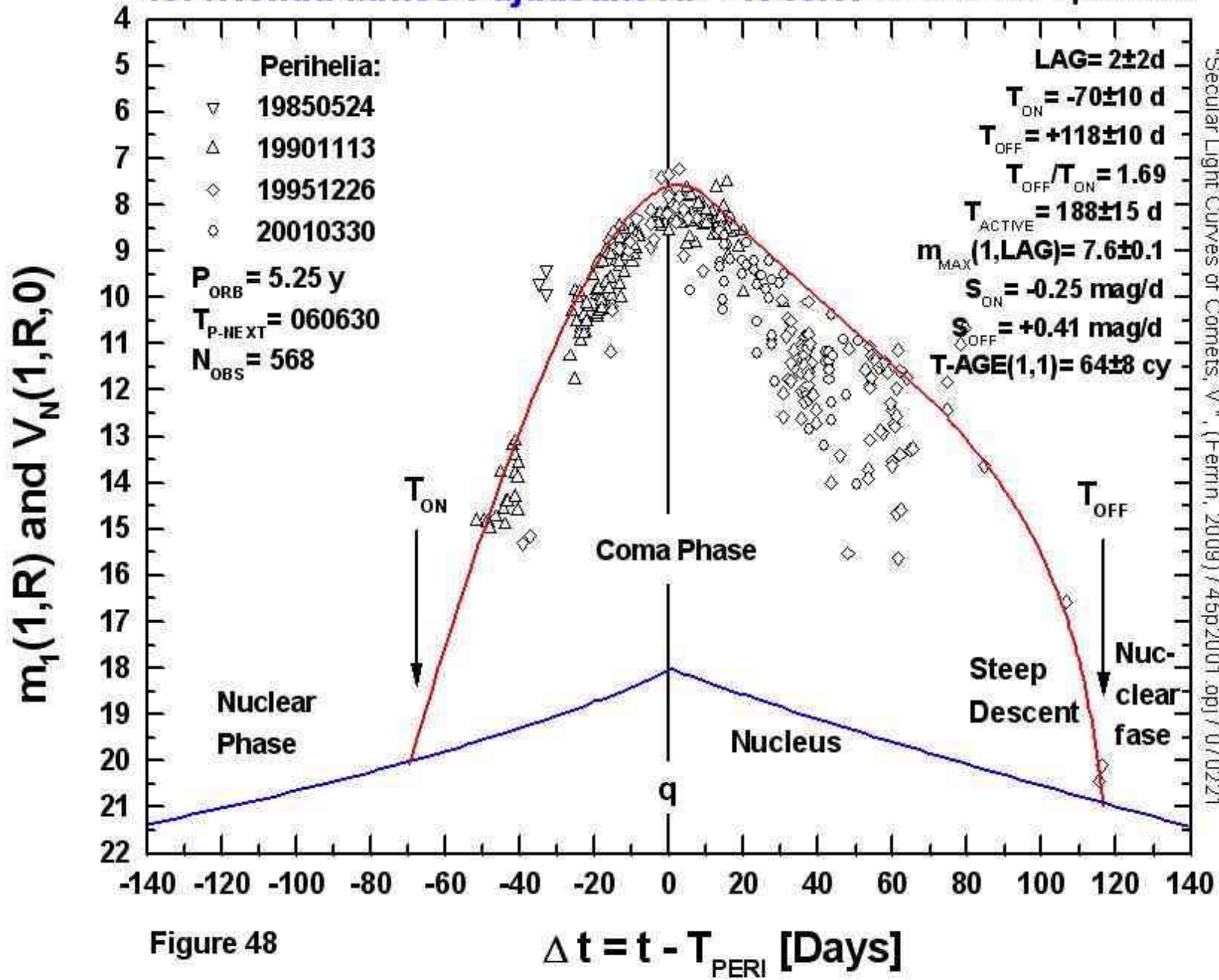

Figure 48

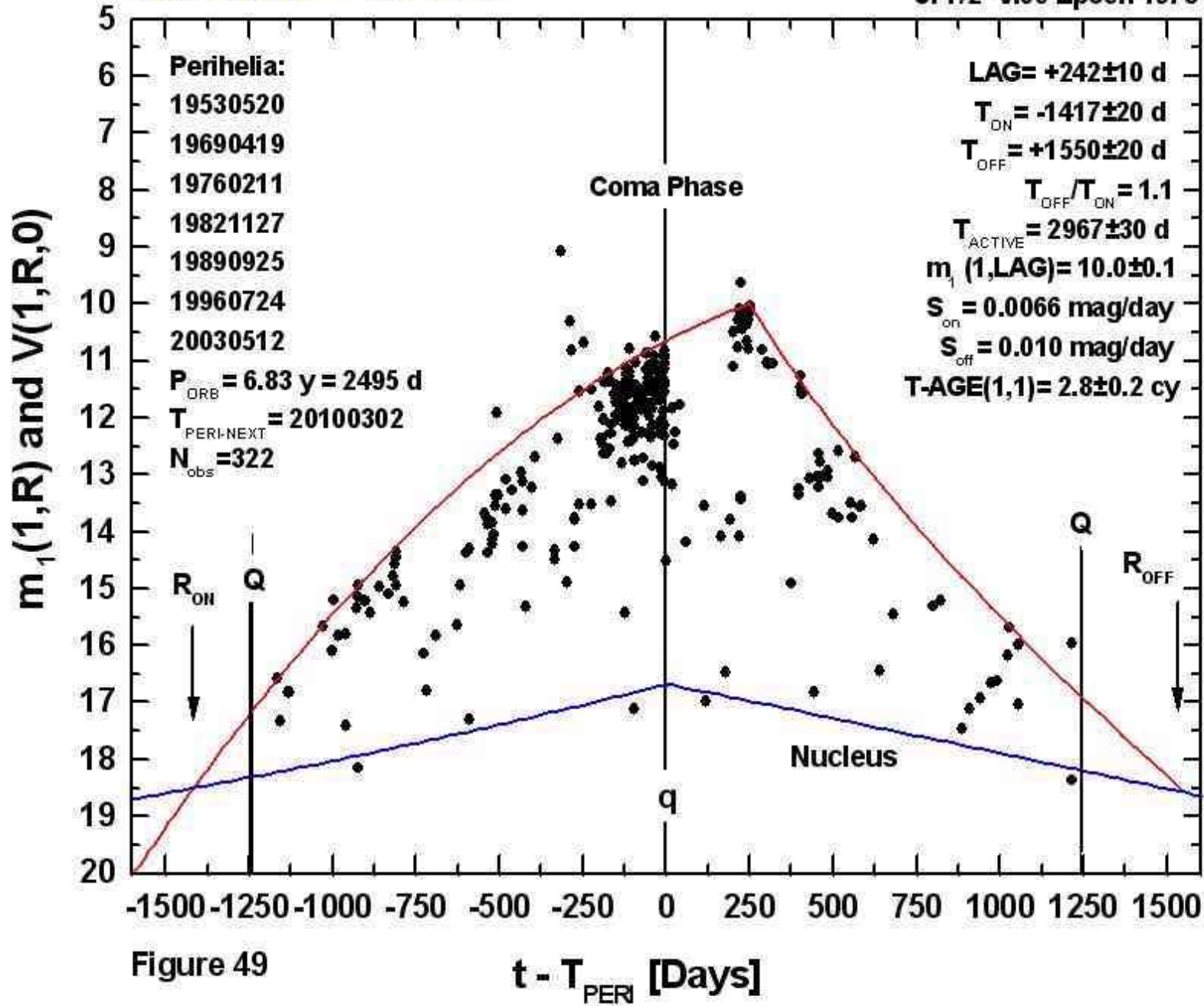

Figure 49

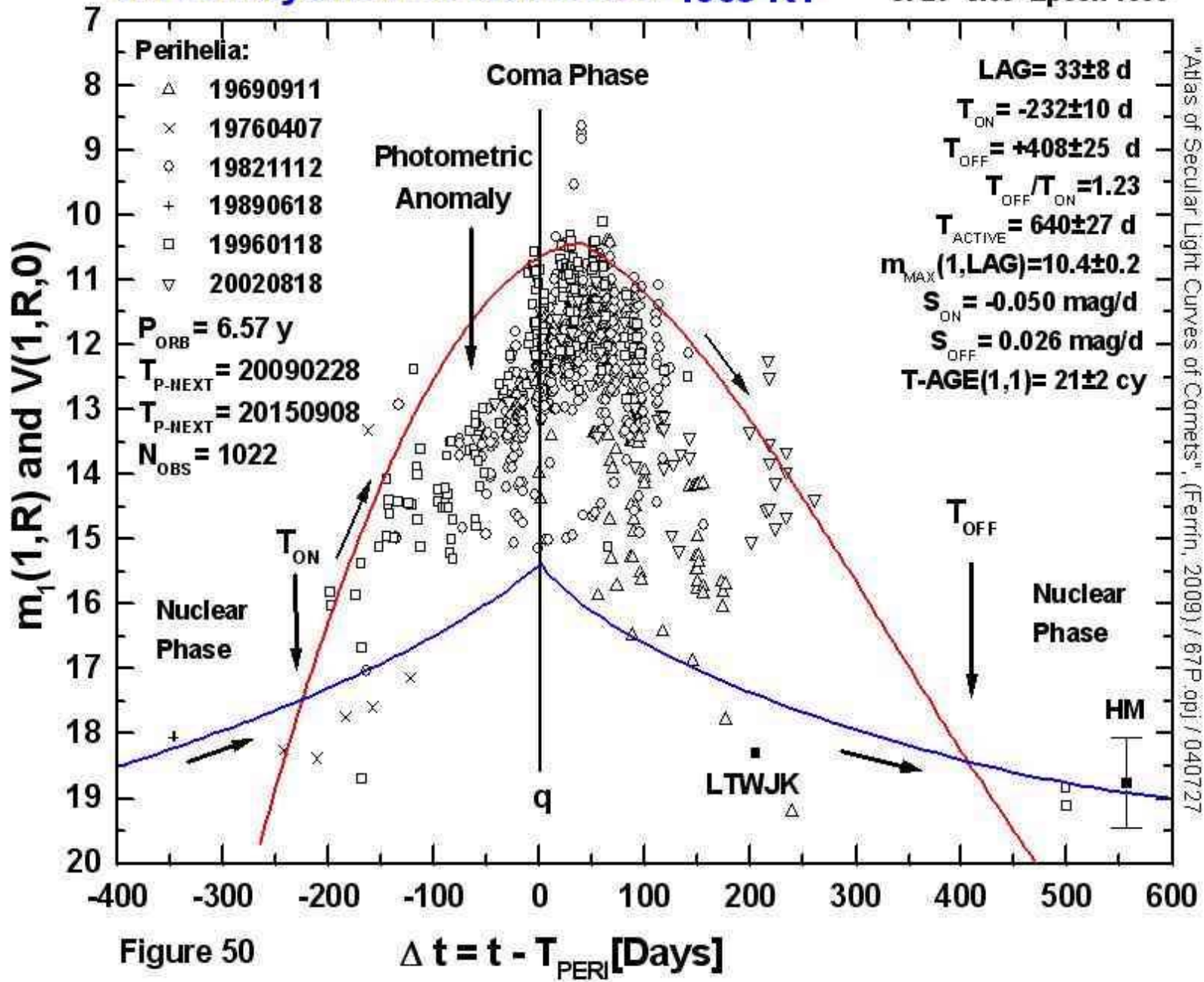

Figure 50

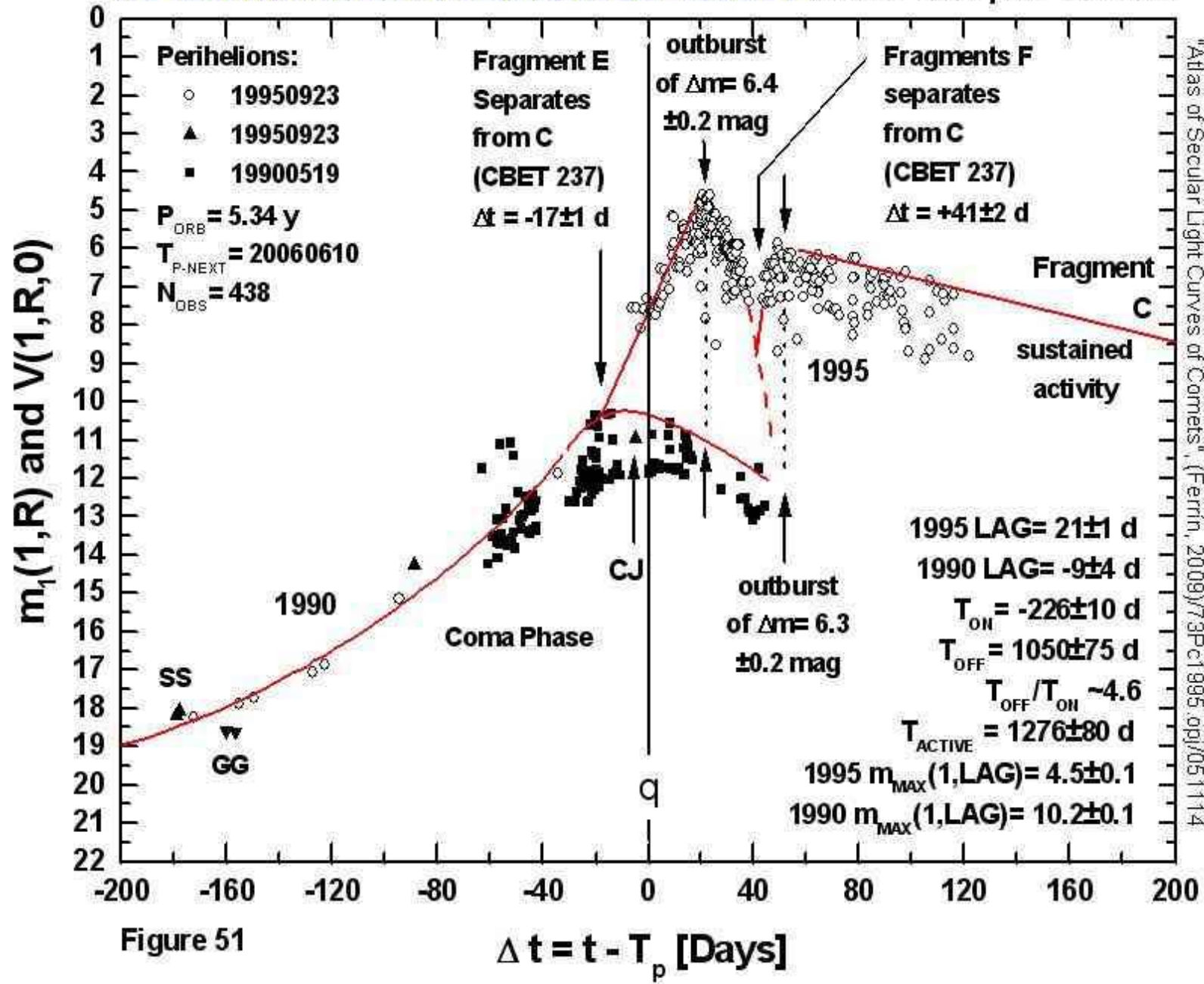

Figure 51

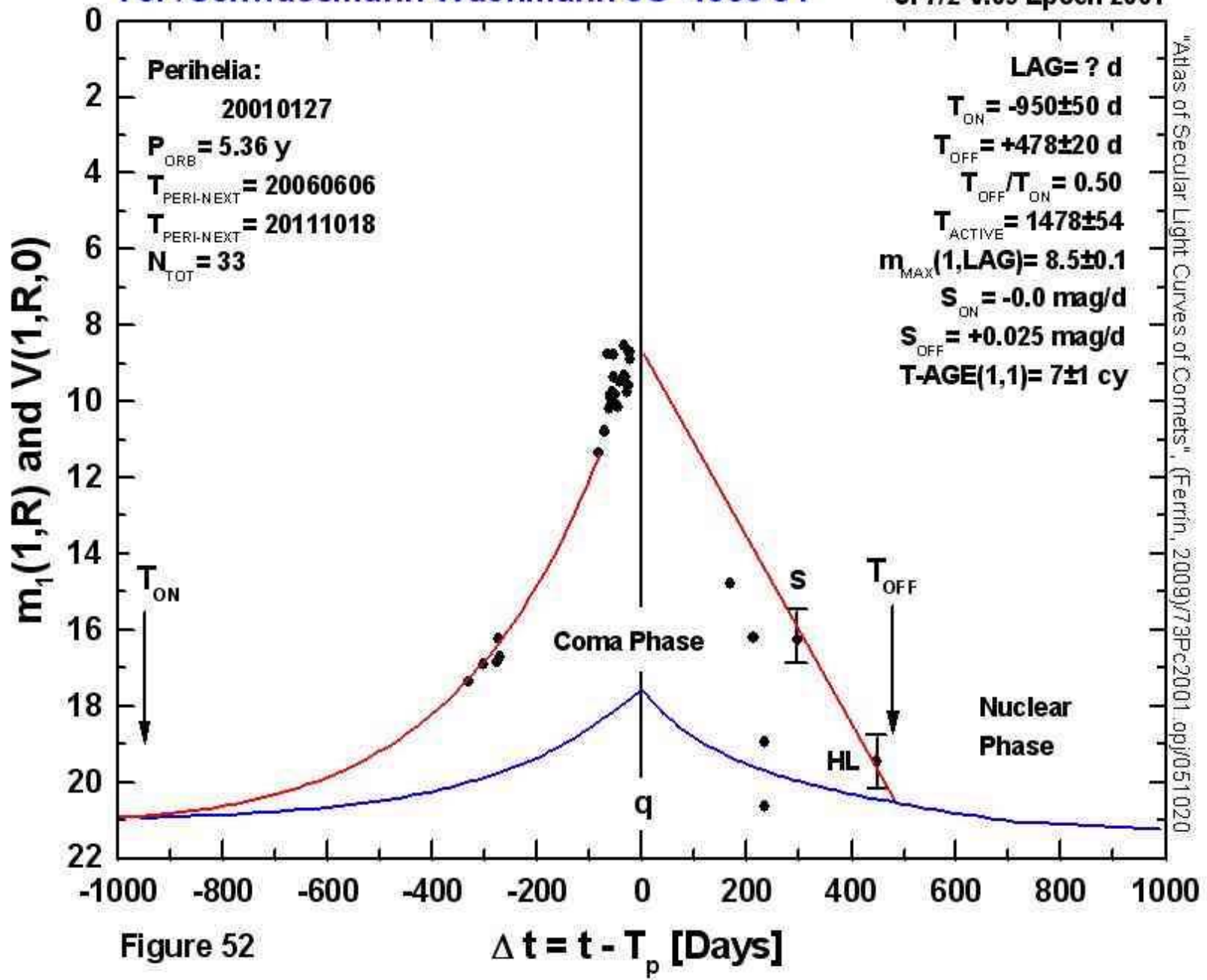

Figure 52

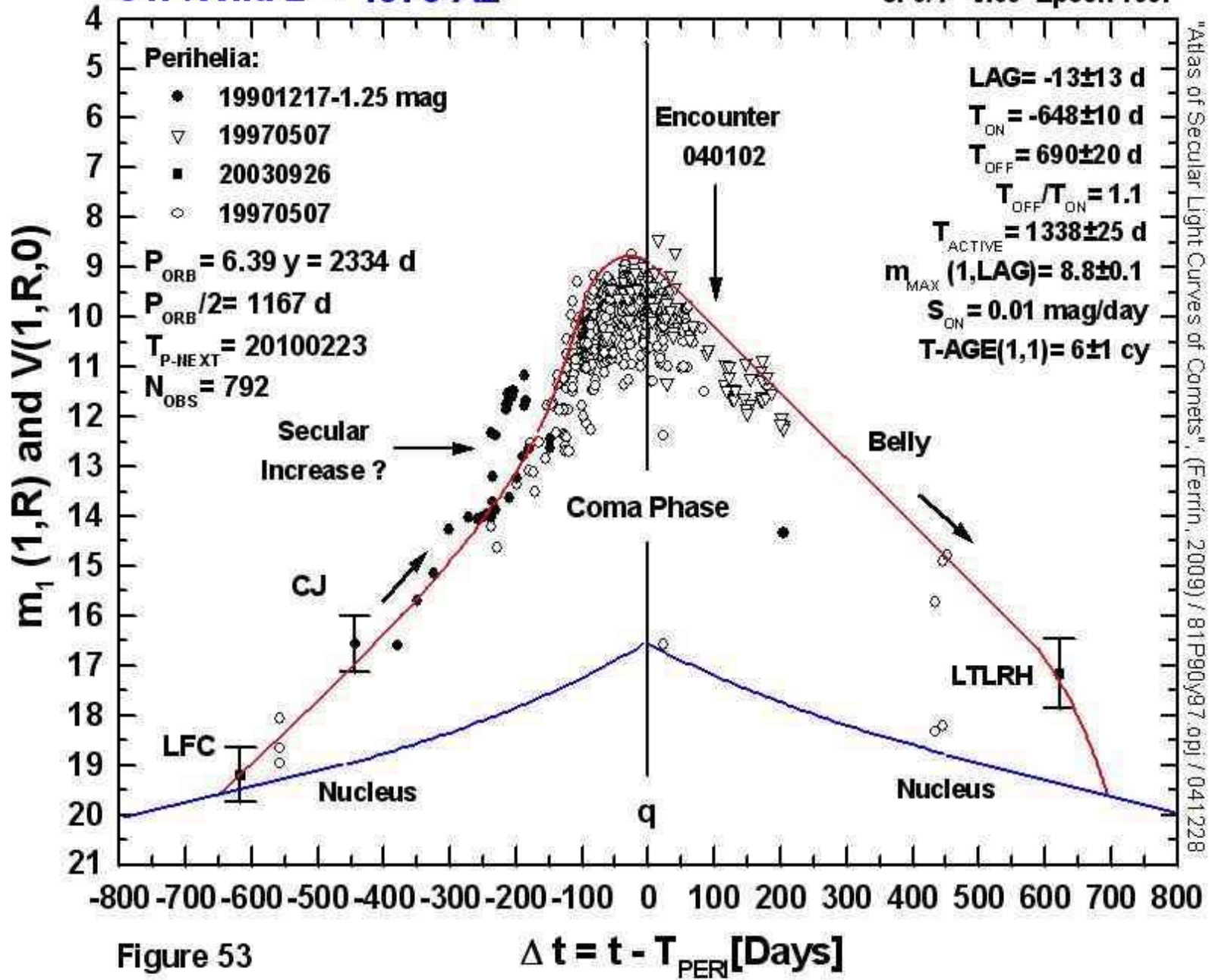

Figure 53

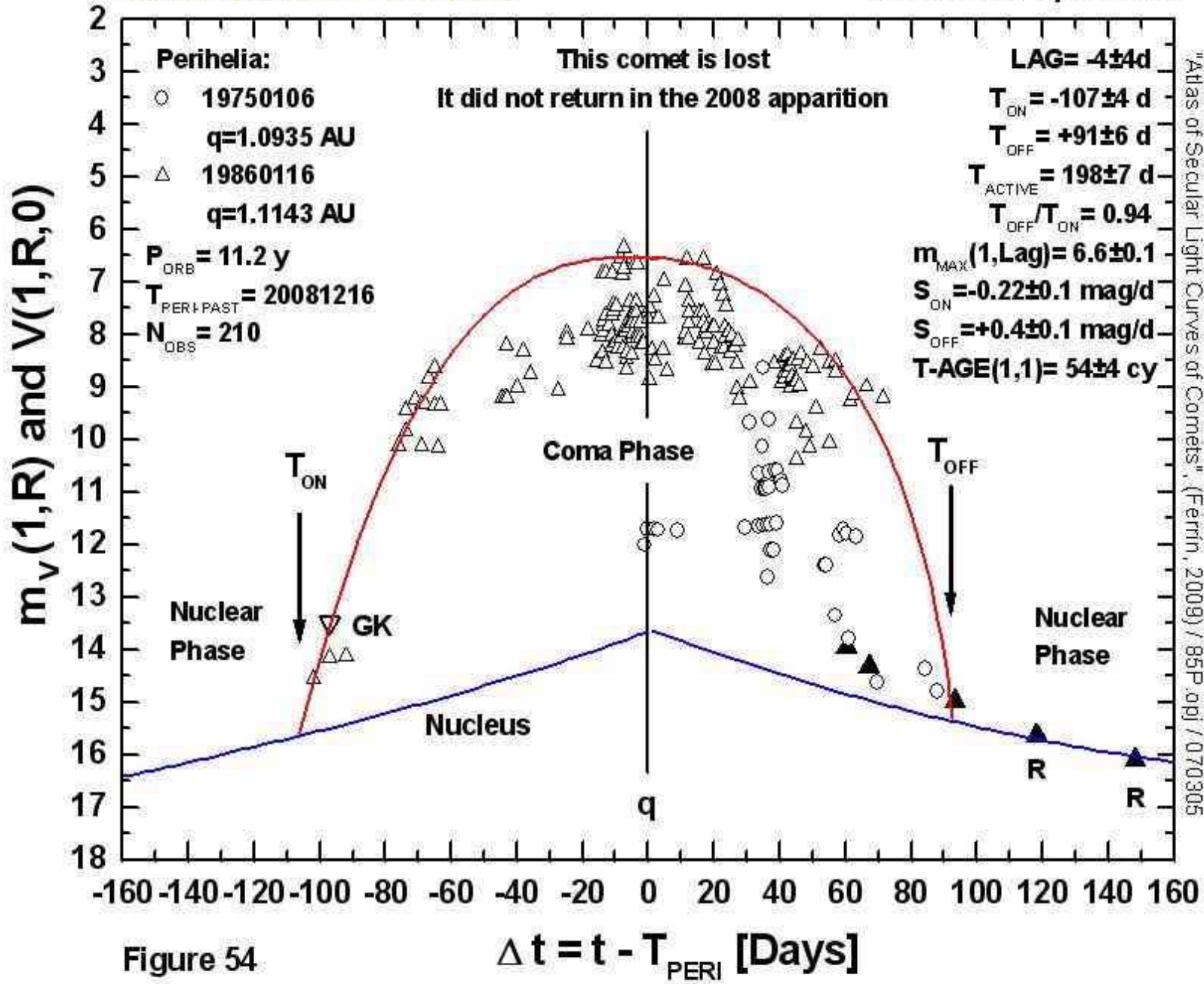

Figure 54

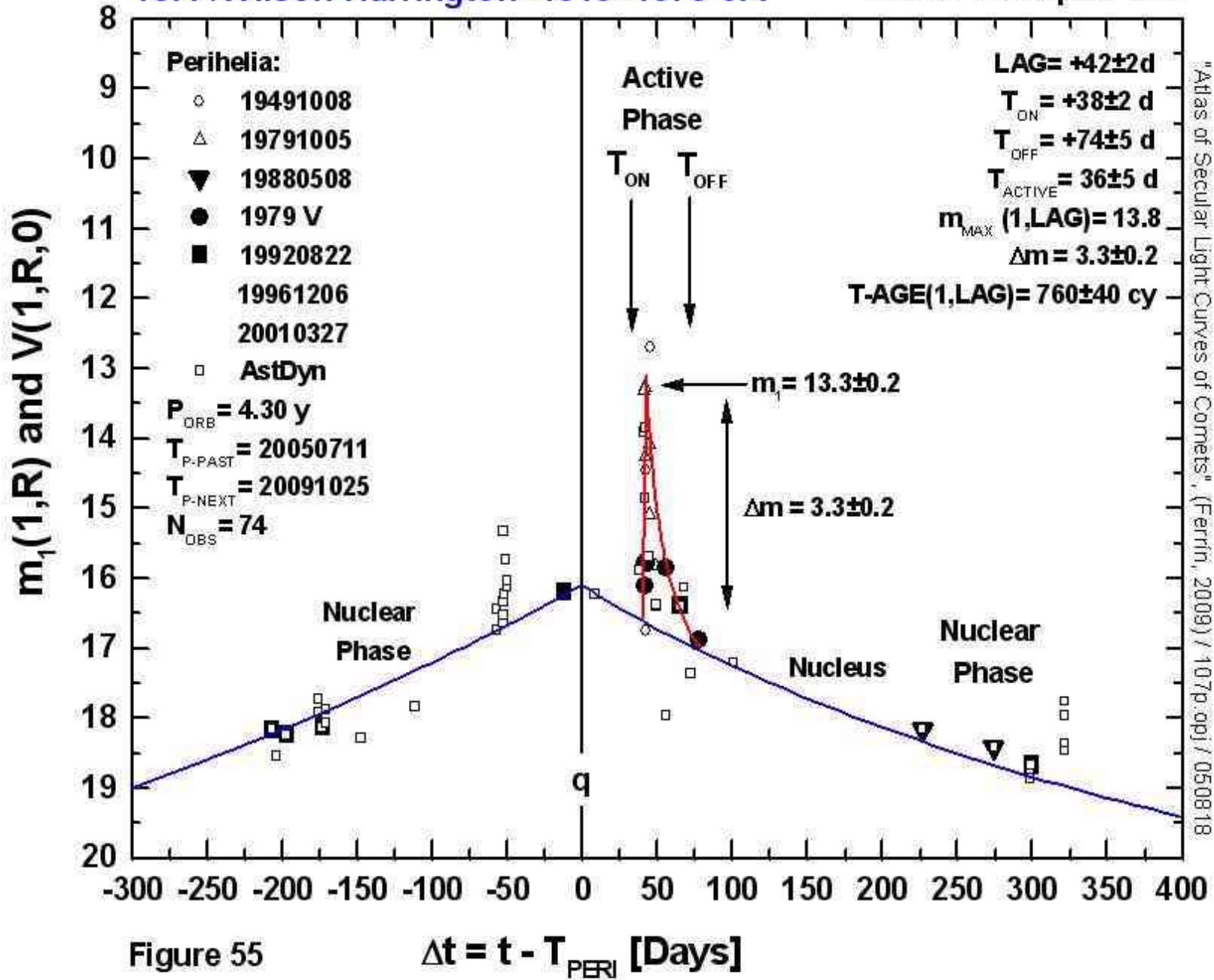

Figure 55

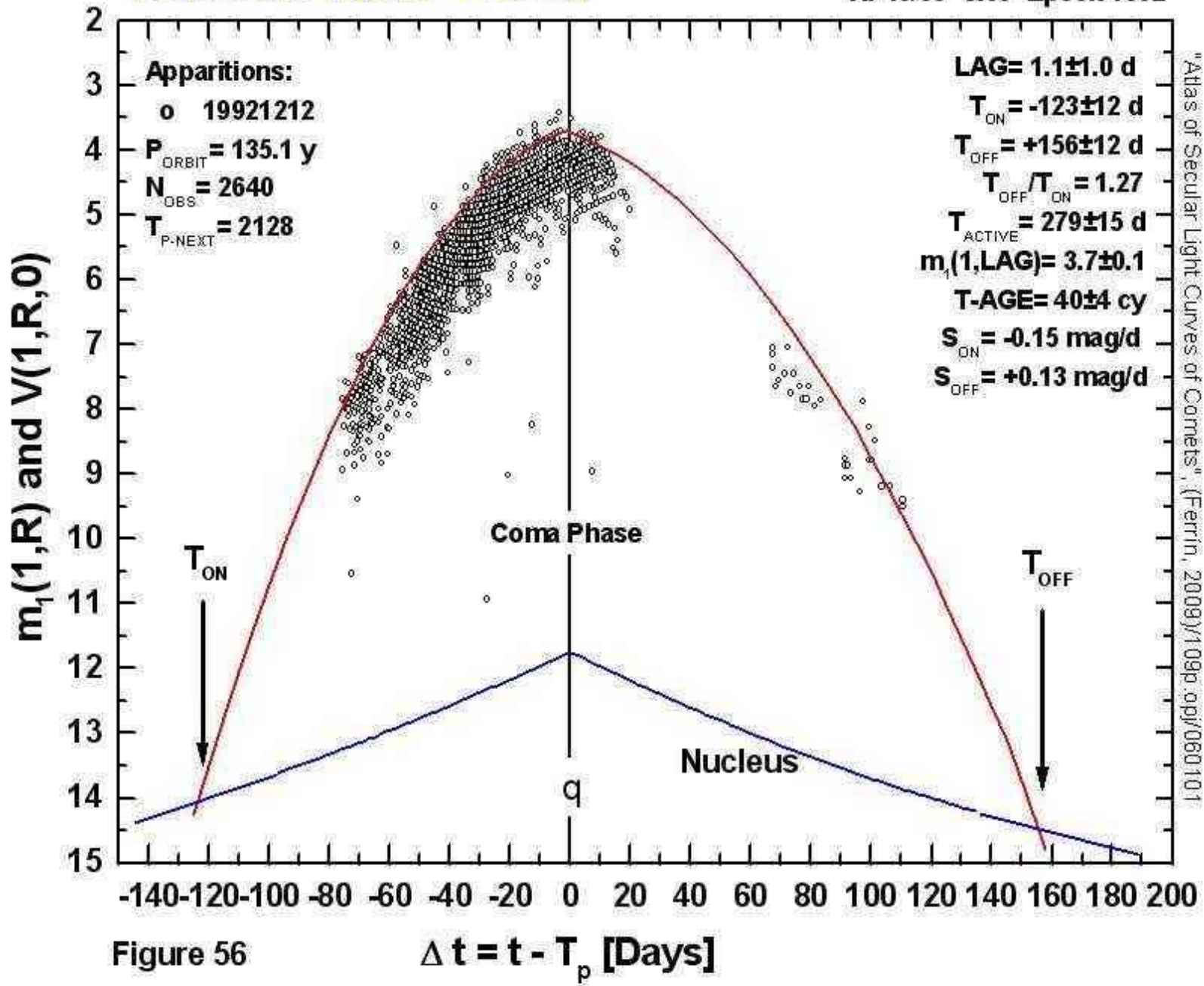

Figure 56

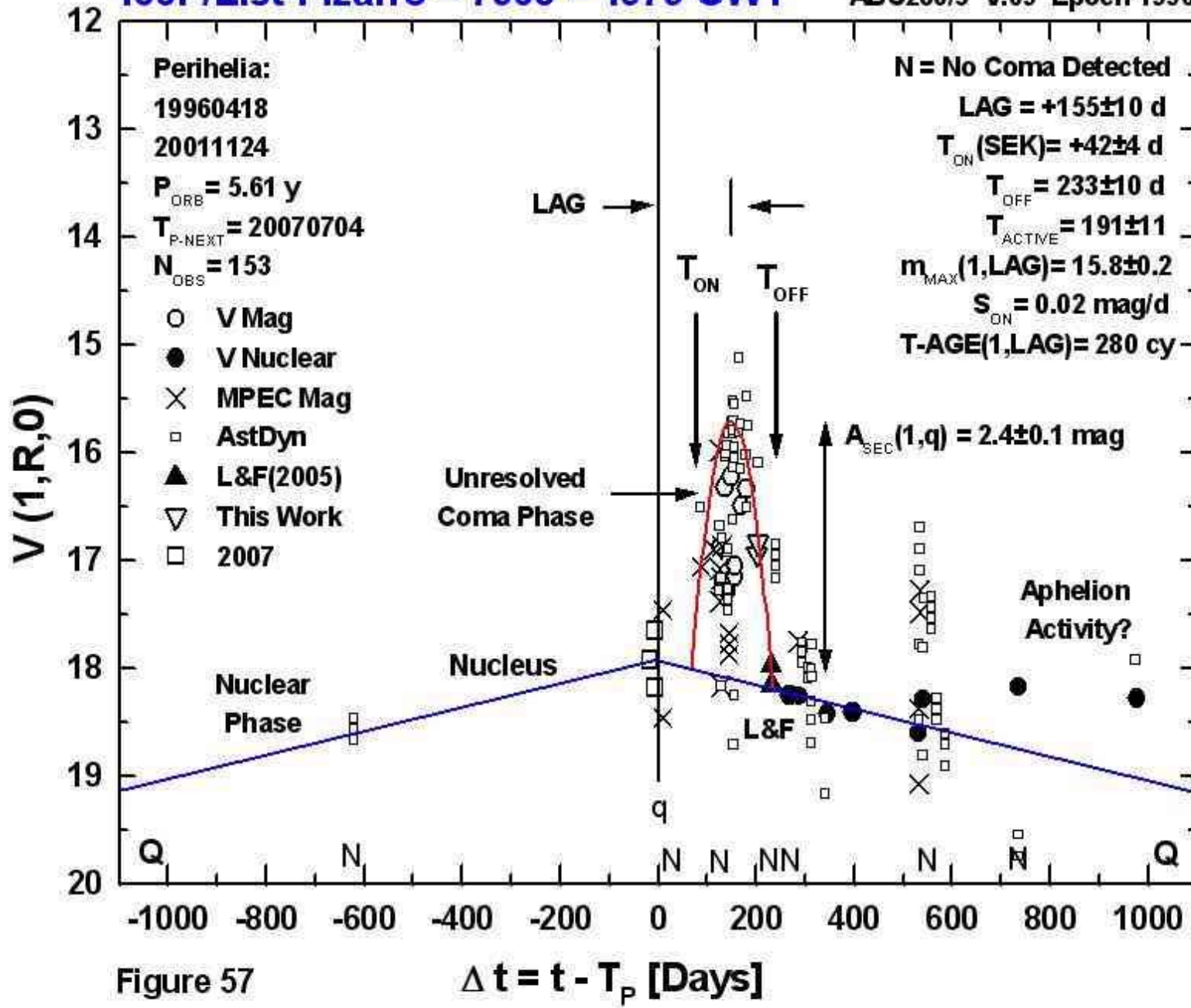

Figure 57

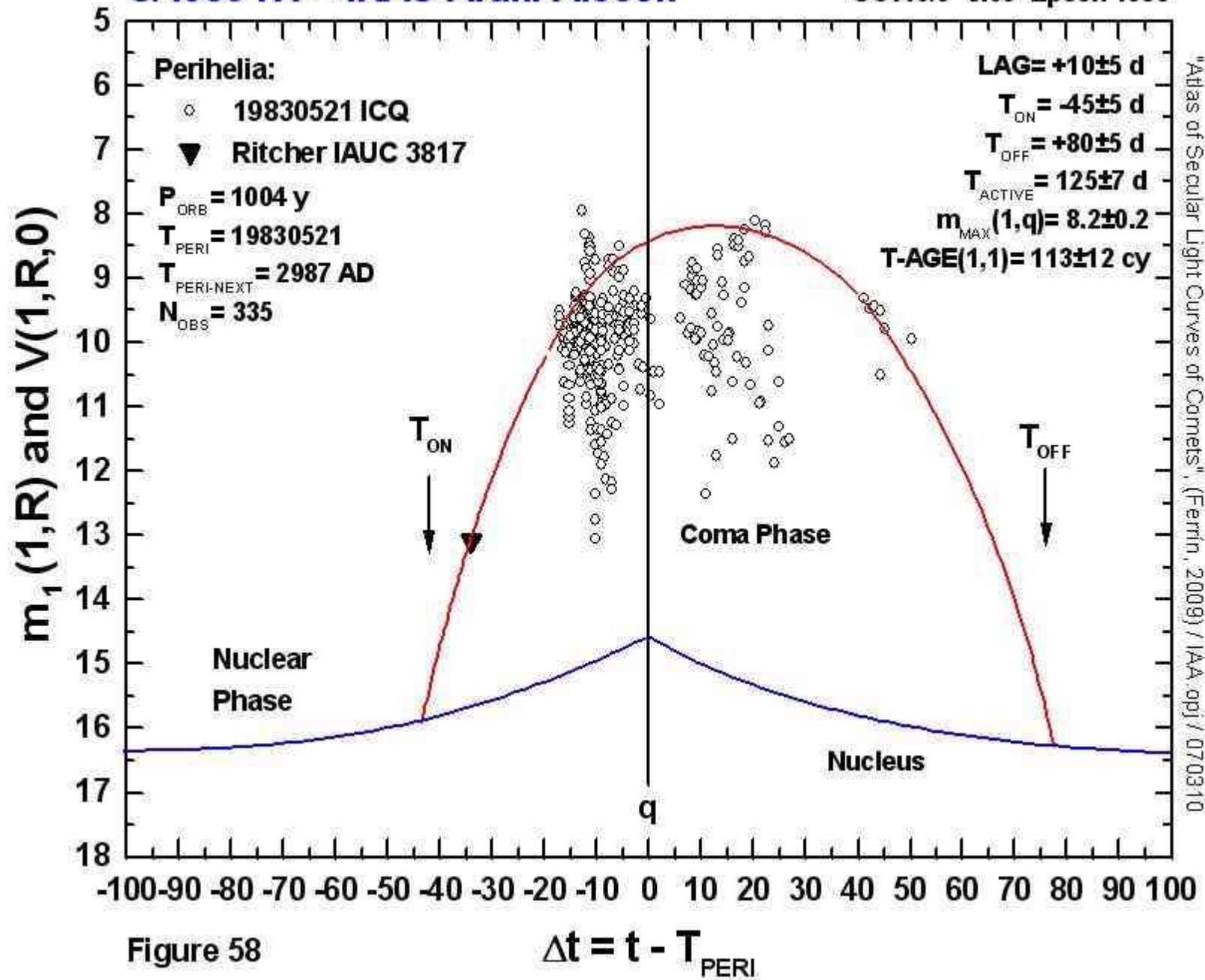

Figure 58

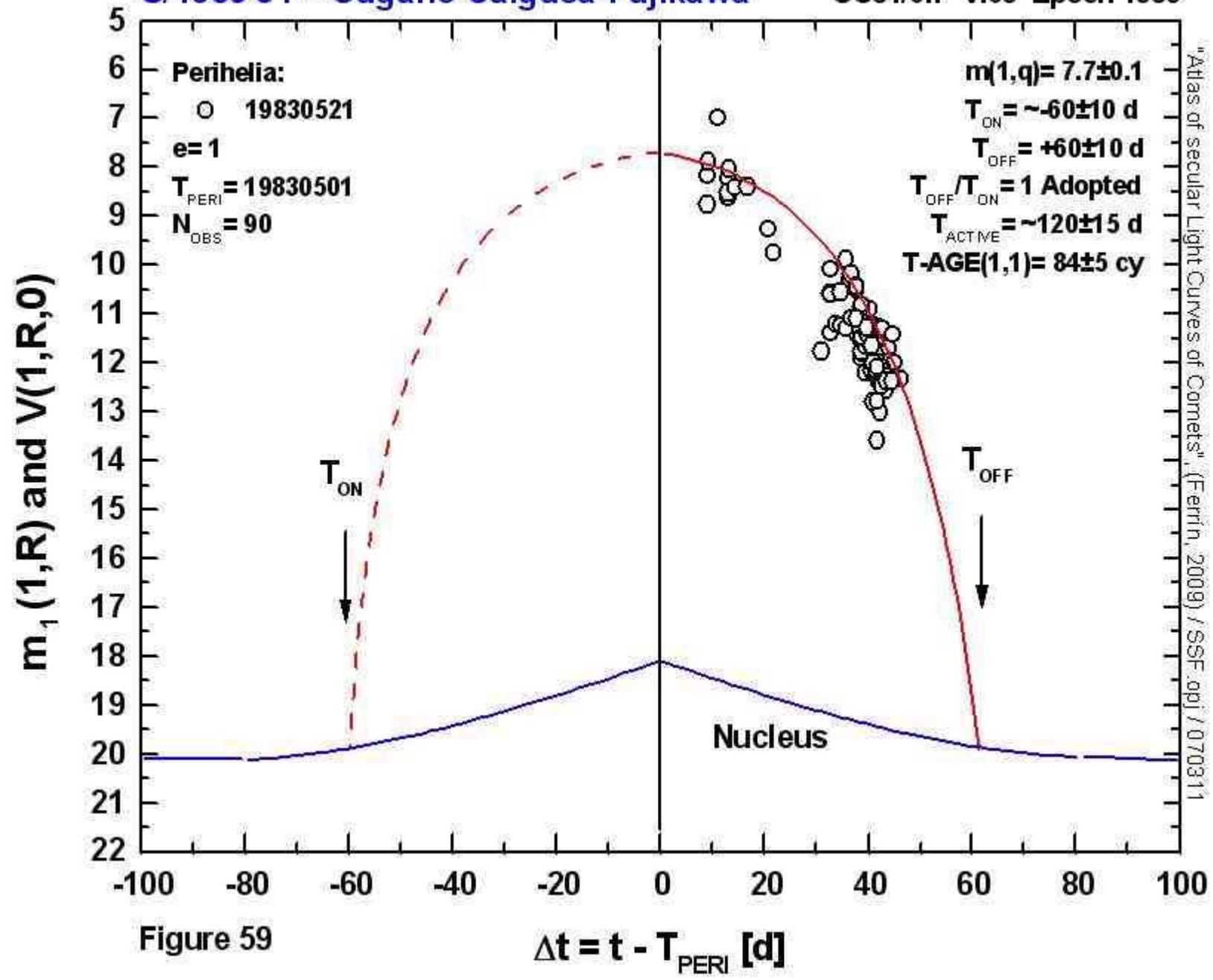

Figure 59

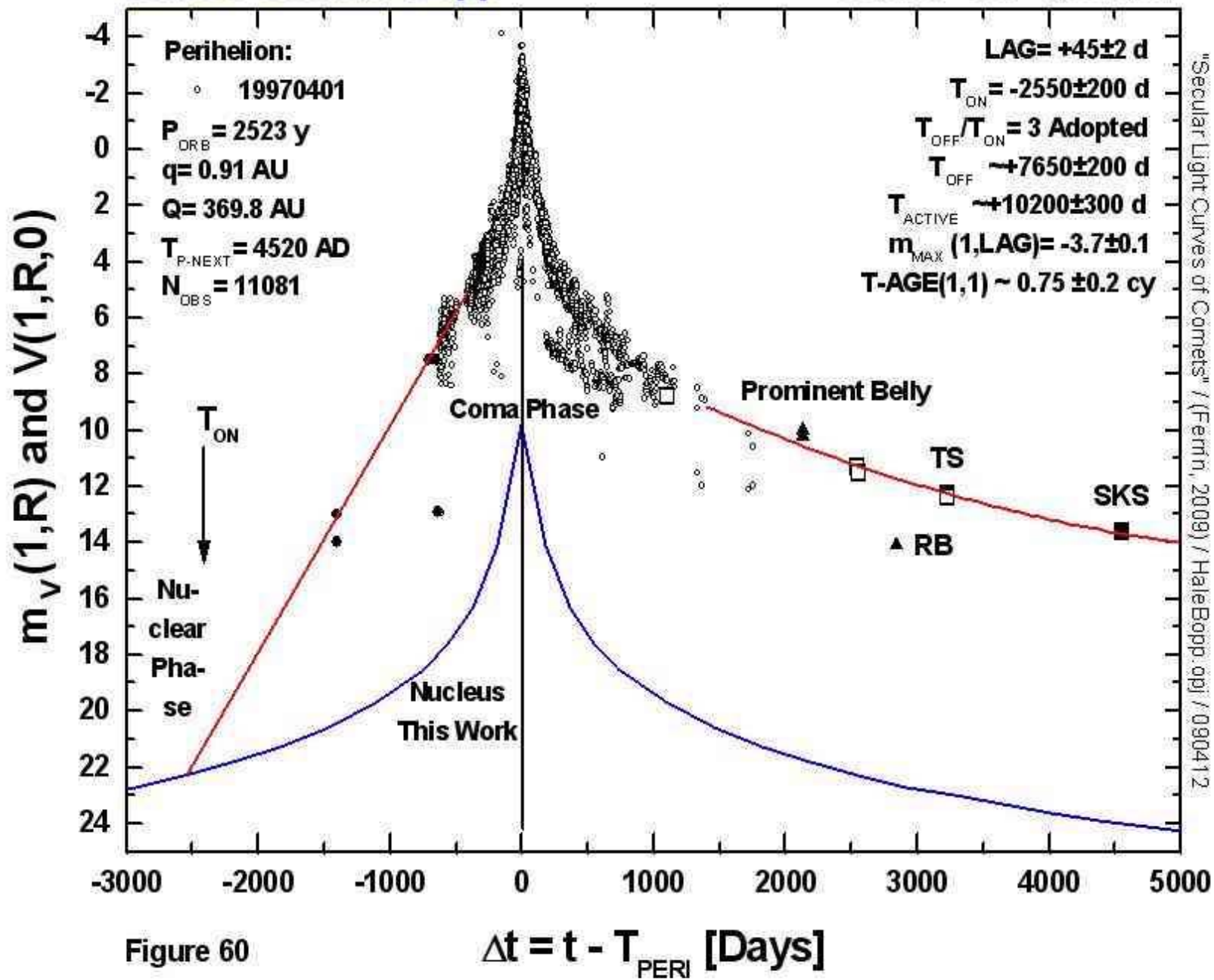

Figure 60

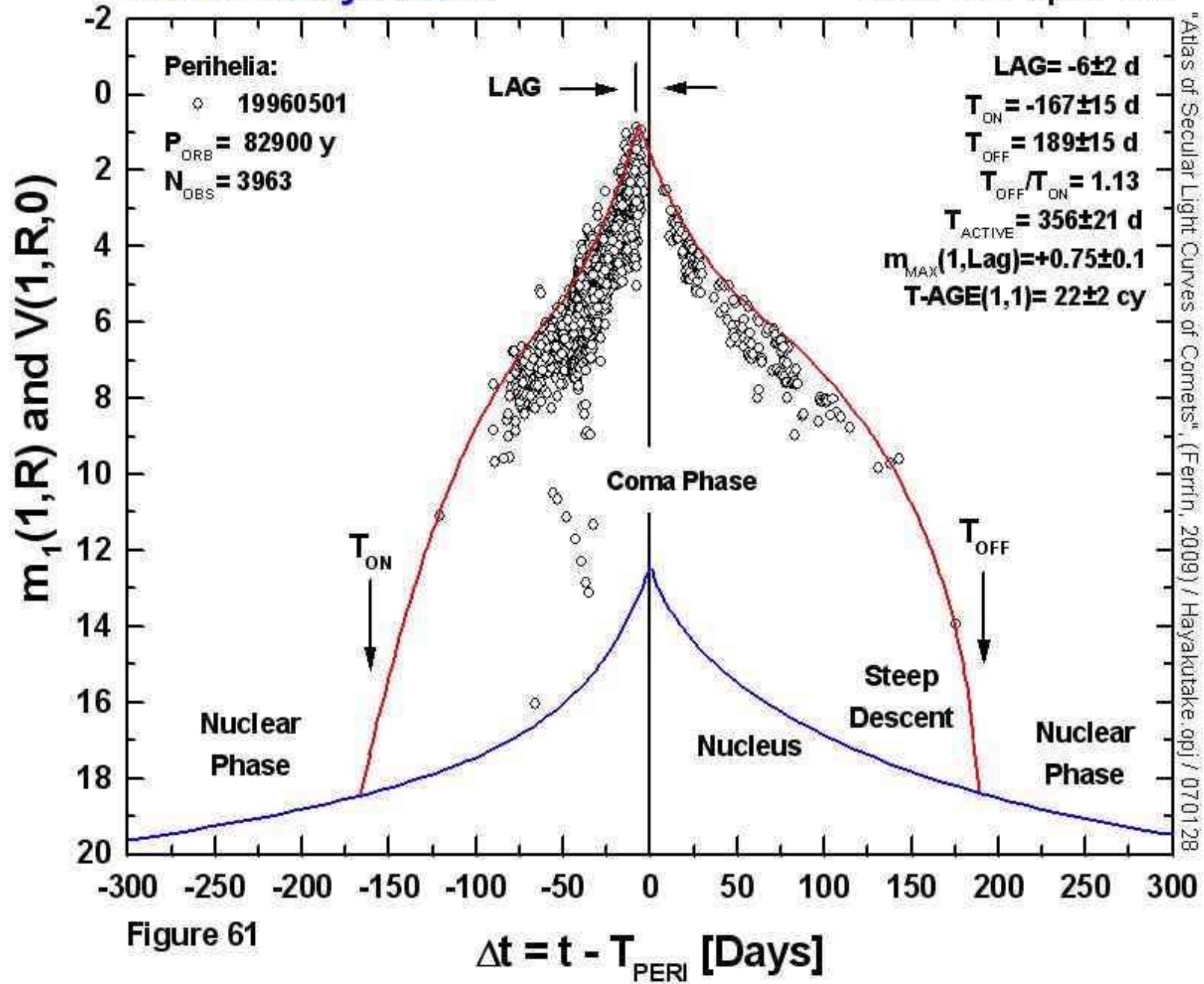

Figure 61

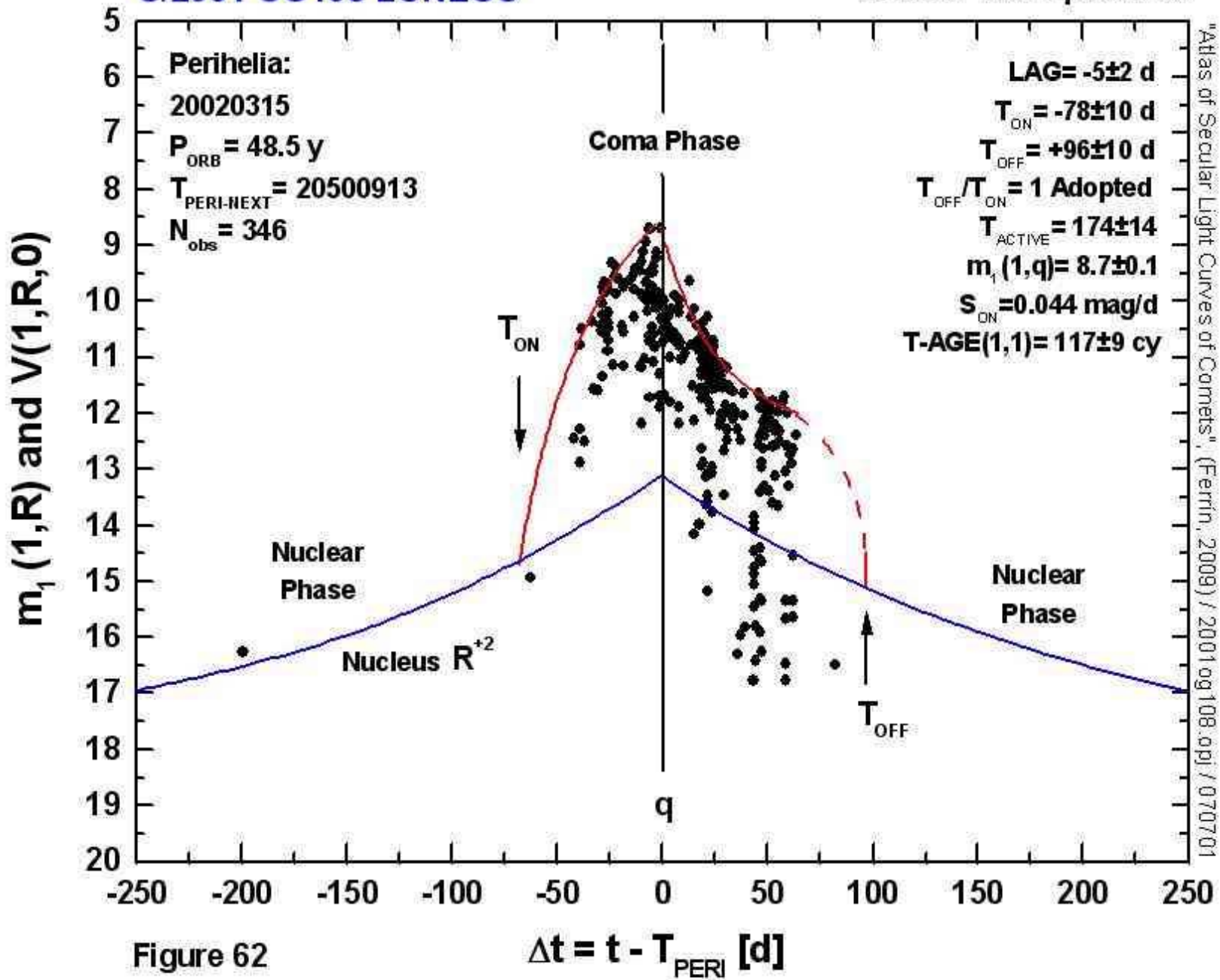

Figure 62

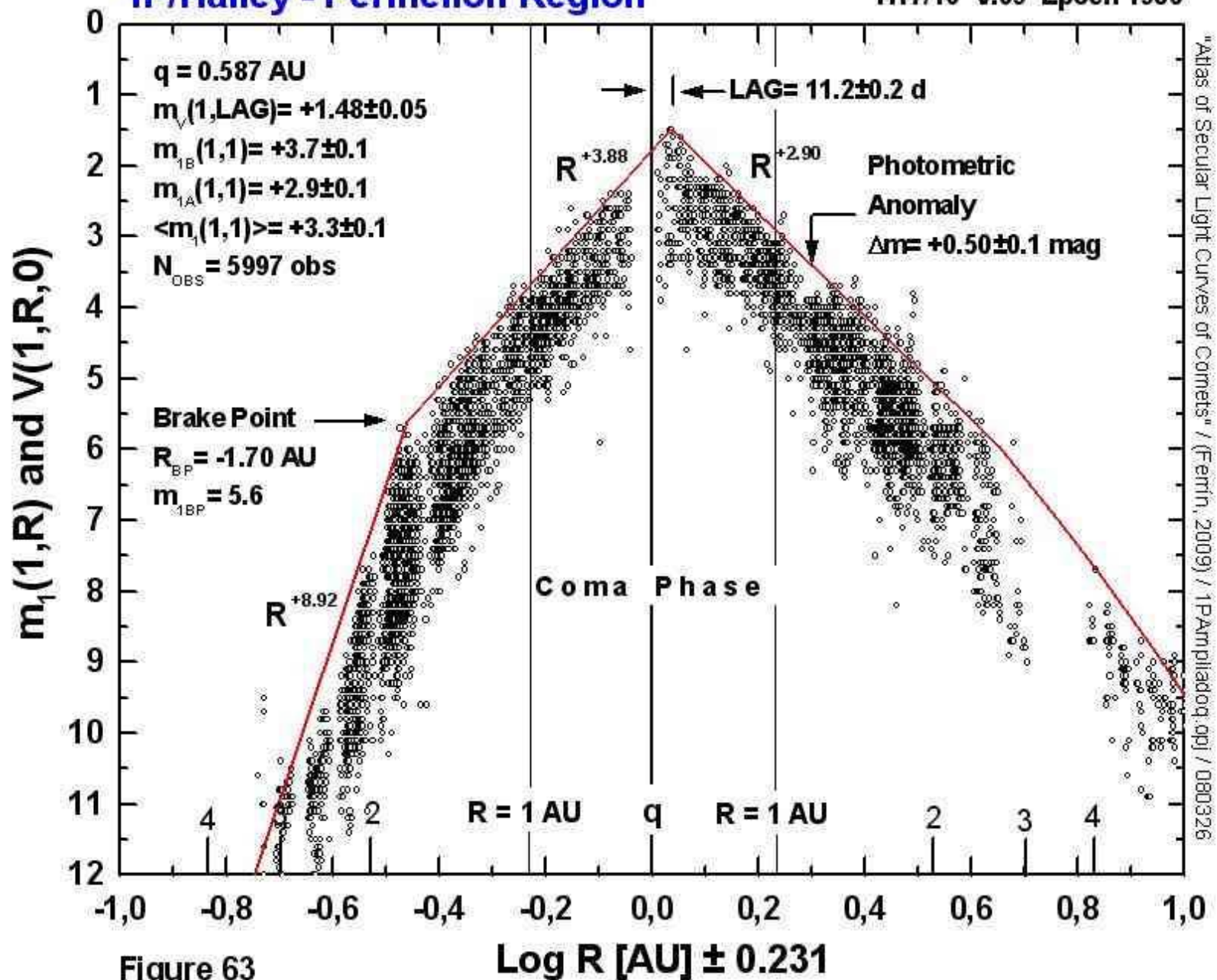

Figure 63

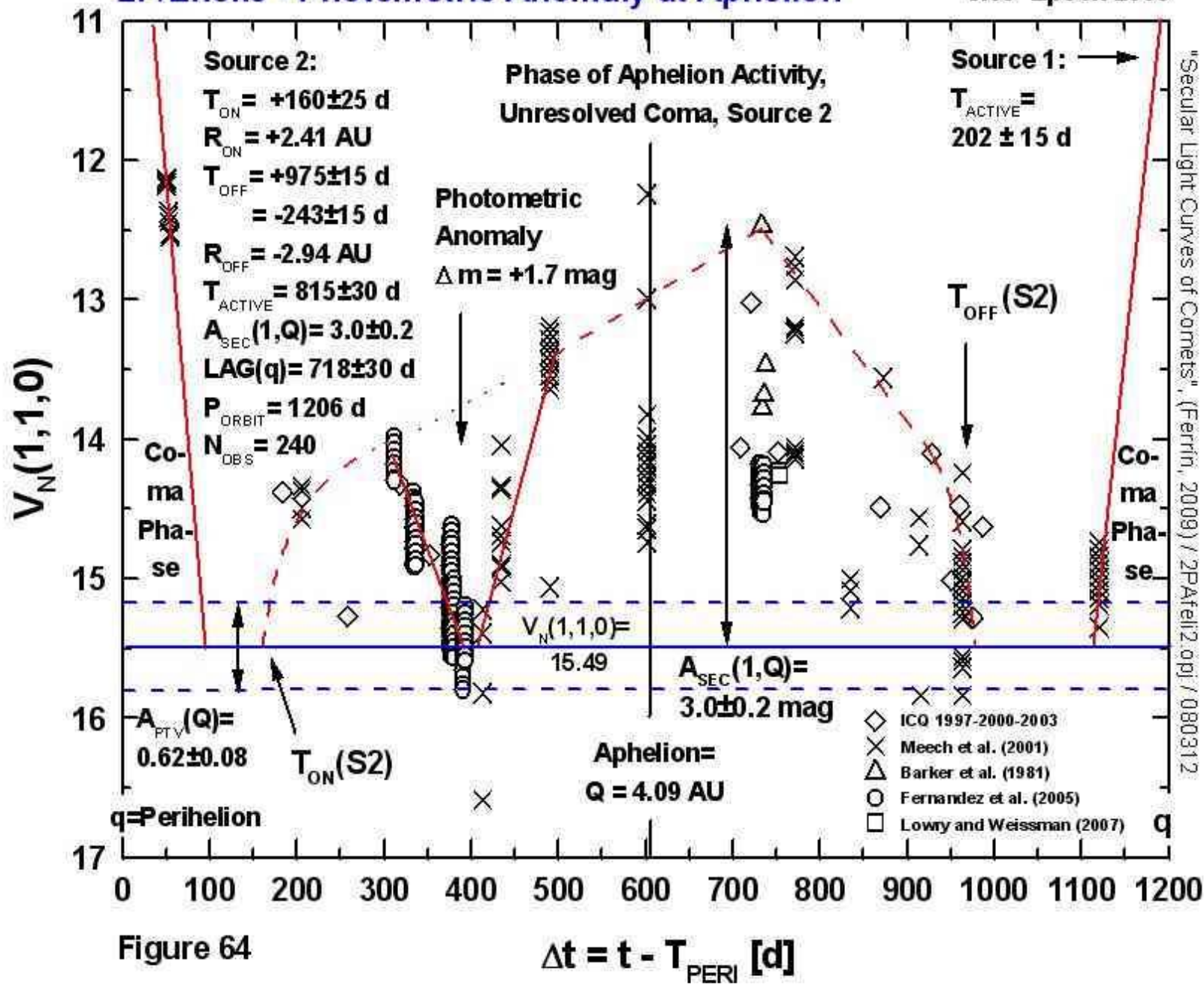

Figure 64

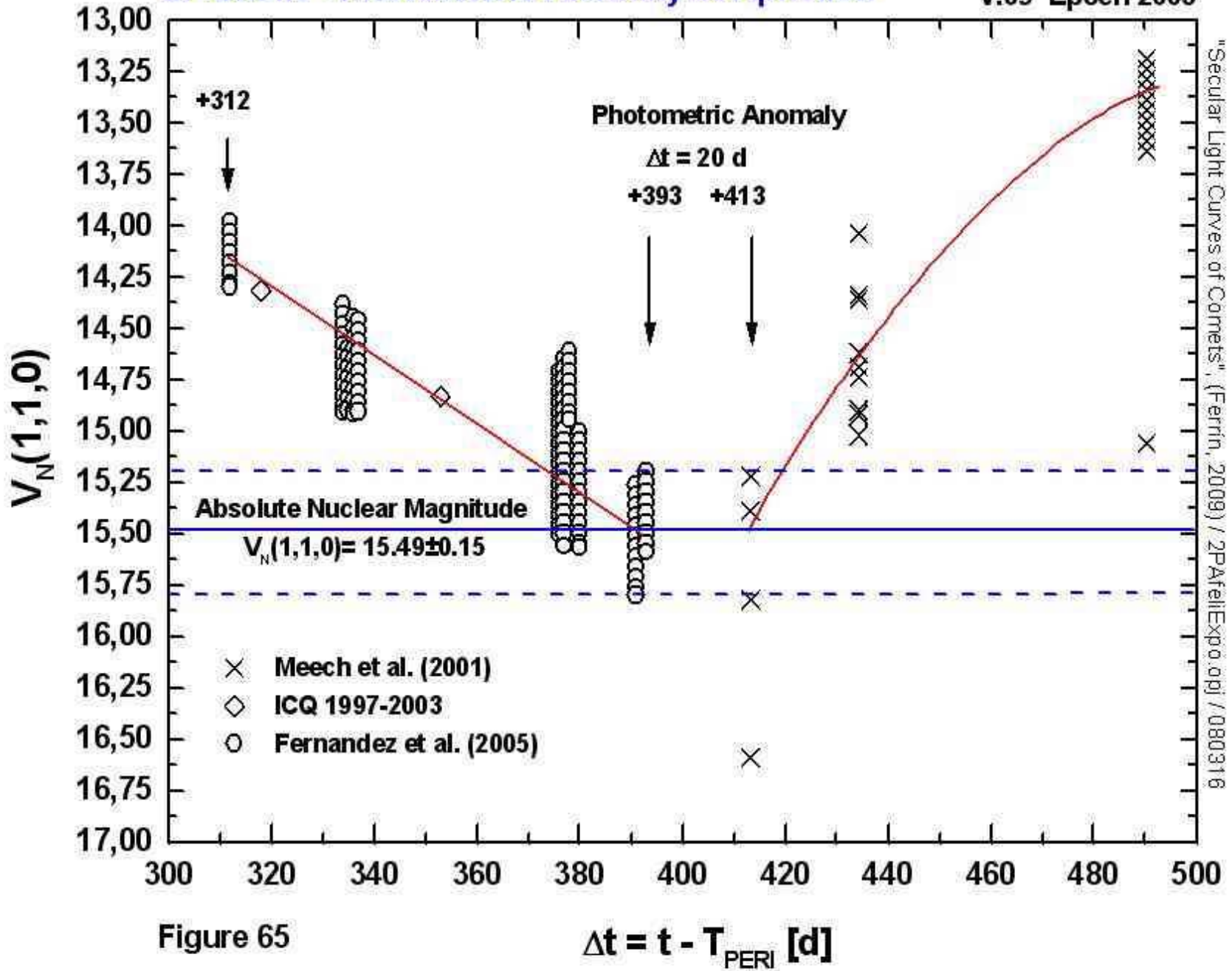

Figure 65

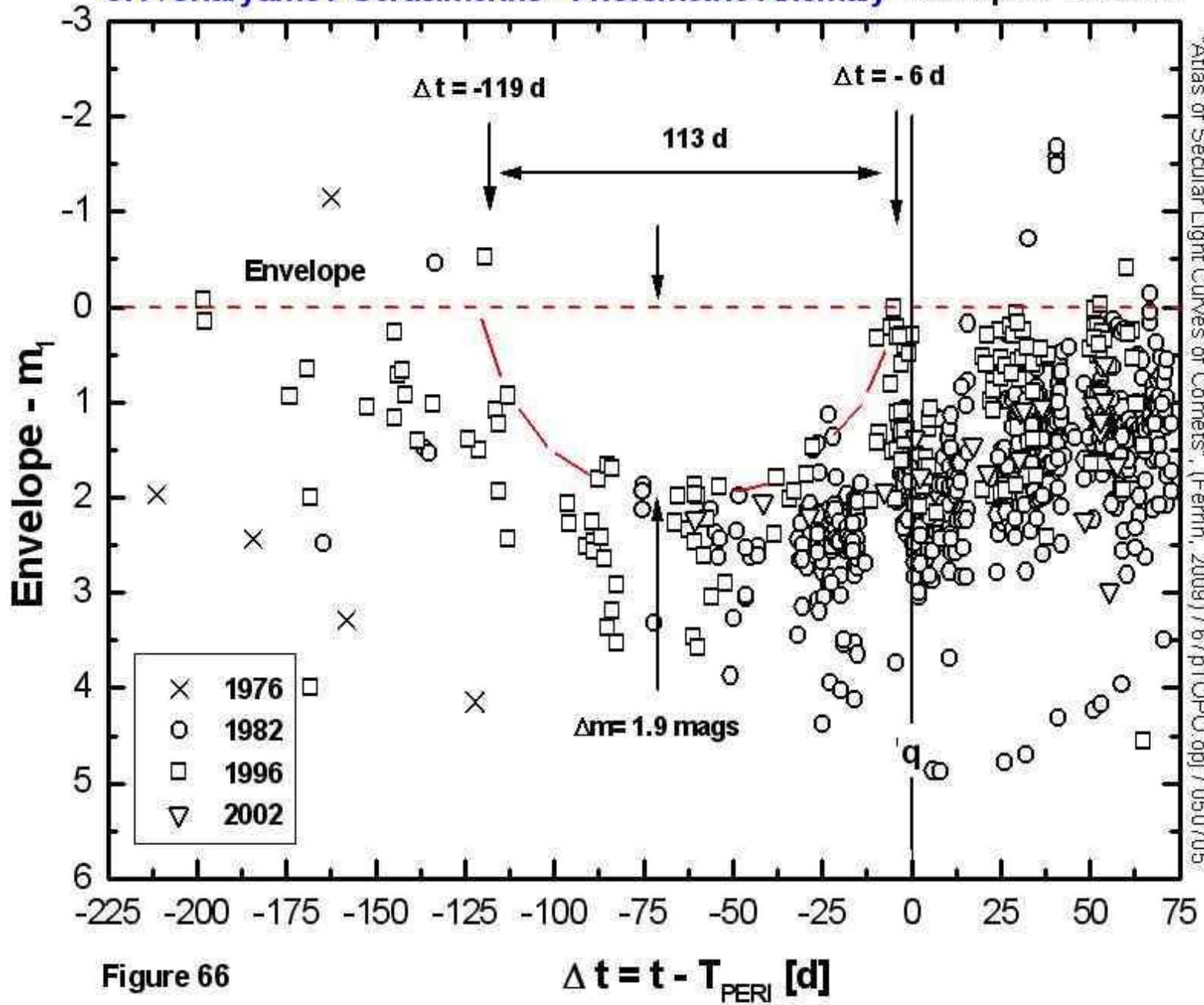

Figure 66

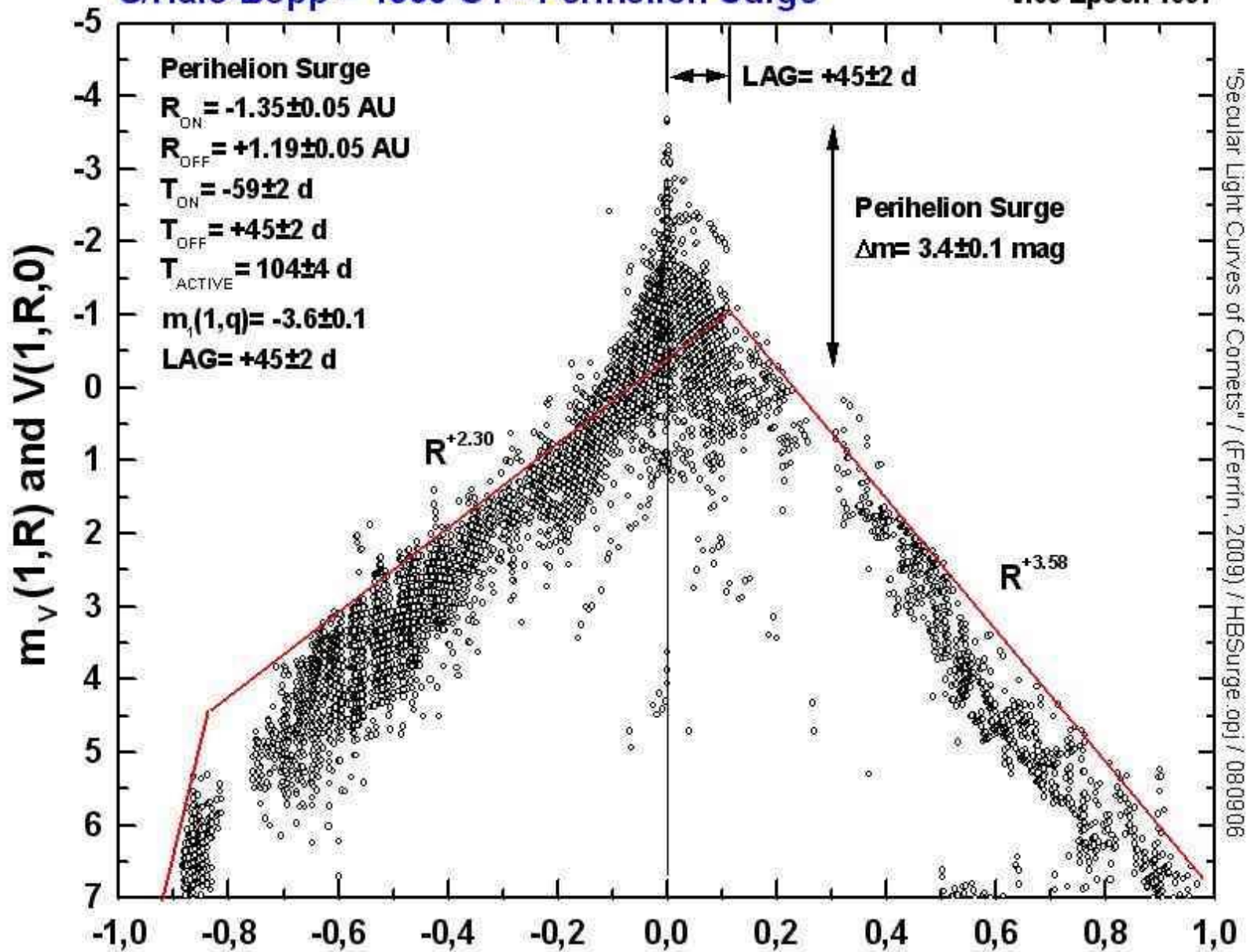

Figure 67

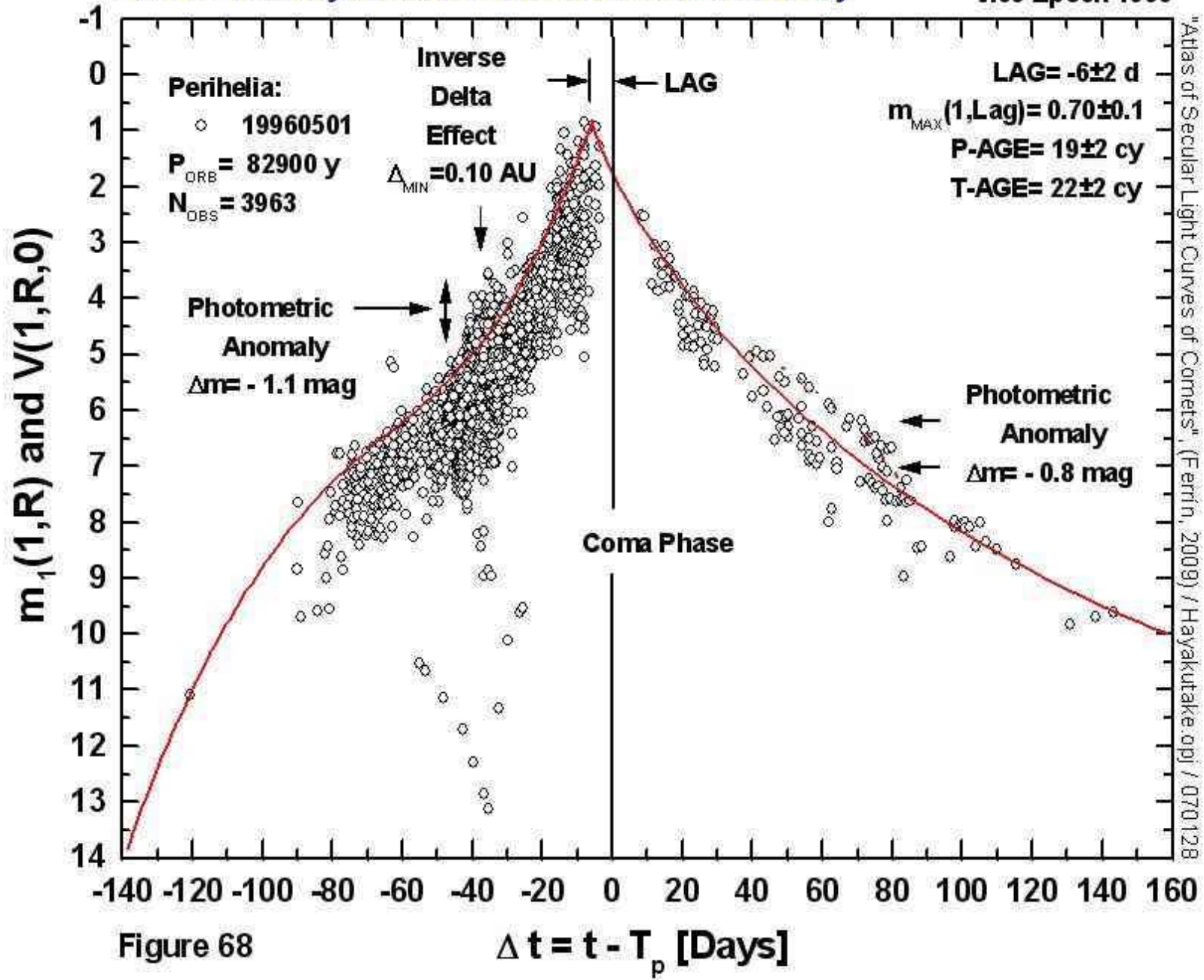

Figure 68